\documentclass[fleqn,usenatbib]{mnras}

\usepackage{newtxtext,newtxmath}

\usepackage[T1]{fontenc}

\DeclareRobustCommand{\VAN}[3]{#2}
\let\VANthebibliography\thebibliography
\def\thebibliography{\DeclareRobustCommand{\VAN}[3]{##3}\VANthebibliography}

\usepackage{graphicx}	
\usepackage{amsmath}	
\usepackage[toc,title,page]{appendix}

\usepackage{caption}
\usepackage{subcaption}
\usepackage{epstopdf}
\usepackage[english]{babel}
\usepackage{wrapfig}
\usepackage{pdflscape}
\usepackage{lscape}
\usepackage{longtable}
\usepackage{appendix}
\usepackage{textcomp}
\usepackage{multirow}
\bibpunct{(}{)}{;}{a}{}{,}
\usepackage{array}
\usepackage{ragged2e}
\usepackage{adjustbox}
\usepackage{gensymb}
\usepackage{newtxtext,newtxmath}

\title[Extreme radio variability in NLS1s]{Unprecedented extreme high-frequency radio variability in early-stage active galactic nuclei}

\author[E. Järvelä et al.]{
E. Järvelä,$^{1,2}$\thanks{E-mail: ejarvela@ou.edu}\thanks{Dodge Family Prize Fellow in The University of Oklahoma}
T. Savolainen,$^{3,4,5}$
M. Berton,$^{6}$
A. Lähteenmäki,$^{3,4}$
S. Kiehlmann,$^{7,8}$
T. Hovatta,$^{3,9}$ 
\newauthor
I. Varglund,$^{3,4}$
A. C. S. Readhead,$^{10}$
M. Tornikoski,$^{3}$
W. Max-Moerbeck,$^{11}$
R. A. Reeves,$^{12}$
S. Suutarinen$^{3,4}$
\\
$^{1}$Homer L. Dodge Department of Physics and Astronomy, The University of Oklahoma, 440 W. Brooks St., Norman, OK 73019, USA\\
$^{2}$ European Space Agency (ESA), European Space Astronomy Centre (ESAC), Camino Bajo del Castillo s/n, 28692 Villanueva de la Cañada, Madrid, Spain\\
$^{3}$ Aalto University Metsähovi Radio Observatory, Metsähovintie 114, 02540 Kylmälä, Finland \\
$^{4}$ Aalto University Department of Electronics and Nanoengineering, PO Box 15500, 00076 Aalto, Finland \\
$^{5}$ Max-Planck-Institut f\"ur Radioastronomie, Auf dem H\"ugel 69, 53121 Bonn, Germany \\
$^{6}$ European Southern Observatory (ESO), Alonso de Córdova 3107, Casilla 19, Santiago 19001, Chile \\
$^{7}$ Institute of Astrophysics, Foundation for Research and Technology-Hellas, GR-70013 Heraklion, Greece \\
$^{8}$ Department of Physics, Univ. of Crete, GR-70013 Heraklion, Greece \\
$^{9}$ Finnish Centre for Astronomy with ESO, University of Turku, Vesilinnantie 5, FI-20014, Finland \\
$^{10}$ Owens Valley Radio Observatory, California Institute of Technology, Pasadena, CA 91125, USA \\
$^{11}$ Departamento de Astronomía, Universidad de Chile, Camino El Observatorio 1515, Las Condes, Santiago, Chile \\
$^{12}$ Departamento de Astronomía, Universidad de Conceptión, Concepción, Chile
}

\date{Accepted XXX. Received YYY; in original form ZZZ}

\pubyear{2023}

\begin{document}
\label{firstpage}
\pagerange{\pageref{firstpage}--\pageref{lastpage}}
\maketitle

\begin{abstract}
We report on the discovery of one of the most extreme cases of high-frequency radio variability ever measured in active galactic nuclei (AGN), observed on timescales of days and exhibiting variability amplitudes of three to four orders of magnitude. These sources, all radio-weak narrow-line Seyfert 1 (NLS1) galaxies, were discovered some years ago at Aalto University Metsähovi Radio Observatory (MRO) based on recurring flaring at 37~GHz, strongly indicating the presence of relativistic jets. In subsequent observations with the Karl G. Jansky Very Large Array (JVLA) at 1.6, 5.2, and 9.0~GHz no signs of jets were seen. To determine the cause of their extraordinary behaviour, we observed them with the JVLA at 10, 15, 22, 33, and 45~GHz, and with the Very Long Baseline Array (VLBA) at 15~GHz. These observations were complemented with single-dish monitoring at 37~GHz at MRO, and at 15~GHz at Owens Valley Radio Observatory (OVRO). Intriguingly, all but one source either have a steep radio spectrum up to 45~GHz, or were not detected at all. Based on the 37~GHz data the timescales of the radio flares are a few days, and the derived variability brightness temperatures and variability Doppler factors comparable to those seen in blazars. We discuss alternative explanations for their extreme behaviour, but so far no definite conclusions can be made. These sources exhibit radio variability at a level rarely, if ever, seen in AGN. They might represent a new type of jetted AGN, or a new variability phenomenon, and thus deserve our continued attention.


\end{abstract}




\section{Introduction}

Approximately 10 per cent of active galactic nuclei (AGN) are capable of launching and maintaining relativistic jets \citep{2017padovani1}. Traditionally, these jetted AGN have been often identified using the radio loudness parameter{\footnote{Radio loudness parameter, $\emph{R}$, is defined as the ratio between 5~GHz flux density and optical $B$ band flux density. Sources with $\emph{R} >$ 10 are considered radio-loud, $\emph{R} <$ 10 radio-quiet \citep{1989kellermann1}.} as a proxy for the jet activity: all the jetted AGN were believed to be found among the radio-loud population. Whereas the radio loudness parameter might still serve a purpose when considering bright, high-redshift AGN with steady, powerful jets, and negligible host galaxy contribution, recent studies have shown that it utterly fails when faced with the true diversity of AGN jet phenomenon and variability \citep{2017padovani1, 2018lahteenmaki1}. This is especially problematic in the local Universe, where we are able to detect also lower-power jets and outflows in AGN, and where the host galaxy can have a major contribution to the low-frequency radio emission, such that disentangling different sources of radio emission poses a problem \citep{2015caccianiga1, 2017jarvela1, 2022jarvela1}. This can lead to AGN with low-power relativistic jets to be classified as radio-quiet, or non-jetted AGN with strong star formation to be classified as radio-loud \citep{2015caccianiga1}, making radio loudness a problematic proxy for the jet power and activity.

Especially one class of AGN, the narrow-line Seyfert 1 (NLS1) galaxies, have played a major role in revealing the diversity seen in AGN activity, and have revolutionised some long-standing assumptions held about AGN. NLS1s are identified based on the optical spectrum: the full-width at half maximum (FWHM) of their broad H$\beta$ emission line is $<$2000 km s$^{-1}$, and their [O~III] emission is weak compared to the broad H$\beta$: $S$([O~III])/$S$(H$\beta$)$<$3 \citep{1985osterbrock1, 1989goodrich1}. They often also exhibit strong Fe~II emission, confirming the unobstructed view of the central engine. The narrow FWHM(H$\beta$) can be attributed to low rotational velocity around a low-mass supermassive black hole (10$^{6}$--10$^{8} M_{\odot}$, \citealp{2011peterson1,2018komossa1}). The low-mass hypothesis is supported by reverberation mapping studies \citep{2016wang1,2018du1}, predominantly turbulence-dominated Lorentzian emission line profiles \citep[e.g.,][]{2011kollatschny1, 2000sulentic1, 2020berton1}, the existence of tidal disruption events in NLS1s \citep[e.g.,][]{2021frederick1}, and the prevalence of disk-like host galaxies with pseudo-bulges \citep[e.g.,][]{2017jarvela1, 2020olguiniglesias1, 2022varglund1}. The luminosities of NLS1s, comparable to those of higher black hole mass AGN, such as broad-line Seyfert 1 (BLS1) galaxies, combined with their lower black hole masses indicate that a considerable fraction of NLS1s are accreting close to or even above the Eddington limit \citep{1992boroson1}. This ensemble of properties has led to the conclusion that they are fast-growing, early-stage AGN \citep{2000mathur1}, possibly experiencing one of their first activity cycles.

Based on their properties, NLS1s were not expected to show prominent jet activity, as the ability to launch and maintain powerful relativistic jets was considered to be exclusively a property of massive elliptical galaxies, hosting the most massive black holes \citep{2000laor1}. However, contradictory to this jet paradigm several NLS1s were found to exhibit blazar-like properties in radio band \citep{2006komossa1, 2008yuan1}, and finally the first NLS1 was detected at gamma-rays -- indisputably produced by relativistic jets -- in 2009 \citep{2009abdo2}. Since then $\sim$20 NLS1s have been detected at gamma-rays \citep{2018romano1, 2019paliya1}, and several dozen new candidates have been identified \citep{2021foschini1, 2022foschini1}. Furthermore, additional $\sim$50 NLS1s have been confirmed to host jets via radio imaging \citep[e.g.,][]{2015richards1, 2016lister1, 2018berton1, 2020chen1, 2022chen1}. NLS1s with relativistic jets share similar properties with the non-jetted NLS1 population, and thus broke the jet paradigm beyond any doubt. These jetted NLS1s are also the first AGN with systematically high Eddington ratios to host relativistic jets. Blazars, in general, have Eddington ratios $<$ 0.1 \citep{2014heckman1}, and it was believed that AGN with Eddington ratios significantly higher than that are very rarely capable of launching jets, though some exceptions exist \citep[e.g.,][]{2022belladitta1}. Recently, also general relativistic (radiative) magnetohydrodynamics (GRRMHD) simulations have shown that efficient and powerful collimated jets are formed in systems with high Eddington ratios, even exceeding unity, if the state of magnetically-arrested accretion (MAD) is reached \citep{2017mckinney1, 2022liska1}. Thus it seems that our earlier beliefs regarding relativistic jets were mainly a product of observational biases, for example, concentrating the studies only on the brightest or radio-loudest AGN. It has been suggested that jetted NLS1s represent an early stage of the evolution of jetted AGN, and that they will eventually grow into flat-spectrum radio quasars (FSRQ) and radio galaxies \citep{2015foschini1, 2017berton1}. If this is the case, they offer us an unprecedented opportunity to study the very first stages in the evolution of powerful AGN with relativistic jets.

Intriguingly, the radio properties of NLS1s are very diverse: only 15 per cent of them have been detected in radio \citep{2006komossa1, 2015jarvela1}, and include a continuum of sources from host-dominated to relativistic jet-dominated \citep{2022jarvela1}, whereas the majority of 85 per cent seem to be totally radio-silent. However, NLS1 samples often suffer from misclassifications, and include a significant fraction of BLS1s and intermediate-type AGN that affect the population-wise statistics. Indeed, an ongoing investigation utilising a carefully selected sample of NLS1s and new radio surveys, such as the LOw-Frequency ARray (LOFAR) Two-metre Sky Survey (LoTSS) and the National Radio Astronomy Observatory (NRAO) Very Large Array Sky Survey (VLASS), indicates that the radio detection fraction among NLS1s is even lower, around $\sim$8 per cent (Varglund et al. in prep.). To understand the nature of this seemingly heterogeneous class and how different NLS1s are related, it is necessary to study the population as a whole. Most studies have concentrated on the most obvious radio-bright NLS1s, whereas the radio-faint and -silent population has been scarcely investigated. 

\subsection{The road so far}

A different approach was adopted at the Aalto University Metsähovi Radio Observatory (MRO, Finland), where several hundreds of jetted AGN are frequently monitored at 37~GHz. In addition to the usual suspects, that is, NLS1s that are bright in radio, two samples of NLS1s were selected for monitoring based on totally distinct criteria, independent of their radio properties. One sample consisted of NLS1s residing in very dense large-scale (Mpc-scale) environments, such as superclusters \citep{2017jarvela1}, and the other was compiled from NLS1s exhibiting spectral energy distributions (SED) that seemed favourable for 37~GHz observations. Eight NLS1s from these samples, four from each, were detected at flux density levels of several hundred mJy \citep{2018lahteenmaki1}. What makes these sources extraordinary is that most of them had been deemed to be radio-silent or had only very faint previous radio detections. Seven sources have been detected several times, strongly suggesting that these are genuine detections of recurrent radio flares in these sources. The most likely emission mechanism to produce such high-amplitude, rapid variability at a radio frequency this high is the synchrotron emission of a relativistic jet (however, see Sect.~\ref{sec:discussion}). Additional evidence was obtained when one of the sources was identified as a new gamma-ray emitter, and has since been seen brightening in X-rays soon after an MRO-detected flare \citep{2023romano1}.

Only two of these sources had previous radio detections, and only at mJy levels in the Faint Images of the Radio Sky at Twenty-Centimeters survey (FIRST) and the NRAO Very Large Array (VLA) Sky Survey (NVSS), while the rest were non-detections. To decipher this puzzling behaviour and to discriminate between the different hypotheses of their nature, the sources with several MRO detections were observed with the Karl G. Jansky VLA (JVLA) in A-configuration in L, C, and X bands. Instead of clarifying the situation, these observations raised more questions. Two of the sources were non-detections and the remaining sources had flux densities ranging from a few tens of $\mu$Jy to a few mJy, all of them consistently showing steep spectra below 9~GHz \citep[see Fig.~6 in][]{2020berton2}. Three of them showed slightly extended radio morphology. In a closer inspection, exploiting spatially resolved spectral index maps, it was found that at least two of these sources show signs of flat core spectrum \citep{2021jarvela1} and thus the presence of a partially optically thick radio core. The JVLA and the MRO observations are not simultaneous, but such an extreme, similar behaviour observed in several sources indicates that it is real, not just a curiosity.

However, the beam size of MRO ($\sim$2 arcmin) is considerably larger than the beam size of the JVLA in A-configuration ($\sim$arcsec-scale). It is therefore important to consider the possibility that the discrepancy between the flux densities of the JVLA and MRO could arise from different beam sizes. This seems improbable when taking into account the properties of the emission. Due to the redshift of these sources the JVLA observations probe kpc-scale structures. The angular sizes of these sources in the optical band are between 2 and 12~arcsec, so we were able to see the whole galaxy in the JVLA observations, in which the smallest field of view -- at 9~GHz -- was 4.7~arcmin. It is hard to explain such strong and variable radio emission in the outskirts of, or even outside, a galaxy. Due to the rapid variability, indicating a small emitting region, it is highly improbable that resolved-out structures could be responsible for this emission. Furthermore, contamination by nearby sources was ruled out in \citet{2018lahteenmaki1}. It can thus be assumed that the JVLA and MRO probe the same phenomenon. The effects of different beam sizes is further discussed in Sect.~\ref{sec:obseffects}.

Since the low frequency flux densities are consistent with FIRST there is no need to assume that these NLS1s have undergone drastic changes, for example, triggering of jets, but it cannot be ruled out either. Thanks to the MRO data we know these sources most likely host relativistic jets, but their radio emission below 9~GHz (X band) seems to be consistent with star formation, with little or no contribution from the AGN. Extrapolating, or even assuming a flat radio spectrum up to 37~GHz would mean that in the quiescent state the flux density would be less than a mJy, which, in the most extreme case, would require a nine-thousand-fold increase during flares. This would be very extreme, and a more plausible explanation is that the spectrum turns inverted at some point above 9~GHz, as indicated by the MRO data. This kind of behaviour is commonly seen in kinematically young AGN, for example, high-frequency peakers and gigahertz-peaked sources \citep{2021odea1}. In these sources the convex radio spectrum is explained by synchrotron self-absorption (SSA) in a young, parsec-scale jet. However, even in these sources the peak frequency does not usually exceed 10-15~GHz, which in contrast seems to be the case in our sources. 

An alternative to SSA could be free-free absorption (FFA), which also allows more inverted spectral indices than SSA \citep{1993rodriguez1}, requiring less extreme variability at 37~GHz. Some cases where the turnover frequency stays consistently high have been found \citep[tens of GHz,][]{2016doi1}, and usually this behaviour is explained by FFA. This could be the case also in these NLS1s: if these sources are kinematically young AGN, FFA could happen in the shocked ionised ambient clouds in front of the jet head \citep{2021odea1}. Alternatively, the required ionised gas could be provided by the enhanced circumnuclear star formation activity often seen in NLS1s \citep{2010sani1, 2022winkel1}. Either way, these NLS1s with jets that are almost totally absorbed at low radio frequencies are the last nail to the coffin of the radio loudness parameter as a universal proxy for the jet activity of AGN, and urge us to expand our horizons when it comes to our understanding of the diversity of AGN jets. 

To discern between these alternatives, we observed seven of these sources with the JVLA in X, Ku, K, Ka, and Q bands. These observations were complemented by Very Long Baseline Array (VLBA) observations at 15~GHz, and single-dish observations at 15 and 37~GHz, using the OVRO 40m telescope and the MRO telescope, respectively. In Sect.~\ref{sec:sample} we introduce the sample, in Sects.~\ref{sec:jvla} through \ref{sec:archival} we describe the performed observations, and the data reduction and analysis, in Sect.~\ref{sec:results} we present our results, in Sect.~\ref{sec:discussion} these results and their implications are discussed, and in Sect.~\ref{sec:summary} we provide a brief summary of this work. Throughout this paper, we adopt a standard $\rm\Lambda CDM$ cosmology, with a Hubble constant $H_0 = 72$~km s$^{-1}$ Mpc$^{-1}$, and $\Omega_\Lambda = 0.73$. 
 
\section{Sample}
\label{sec:sample}

The sample includes seven radio-weak NLS1s repeatedly detected at Jy-level flux densities at 37~GHz at MRO. The eighth such source was dropped because it was detected only once. Originally these sources were selected for the MRO AGN monitoring based on their dense large-scale environments \citep{2017jarvela1} or SEDs that suggested that they could be detectable at high radio frequencies \citep{2015jarvela1}. The basic properties of the sample are summarised in Table~\ref{tab:basicdata}. These sources are very similar to the general NLS1 population: all have a black hole mass less than 10$^8 M_{\odot}$ \citep{2015jarvela1, 2018lahteenmaki1}, and six of them are hosted in a disk-like host galaxy \citep{2018jarvela1, 2020olguiniglesias1, 2022varglund1}, whereas the morphology of the highest-$z$ source is unknown.

\begin{table*}
\caption{Basic properties of the sample. Columns: (1) source name in the SDSS, the superscript indicates the band the coordinates are from, $^\textrm{G}$ stands for Gaia; (2) short name; (3, 4) right ascension and declination (J2000); (5) redshift; (6) scale at the redshift of the source; (7) logarithmic black hole mass, taken from \citet{2018lahteenmaki1}; (8) large-scale environment, taken from \citet{2017jarvela1}; (9) host galaxy morphology, PB = pseudo-bulge, taken from $^a$ \citet{2018jarvela1}, $^b$ \citet{2020olguiniglesias1}, $^c$ \citet{2022varglund1}.}
\centering
\footnotesize
\begin{tabular}{lcccccccc}
\hline\hline
SDSS Name                    & Short alias & RA           & Dec          & $z$   & Scale   & log $M_{\rm{BH}}$ & Large-scale  & Host    \\ 
                             &             & (hh mm ss.s) & (dd mm ss.s) &       & (kpc/") & ($M_{\odot}$)     & environment  &      \\ \hline
J102906.69+555625.2$^{\textrm{G}}$ & J1029+5556  & 10 29 06.69  & +55 56 25.25 & 0.451 & 5.662   & 7.33  & supercluster & --       \\
J122844.81+501751.2$^{\textrm{Ka}}$ & J1228+5017 & 12 28 44.82  & +50 17 51.24 & 0.262 & 3.957   & 6.84  & supercluster & disk$^c$     \\
J123220.11+495721.8$^{\textrm{X}}$ & J1232+4957  & 12 32 20.12  & +49 57 21.82 & 0.262 & 3.957   & 7.30  & supercluster & disk$^c$   \\
J150916.18+613716.7$^{\textrm{G}}$ & J1509+6137  & 15 09 16.17  & +61 37 16.80 & 0.201 & 3.235   & 6.66  & void         & disk$^c$    \\
J151020.06+554722.0$^{\textrm{X}}$ & J1510+5547  & 15 10 20.05  & +55 47 22.11 & 0.150 & 2.550   & 6.66  & intermediate & disk, bar$^a$    \\
J152205.41+393441.3$^{\textrm{Ka}}$ & J1522+3934 & 15 22 05.50  & +39 34 40.45 & 0.077 & 1.420   & 5.97  & void         & disk, bar, PB$^a$, merger \\
J164100.10+345452.7$^{\textrm{Ka}}$ & J1641+3454 & 16 41 00.10  & +34 54 52.67 & 0.164 & 2.746   & 7.15  & intermediate & disk$^b$     \\ \hline 
\end{tabular} \label{tab:basicdata}
\end{table*}

\section{Data}

\subsection{Karl G. Jansky Very Large Array}
\label{sec:jvla}

\subsubsection{Observations and pre-processing}

We observed our sample with the JVLA in A-configuration in five different bands, X, Ku, K, Ka, and Q, centred at 10, 15, 22, 33, and 45~GHz, respectively (Project VLA/22A-002, PI Järvelä). The dates and integration times of the JVLA observations are given in Table~\ref{tab:jvlaobssummary}. The total bandwidth was 4~GHz in X, 6~GHz in Ku, and 8~GHz in K, Ka, and Q band, each band divided to 128~MHz subbands, consisting of 64 channels of 2~MHz. The NLS1 \citep{2017berton1} 3C 286 was used as the bandpass and flux density calibrator for each source, and each source had an individual nearby, bright source that was used as the complex gain calibrator. The pointing offset calibration was done either at 3C 286 or the current complex gain calibrator. The expected thermal noise levels were 7, 7, 12, 12, and 25~$\mu$Jy beam$^{-1}$ in X, Ku, K, Ka, and Q, respectively. We were able to reach these levels in most cases.

We used the Science Ready Data Products (SRDP) provided by the NRAO. The data were calibrated using the VLA Imaging Pipeline 2022.2.0.64. In addition, the data were checked manually and any remaining bad data were flagged, producing the SRDP measurement set for each source. We also re-checked all the data manually, but no additional flagging was required. In further data processing and analysis we used the Common Astronomy Software Applications (CASA) version 6.2.1-7. We split the data of our sources from the measurement set separately in each band averaging over time (\texttt{timebin} = 10~s) and frequency (\texttt{width} = 64, to average 64 channels to form one output channel per subband). Before the actual imaging of the targets, we produced radio maps of the size of 2.7~arcmin $\times$ 2.7~arcmin, or the whole primary beam, to check the whole beam of the MRO and OVRO telescopes to identify any other sources of radio emission within them. We did not find other strong radio emitters in any of these fields, further supporting the assumption that the radio emission detected at MRO is coming from the NLS1 nucleus.

\subsubsection{Radio maps and measurements}

We used the \texttt{tclean} algorithm with interactive cleaning in CASA to produce the radio images of our sources. The cell size was chosen so that the synthesised beam is properly sampled, meaning a cell size of 250, 150, 100, 70, and 50~mas in X, Ku, K, Ka, and Q-bands, respectively. The image size was chosen so that the whole galaxy fits into the image, taking into account the varying cell sizes in different bands, and the redshifts of our sources. We used Briggs weighting, with \texttt{robust} = 1.8, in all cases. Some sources appear to be slightly hexagonal (for example, J1522+3439), possibly due to the sidelobes. In these cases we trialled with \texttt{robust} values closer to uniform weighting to suppress the sidelobes but there was no visible difference, so we decided to maximise the sensitivity and use the same robustness value for all sources. No source was bright enough to be self-calibrated. We used the \texttt{mtmfs} deconvolver with \texttt{nterms} = 2 and \texttt{scales} = 0 in case some sources would be bright and extended enough to produce spatially resolved in-band spectral index maps, which turned out not to be the case. However, due to this we did the wide-band primary beam correction separately with \texttt{widebandpbcor}.

We fitted each detected source using a 2D Gaussian to obtain the central coordinates and the peak flux density and its error. In cases of extended sources we measured the emission inside the 3$\sigma$ contour, and estimated its error by multiplying the rms by the square root of the emitting region area expressed in beams. The rms for each map was measured in an empty region of sky far from the central source. In case the source was not detected, we report 3$\sigma$ upper limits. The results are given in Sect.~\ref{sec:j1029}--~\ref{sec:1641}, and the radio maps shown in App.~\ref{app:radiomaps}.

\subsection{Very Long Baseline Array}
\label{sec:vlba}

\subsubsection{Observations}

We observed our sample also on milliarcsecond scale using the VLBA in the Ku band, centred at 15.1~GHz (Project BJ~109, PI J\"arvel\"a). The observations were carried out during one 10~hr long experiment on 2022-02-08. The recording setup used the Digital Downconverter (DDC) system of the Roach Digital Backend (RDBE) with four 128~MHz wide subbands -- giving a total bandwidth of 512~MHz -- two circular polarisations, and two-bit sampling, resulting in a total recording rate of 4~Gbps.

Due to the potentially low compact flux densities of the target sources, the observations were carried out using the standard phase-referencing technique, i.e. a rapid switching between the target and a nearby calibrator. The phase-reference calibrators together with their distances from the targets, their VLBI scale flux densities, and the used source-switching duty cycles are given in Table~\ref{tab:vlbaobssummary}. Each target source had 38~min total on-source integration time. Bright flat-spectrum radio quasars 3C\,279 and 3C\,345 were observed for two 5~min long scans and for three 3~min long scans, respectively. They were used as fringe finders and, more importantly, as calibrator sources for determining instrumental delays and bandpass shapes. Nine out of ten VLBA antennas participated in the observations, since Hancock was out due to a frozen focus/rotation mount.

\subsubsection{Data reduction}

The recorded station data were correlated with the VLBA DiFX correlator in Socorro using 0.5~MHz wide spectral channels and 1~s correlator integration time. This allows a relatively wide field of view, $>$4'' from the phase centre, to be searched for compact sources. 

The data were calibrated in the Astronomical Image Processing System \citep[AIPS;][]{2003greisen} using standard procedures for phase-referencing observations. The calibration started with a priori corrections to the station parallactic angle, updates to the Earth Orientation Parameters, and first-order removal of dispersive ionospheric delays using total electron content maps derived from the Global Navigation Satellite System (GNSS) data. Instrumental delays and phase offsets between subbands were removed by fringe-fitting a single scan of the bright calibrator 3C\,279. A priori amplitude calibration included corrections to sampler threshold levels by using autocorrelations, bandpass calibration using again a scan on 3C\,279, and conversion of raw correlator coefficients to Janskys by applying measured system temperatures and gain curves. 

The phase reference calibrators as well as the bright calibrators 3C 279 and 3C 345 were fringe-fitted using the AIPS task \textsc{fring} and combining subbands and using an integration time of either 2~min or the scan length, whichever was shorter. The fringe-fitting gave excellent results; the percentage of failed solutions was typically $\sim$1 per cent. The fringe-fitting solutions from the phase-reference calibrators were applied to both the calibrators and the target sources. The relative R-L delays were corrected by cross-hand fringe-fitting of a single scan of 3C 279. After this step, we imaged the calibrator data in Difmap \citep{1997shepherd} and loaded the images back to AIPS. The calibrator images were used to derive phase self-calibration solutions for the calibrator data using the AIPS task \textsc{calib} and 10~s integration times. These phase solutions were then applied to the target sources. As the last correction, we also used the amplitude self-calibration solutions from imaging the bright calibrators 3C\,279 and 3C\,345 to fine-tune the amplitude calibration for those antennas and subbands that had an average amplitude self-calibration solution deviating more than 5 per cent from unity. After this step, the target data were ready for imaging. 

\subsubsection{Imaging and searching for the target sources}

While we had quite accurate a priori positions of the target sources based on the previous JVLA data (positional uncertainties less than 10~mas), we still wanted to search for an area that covers most of the galaxy in case the variable emission seen in the single-dish data does not come from the JVLA core. To achieve this, for each target source we generated a set of naturally weighted images with a field-of-view of 820$\times$820~mas that covered an area of 7.4''$\times$7.4'' centred on the JVLA position using the multifield option of the AIPS task \textsc{imagr}. The image rms was $\sim 60$\,$\mu$Jy/beam for all the target sources which is at the expected thermal noise level. Since we searched for a large area covering one million synthesised beam areas per image, we set the detection threshold to 6$\sigma$ to avoid picking noise spikes. No sources were detected, and in Tables~\ref{tab:j1029interf}, \ref{tab:j1228interf}, \ref{tab:j1232interf}, \ref{tab:j1509interf}, \ref{tab:j1510interf}, \ref{tab:j1522interf}, and  \ref{tab:j1641interf} we quote 6$\sigma$ upper limits for the VLBA data.

\subsection{Single-dish data}
\label{sec:singledish}

In addition to radio interferometric data, we obtained non-simultaneous single-dish monitoring data for all of these sources from MRO and OVRO; these data will be published here. We also have 1-3 epochs of single-dish observations per source from the Effelsberg 100-m Radio Telescope between 4.5 and 45~GHz, and one epoch of 2 and 1.15~mm observations with the New IRAM Kids Arrays (NIKA2) instrument on the Institut de Radioastronomie Millimétrique (IRAM) 30-m Radio Telescope on Pico Veleta for five sources. The Effelsberg and IRAM data, complemented by MRO and OVRO data from the same time period, will be published in an upcoming paper.

\begin{table}
\caption{Summary of the single-dish observations published here. Columns: (1) source name; (2) date the MRO observations were started; (3) number of detections and observation at MRO; (4) date the OVRO observations were started, (5) number of detections and observations at OVRO.}
\label{tab:source_list}
\centering
\footnotesize
\begin{tabular}{lcccc}
\hline\hline
Name       & \multicolumn{2}{c}{MRO} & \multicolumn{2}{c}{OVRO}  \\ 
           & start date & $N_{\mathrm{det}}$ / $N_{\mathrm{obs}}$ & start date & $N_{\mathrm{det}}$ / $N_{\mathrm{obs}}$  \\ \hline\hline
J1029+5556 & 2014-09-28 & 3 / 49                                  & 2020-06-30 & 1 / 81   \\
J1228+5017 & 2014-09-08 & 7 / 46                                  & 2020-06-18 & 0 / 93   \\
J1232+4957 & 2014-04-17 & 7 / 66                                  & 2020-06-20 & 0 / 83   \\
J1509+6137 & 2014-09-08 & 23 / 91                                 & 2020-07-09 & 0 / 82   \\
J1510+5547 & 2014-03-19 & 19 / 107                                & 2020-06-19 & 0 / 75   \\
J1522+3934 & 2014-05-07 & 5 / 129                                 & 2020-06-06 & 4 / 88   \\
J1641+3454 & 2014-04-01 & 12 / 821                                & 2020-06-05 & 1 / 87   \\ \hline 
\end{tabular}
\end{table}

\subsubsection{Metsähovi Radio Observatory}

The measurements included in this study are part of the large ongoing AGN monitoring programme at 37~GHz with the 13.7-m radio telescope at MRO. The observations are made with a 1 GHz-band dual beam receiver centred at 36.8~GHz. The beam full-width at half power is 144~arcsec. The observations are on–on observations, alternating the source and the sky in each feed horn. A typical integration time to obtain one flux density data point of a faint source is 1800~s. The sensitivity is limited by sky noise due to the location of the telescope, and it has been experimentally shown that the results do not significantly improve after the used maximum integration time of 1800~s. The detection limit of the telescope at 37~GHz is of the order of 200~mJy under optimal conditions. Data points with a S/N $<$ 4 are handled as non-detections. The flux density scale is set by observations of DR 21. Sources NGC 7027, 3C 274, and 3C 84 are used as secondary calibrators. A detailed description of the data reduction and analysis is given in \citet{1998terasranta1}. The error estimate in the flux density includes the contribution from the measurement rms and the uncertainty of the absolute calibration. The upper limits are 4$\sigma$ upper limits based on the measurement rms. Additional details regarding the MRO observations are given in App.~\ref{app:mrodetails}. The data included in this work have been taken between March 2014 and June 2022. 

\subsubsection{Owens Valley Radio Observatory}

The 15~GHz observations were carried out as part of the general Owens Valley Radio Observatory (OVRO) 40~m radio telescope AGN monitoring programme. This telescope uses off-axis dual-beam optics and a cryogenic receiver with a 15.0~GHz centre frequency and 3~GHz bandwidth. The beam full-width at half power is 157~arcsec. The observations are carried out in on-on fashion to remove atmospheric and ground contamination. In May 2014 a new pseudo-correlation receiver was installed on the 40~m telescope and the fast gain variations are corrected using a 180 degree phase switch. Calibration is achieved using a temperature-stable diode noise source to remove receiver gain drifts and the flux density scale is derived from observations of 3C 286 assuming the \citet{1977baars1} value of 3.44~Jy at 15.0~GHz. The systematic uncertainty of about 5 per cent in the flux density scale is included in the error bars. The upper limits are 4$\sigma$ upper limits based on the measurement rms. Complete details of the reduction and calibration procedure are found in \citet{2011richards1} and more details specific to the NLS1 observations are given in App.~\ref{app:ovrodetails}.

These seven sources were added to the OVRO AGN monitoring programme in July 2020, and since then three of them have been detected with S/N $>$ 4. This paper includes OVRO data until June 2022. 

\subsection{Archival data} 
\label{sec:archival}

In addition to the new data obtained we also used already published data of these sources. We included the JVLA A-configuration L, C and X band data from \citet{2020berton2} taken in September 2019. We also included the LOFAR LoTSS Data Release 2 (DR2) data with a central frequency of 144~MHz (band 120-168~MHz) \citep{2022shimwell1}. All of our sources reside within the published region of the sky. The resolution of LOFAR LoTSS DR2 is 6~arcsec, the median rms sensitivity is 83~$\mu$Jy beam$^{-1}$, the flux density scale accuracy is $\sim$10 per cent, and the astrometric accuracy is 0.2~arcsec. We used a 1.2~arcmin search radius to check the whole MRO beam area. In addition, we checked the Stokes I continuum radio maps to correctly identify the NLS1, and any other possible radio sources, and to visually cross-match the radio sources with any optical sources. Last, we included NRAO VLASS Epoch 1 and 2 data. The angular resolution of VLASS is $\sim$2.5~arcsec, and it covers the entire sky north of $\delta = -40\deg$. In this paper we use data based on the Quick Look and single epoch imaging, which have a systematic $\sim$15 per cent underestimation of the flux density values at $S_{\textrm{peak}} > 3$~mJy beam$^{-1}$. We used the same search radius as for LOFAR. These data are discussed in detail in the individual source sections.

\section{Results}
\label{sec:results}

The results for each source are given in the following Sections. In addition to the radio map measurements, we calculated the redshift- and k-corrected radio luminosities as:
\begin{equation}
\nu L_{\nu} = \frac{4 \pi \nu S_{\nu} d_L^2}{(1+z)^{(1+\alpha)}} \quad [\mathrm{erg\,s^{-1}}],
\end{equation}
where $\nu$ is the central frequency of the band in Hz, $S_{\nu}$ the observed flux density in erg s$^{-1}$ cm$^{-2}$ Hz$^{-1}$, $d_L^2$ the luminosity distance in cm, and $\alpha$ the spectral index of the emission. For simplicity we used $\alpha$ = 0 in all calculations. Even drastic changes in $\alpha$ do not significantly affect the luminosity, i.e. the order of magnitude remains the same. Furthermore, since our sources are variable they do not have a characteristic spectral index. The luminosities are given in the Tables in the following Sections for individual sources.

Since our sources are only marginally extended or point-like and their JVLA spectra show a consistent slope throughout the detected bands, it is unlikely that in-band spectral index maps could yield significant new information regarding their spectral properties. Thus we calculated only the traditional spectral indices between new detections with interferometric arrays using both the peak flux densities and the integrated flux densities.

Additionally, we used temporally close consecutive 37~GHz detections to estimate the properties of the flares. The details of the calculations and the results are given in Sect.~\ref{sec:flarecharacteristics}, but referred to in the following sections for the individual sources.

\subsection{SDSS J102906.69+555625.2}
\label{sec:j1029}

So far J1029+5556 has been detected at 37~GHz at MRO and at 15~GHz at OVRO (Table~\ref{tab:j1029singledish}). It has not been detected in any radio interferometric observations (see Table~\ref{tab:j1029interf}). J1029+5556 has the highest redshift, $z$ = 0.451, in this sample, and due to this it is also the only source that is missing the host galaxy morphology information. Interestingly, it was detected at MRO only three times in 2016-2017, with moderate flux densities around 500~mJy and below, and has not been detected after that, though it has not been observed very frequently in the past few years. Its overall detection percentage at 37~GHz is 6.1 per cent, and the mean luminosity $\nu L_{\nu}$ = 9.5 $\times$ 10$^{43}$ erg s$^{-1}$. The lack of recent detections might indicate a change in the activity level of the nucleus, though it was detected by OVRO in 2020, indicating that the activity has not totally halted. Whether the amplitude of the variability has changed or if the most drastic variability has moved to lower frequencies cannot be determined based on these data. J1029+5556 is not present in LOFAR maps, but there is one radio source, which lacks an optical counterpart, in the LOFAR map within the MRO beam. However, the source is faint, with a flux density of $\sim$1~mJy, and we do not see signs of it in the JVLA data. The non-simultaneous radio spectrum of J1029+5556 is shown in Fig.~\ref{j1029spect} and the light curves in Figs.~\ref{j1029lc} and ~\ref{fig:J1029lc-ul}. 

\begin{figure}
\begin{center}
\includegraphics[width=\columnwidth]{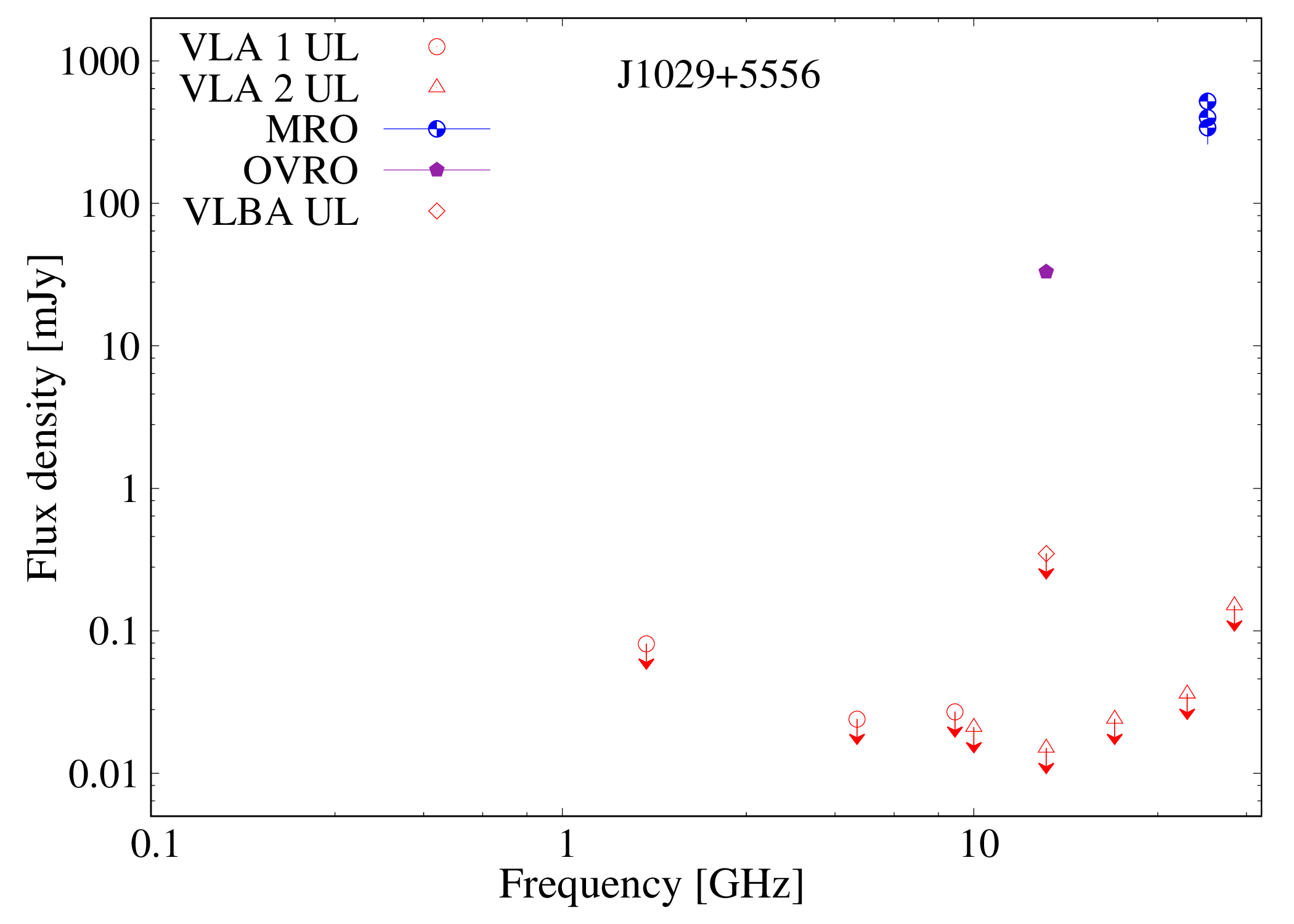}
\caption{Non-simultaneous radio spectrum of J1029+5556. Symbols and colours explained in the figure. Filled symbols denote integrated flux densities and empty symbols mark peak flux densities, except empty red symbols with downward arrows that are used for upper limits. VLA 1 data from \citet{2020berton2} and VLA 2 data from this paper. \label{j1029spect}}
\end{center}
\end{figure}

\begin{table*}
\caption{Interferometric data for J1029+5556. Columns: (1) array; (2); band or central frequency used for the observation; (3) peak flux density, or an upper limit (3$\sigma$ for the JVLA and 6$\sigma$ for the VLBA); (4) rms level of the observation; (5) clean beam size; (5) beam position angle.}
\label{tab:j1029interf}
\centering
\footnotesize
\begin{tabular}{lccccr}
\hline\hline
array & band    & $S_{\mathrm{peak}}$ & rms                   & beam size            & beam PA \\
      &         & (mJy beam$^{-1}$)   &  ($\mu$Jy beam$^{-1}$) & (" $\times$ ")      & (deg)\\  \hline\hline
JVLA  & X       & $<$ 0.021           & 7                     & 0.321 $\times$ 0.213 & $-85.0$  \\ 
JVLA  &  Ku     & $<$ 0.015           & 5                     & 0.222 $\times$ 0.150 & $-82.3$ \\
JVLA  &  K      & $<$ 0.024           & 8                     & 0.145 $\times$ 0.104 & $-84.3$ \\
JVLA  &  Ka     & $<$ 0.036           & 12                    & 0.097 $\times$ 0.082 & 74.0 \\
JVLA  &  Q      & $<$ 0.150           & 50                    & 0.109 $\times$ 0.089 & 16.7 \\
      &         &                     &                       &                      &        \\
VLBA  &  Ku     & $<$ 0.348           & 58                    & 0.00102 $\times$ 0.00055 & $-1.8$        \\   \hline 
\end{tabular}
\end{table*}

\begin{table}
\caption{Single-dish detections for J1029+5556. Columns: (1) telescope; (2) central frequency of the observation; (3) flux density and its error; (4) date of the observation.}
\label{tab:j1029singledish}
\centering
\footnotesize
\begin{tabular}{lccc}
\hline\hline
telescope & frequency & $S_{\mathrm{int}}$ & date \\
          & (GHz)     & (mJy)              & (dec. year)\\  \hline\hline
MRO       & 37        & 520 $\pm$ 80   & 2016.363276        \\ 
MRO       & 37        & 340 $\pm$ 80   & 2017.402475       \\
MRO       & 37        & 400 $\pm$ 80   & 2017.413284         \\
OVRO      &  15       & 33.2 $\pm$ 3.3 & 2020.569700               \\ \hline 
\end{tabular}
\end{table}


\subsection{SDSS J122844.81+501751.2}
\label{sec:1228}

J1228+5017 is detected with the JVLA in all other bands except the Q band, and it is also detected by LOFAR at 144~MHz (Table~\ref{tab:j1228interf}). It is not properly resolved in any JVLA band (Figs.~\ref{j1228x}-\ref{j1228q}). In the 144~MHz radio map it seems to be extended toward north-west, but upon closer inspection the extended part turns out to be a nearby galaxy that can also be seen in optical images. The radio spectrum, shown in Fig.~\ref{j1228spect}, has a constant slope of $-0.7$ from 144~MHz to X band, above which the slope flattens considerably (Table~\ref{tab:j1228spind}). The low-frequency spectral index is consistent with the characteristic star formation activity spectral index of $-0.7$ and the flux density levels could be explained by star formation \citep{2020berton2}. Though it should be noted that also the spectral index of optically thin synchrotron emission by shock-accelerated electrons in jets is around $-0.7$. The spectrum shows the characteristic spectral turnover, or spectral index flattening, toward lower frequencies where the emitting medium starts to become opaque to radio emission \citep{1992condon1}. In principle, the high-frequency spectral index is very close to the thermal free-free emission spectral index of $-0.1$, that in star forming galaxies has an increasing contribution toward higher frequencies, whereas the steep synchrotron emission from supernovae becomes less important. However, the change in the slope between the non-thermal and thermal emission dominated spectral regions should not be this drastic \citep{2018klein1}. Instead, the flattening could be due to a third component, the flat radio core of the AGN that becomes detectable when the emission produced by star formation weakens. Spatially resolved spectral index maps in L, C, and X bands support this scenario since despite the overall steep spectral index, the core spectral index in these bands is significantly flatter \citep{2021jarvela1}. X band also shows a peak flux density decrease from 0.184 $\pm$ 0.008~mJy beam$^{-1}$ in \citet{2020berton2} to 0.128 $\pm$ 0.005~mJy beam$^{-1}$ in these observations. The JVLA configuration and the rms of the maps are the same for both observations, but the central frequencies are slightly different (9 vs 10~GHz), thus the difference could be due to the slightly different beam sizes, since the source is partially resolved.

J1228+5017 has been detected at MRO seven times, with the last detection in 2019, and has a detection percentage of 15.2 per cent and a mean luminosity of $\nu L_{\nu}$ = 2.6 $\times$ 10$^{43}$ erg s$^{-1}$. The single-dish detections are listed in Table~\ref{tab:j1228singledish} and the light curves are shown in Figs.~\ref{j1228lc} and ~\ref{fig:J1228lc-ul}. However, the source does not seem to have totally gone into slumber as it has been detected again recently (Järvelä et al. in prep.).

\begin{figure}
\begin{center}
\includegraphics[width=\columnwidth]{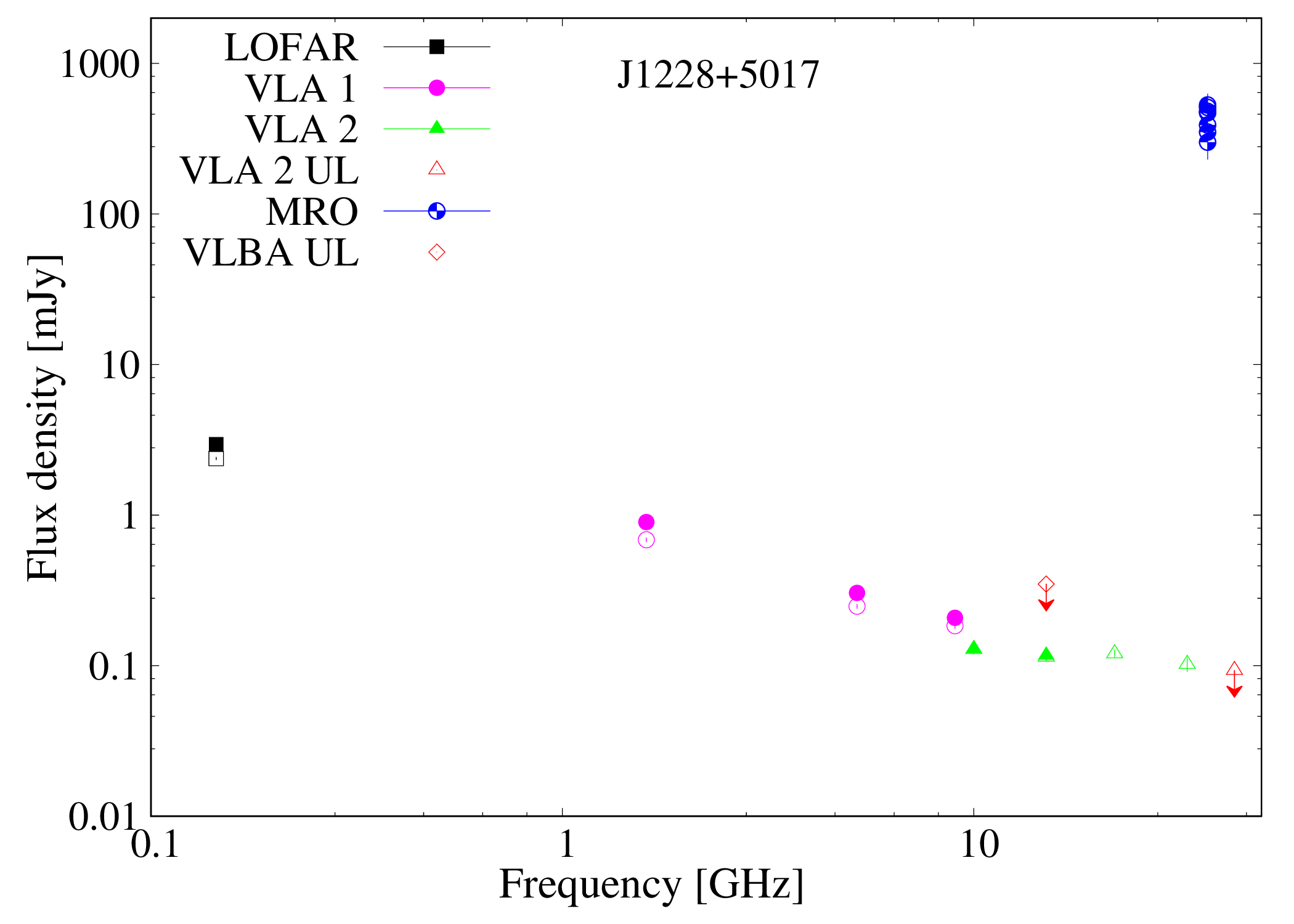}
\caption{Non-simultaneous radio spectrum of J1228+5017. Symbols and colours explained in the figure. Filled symbols denote integrated flux densities and empty symbols mark peak flux densities, except empty red symbols with downward arrows that are used for upper limits. VLA 1 data from \citet{2020berton2} and VLA 2 data from this paper. \label{j1228spect}}
\end{center}
\end{figure}

\begin{table*}
\caption{Interferometric data for J1228+5017. Columns: (1) array; (2); band or central frequency used for the observation; (3) peak flux density, or an upper limit (3$\sigma$ for the JVLA and 6$\sigma$ for the VLBA); (4) integrated flux density; (5) rms level of the observation; (6) peak radio luminosity; (7) integrated radio luminosity; (8) clean beam size; (9) beam position angle.}
\label{tab:j1228interf}
\centering
\footnotesize
\begin{tabular}{lcccccccr}
\hline\hline
array & band    & $S_{\mathrm{peak}}$ & $S_{\mathrm{int}}$ & rms   & $\nu L_{\nu\textrm{, peak}}$ & $\nu L_{\nu\textrm{, int}}$ & beam size & beam PA \\
      &  & (mJy beam$^{-1}$)   & (mJy)  & ($\mu$Jy beam$^{-1}$) & (10$^{39}$ erg s$^{-1}$) & (10$^{39}$ erg s$^{-1}$)  & (" $\times$ ")  & (deg)\\ \hline\hline
JVLA  & X       & 0.128 $\pm$ 0.005   & 0.129 $\pm$ 0.010  & 7                     & 2.027          & 2.043        & 0.294 $\times$ 0.223 & 88.2 \\ 
JVLA  &  Ku     & 0.114 $\pm$ 0.006   & 0.117 $\pm$ 0.009  & 6                     & 2.708          & 2.779        & 0.191 $\times$ 0.142 & $-83.4$ \\
JVLA  &  K      & 0.120 $\pm$ 0.007   &                    & 9                     & 4.181          &              & 0.129 $\times$ 0.094 & $-82.0$ \\
JVLA  &  Ka     & 0.102 $\pm$ 0.011   &                    & 12                    & 5.331          &              & 0.083 $\times$ 0.064 & $-80.2$ \\
JVLA  &  Q      & $<$ 0.093           &                    & 31                    &                &              & 0.060 $\times$ 0.050 & $-81.9$ \\
      &         &                     &                    &                       &                &              &        &        \\
VLBA  &  Ku    & $<$ 0.348    &                            &  58                   &                &              & 0.00103 $\times$ 0.00057   & 16.3  \\ 
      &         &                     &                    &                       &                &              &        &        \\
LOFAR & 144~MHz & 2.374 $\pm$ 0.056   &	2.948 $\pm$ 0.112  & 53                    &                &               & 6 $\times$ 6      &         \\ \hline 
\end{tabular}
\end{table*}

\begin{table}
\caption{Spectral indices for J1228+5017. Columns: (1) bands used; (2) spectral index using peak flux densities; (3) spectral index using integrated flux densities.}
\label{tab:j1228spind}
\centering
\footnotesize
\begin{tabular}{lcc}
\hline\hline
bands     & $\alpha_{\textrm{peak}}$ & $\alpha_{\textrm{int}}$  \\  \hline\hline
LOFAR - X & $-0.69$ $\pm$ 0.01         & $-0.74$ $\pm$ 0.03                 \\
X - Ku    & $-0.29$ $\pm$ 0.23         & $-0.24$ $\pm$ 0.38       \\ 
Ku - K    & 0.13 $\pm$ 0.29          &                           \\
K - Ka    & $-0.40$ $\pm$ 0.41         &                          \\ \hline 
\end{tabular}
\end{table}

\begin{table}
\caption{Single-dish detections for J1228+5017. Columns: (1) telescope; (2) central frequency of the observation; (3) flux density and its error; (4) date of the observation.}
\label{tab:j1228singledish}
\centering
\footnotesize
\begin{tabular}{lccc}
\hline\hline
telescope & frequency & $S_{\mathrm{int}}$ & date  \\
          & (GHz)     & (mJy)              & (dec. year)\\  \hline\hline
MRO       &	37.0      & 390 $\pm$ 60   &   2015.435445 \\
MRO       &	37.0      &	350 $\pm$ 70   &  2016.412445 \\
MRO       &	37.0      & 480 $\pm$ 70   & 2016.415172 \\
MRO	      & 37.0      & 300 $\pm$ 70   & 2016.535148 \\
MRO		  & 37.0      & 510 $\pm$ 100  & 2017.404414 \\
MRO		   & 37.0     &	470 $\pm$ 70	& 2019.248600 \\
MRO		   & 37.0     &	530 $\pm$ 100	  & 2019.369877 \\ \hline 
\end{tabular}
\end{table}


\subsection{SDSS J123220.11+495721.8}
\label{sec:1232}

In the earlier JVLA observations J1232+4957 was detected in L and C bands, but not in X band. In the new observations it is also detected in X and Ku bands, but only at a 3$\sigma$ level (Table~\ref{tab:j1232interf}). Also LOFAR detected J1232+4957 at 144~MHz. It remains unresolved in all interferometric observations (Figs.~\ref{j1232x} and ~\ref{j1232ku}). Its radio spectrum, in Fig.~\ref{j1232spect}, clearly shows a steepening slope toward higher frequencies. The spectral index between 144~MHz and X band is $-0.56$ $\pm$ 0.08, and between X and Ku bands $-1.49$ $\pm$ 0.59. (Table~\ref{tab:j1232spind}). The interferometric flux densities and the spectral properties of J1232+4957 can be explained by star formation activities, and AGN contribution does not seem necessary.

On the other hand, J1232+4957 has been detected at MRO several times with an overall detection percentage of 10.6 per cent. The mean luminosity of the detections is $\nu L_{\nu}$ = 2.8 $\times$ 10$^{43}$ erg s$^{-1}$. The last detection, however, is from 2019 (Table~\ref{tab:j1232singledish}). The 37~GHz flux densities are quite modest, never exceeding 600~mJy. The light curves of J1232+4957 are shown in Figs.~\ref{j1232lc} and ~\ref{fig:J1232lc-ul}.

\begin{figure}
\begin{center}
\includegraphics[width=\columnwidth]{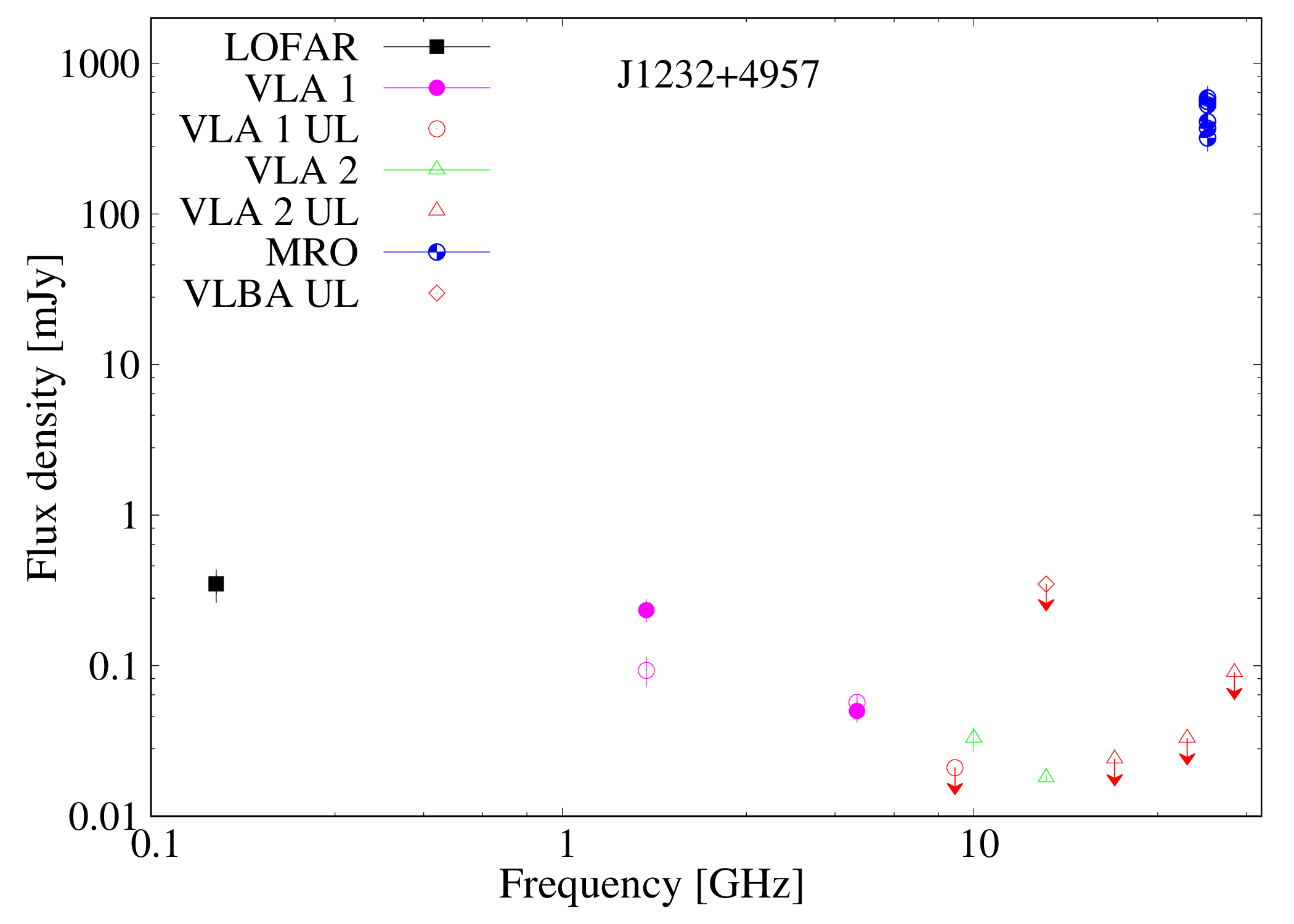}
\caption{Non-simultaneous radio spectrum of J1232+4957. Symbols and colours explained in the figure. Filled symbols denote integrated flux densities and empty symbols mark peak flux densities, except empty red symbols with downward arrows that are used for upper limits. VLA 1 data from \citet{2020berton2} and VLA 2 data from this paper. \label{j1232spect}}
\end{center}
\end{figure}

\begin{table*}
\caption{Interferometric data for J1232+4957. Columns: (1) array; (2); band or central frequency used for the observation; (3) peak flux density, or an upper limit (3$\sigma$ for the JVLA and 6$\sigma$ for the VLBA); (4) integrated flux density; (5) rms level of the observation; (6) peak radio luminosity; (7) integrated radio luminosity; (8) clean beam size; (9) beam position angle.}
\label{tab:j1232interf}
\centering
\footnotesize
\begin{tabular}{lccccccr}
\hline\hline
array & band    & $S_{\mathrm{peak}}$ & $S_{\mathrm{int}}$ & rms                   & $\nu L_{\nu\textrm{, peak}}$ & beam size      & beam PA \\
      &         & (mJy beam$^{-1}$)   & (mJy)              & ($\mu$Jy beam$^{-1}$) & (10$^{39}$ erg s$^{-1}$)     & (" $\times$ ") & (deg)\\  \hline\hline
JVLA  & X       & 0.033 $\pm$ 0.006   &                    & 7                     & 0.523  &   0.240 $\times$ 0.215 & $-66.2$   \\ 
JVLA  &  Ku     & 0.018 $\pm$ 0.001   &                    & 5                     & 0.428  &   0.159 $\times$ 0.141 & $-56.2$ \\
JVLA  &  K      & $<$ 0.024           &                    & 8                     &        &     0.110 $\times$ 0.091 & $-61.1$ \\
JVLA  &  Ka     & $<$ 0.033           &                    & 11                    &        &     0.071 $\times$ 0.065 & $-49.1$ \\
JVLA  &  Q      & $<$ 0.090           &                    & 30                    &        &    0.053 $\times$ 0.052 & 50.5 \\
      &         &                     &                    &                       &        &                           &        \\
VLBA  &  Ku     & $<$ 0.348           &                    & 58                    &        &  0.00104 $\times$ 0.00058 & 15.7        \\ 
      &         &                     &                    &                       &        &        &        \\
LOFAR & 144~MHz & 0.350 $\pm$ 0.050   & 0.348 $\pm$ 0.087  & 51                    &        &  6 $\times$ 6  &        \\ \hline 
\end{tabular}
\end{table*}

\begin{table}
\caption{Spectral indices for J1232+4957. Columns: (1) bands used; (2) spectral index using peak flux densities.}
\label{tab:j1232spind}
\centering
\footnotesize
\begin{tabular}{lc}
\hline\hline
bands     & $\alpha_{\textrm{peak}}$   \\  \hline\hline
LOFAR - X & $-0.56$ $\pm$ 0.08           \\
X - Ku    & $-1.49$ $\pm$ 0.59           \\ \hline 
\end{tabular}
\end{table}

\begin{table}
\caption{Single-dish detections for J1232+4957. Columns: (1) telescope; (2) central frequency of the observation; (3) flux density and its error; (4) date of the observation.}
\label{tab:j1232singledish}
\centering
\footnotesize
\begin{tabular}{lcccc}
\hline\hline
telescope & frequency & $S_{\mathrm{int}}$ & date \\
          & (GHz)     & (mJy)              & (dec. year)\\  \hline\hline
MRO       & 37.0      &	320 $\pm$ 60   & 2014.290645  \\
MRO       & 37.0      &	410 $\pm$ 80   & 2016.125413  \\
MRO       & 37.0      &	530 $\pm$ 70   & 2016.130824  \\
MRO       & 37.0      &	560 $\pm$ 130  & 2017.40169  \\
MRO       & 37.0      &	370 $\pm$ 80   & 2018.907172  \\
MRO       & 37.0      &	560 $\pm$ 90   & 2019.248668  \\
MRO       & 37.0      &	590 $\pm$ 120  & 2019.89896  \\ \hline 
\end{tabular}
\end{table}


\subsection{SDSS J150916.18+613716.7}
\label{sec:1509}

J1509+6137 is an intriguing source as it has clearly the highest detection percentage at 37~GHz -- 25.3 per cent -- but it has not been detected in any JVLA band. The MRO detections have an average luminosity of $\nu L_{\nu}$ = 2.5 $\times$ 10$^{43}$ erg s$^{-1}$. The light curves are shown in Figs.~\ref{j1509lc} and ~\ref{fig:J1509lc-ul}, and the radio data are given in Tables~\ref{tab:j1509interf} and \ref{tab:j1509singledish}. The brightest MRO flares exceed 1~Jy, indicating extreme variability of four orders of magnitude. J1509+6137 also has several double-detections within a week from each other. These detection pairs were used to estimate the flare characteristics (Table~\ref{tab:flaretau}) and are discussed in Sect.~\ref{sec:flarecharacteristics}.

J1509+6137 was not detected by LOFAR, but based on the LOFAR LoTSS DR2 there are two other radio sources within the MRO beam. Neither of these sources have optical counterparts, and both of them are faint, around 0.4 and 0.8~mJy. They are not seen in the JVLA data. J1509+6137 seems to be totally absent in radio -- except during the 37~GHz flares -- and does not even show detectable amounts of radio emission from star formation.

\begin{figure}
\begin{center}
\includegraphics[width=\columnwidth]{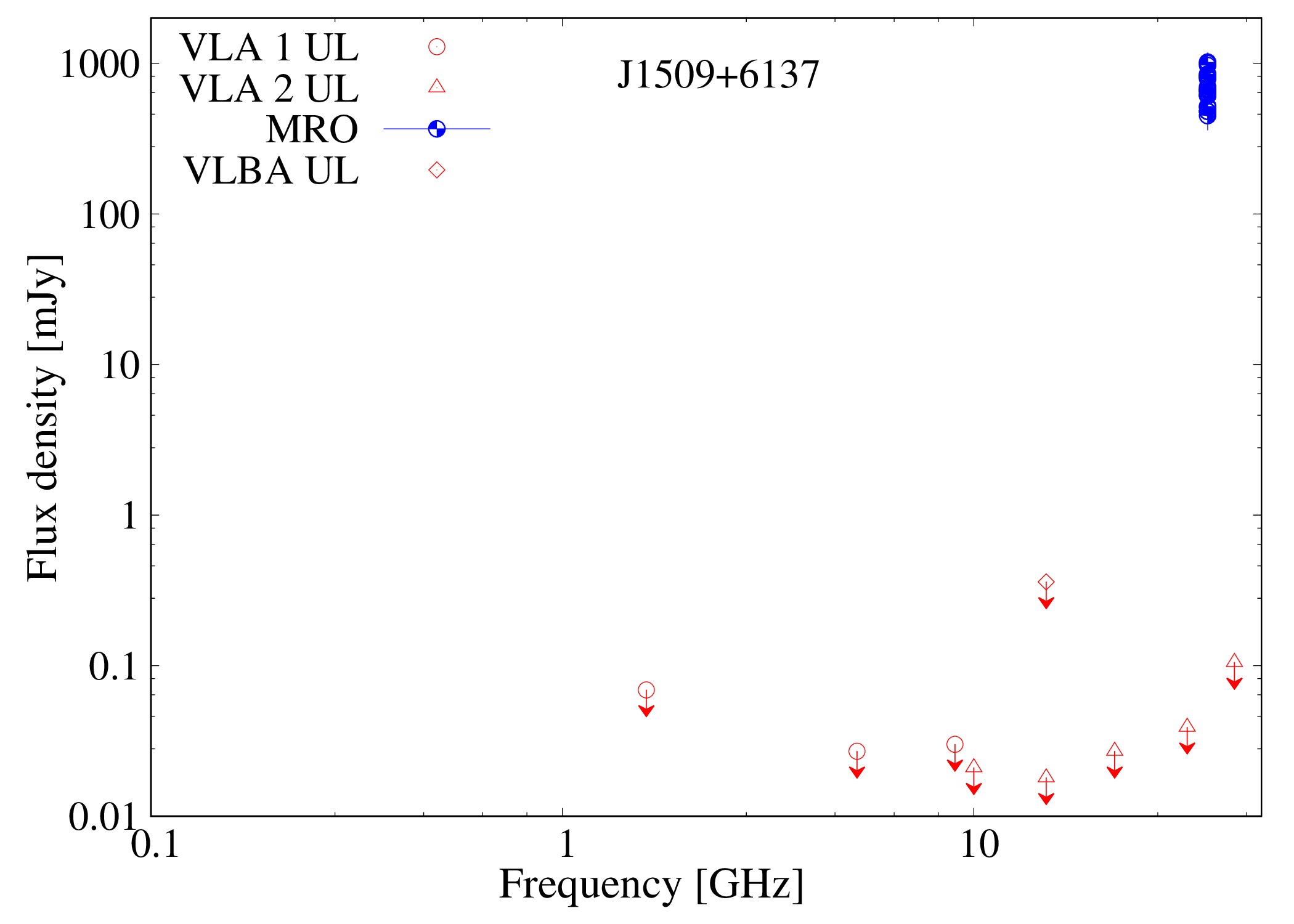}
\caption{Non-simultaneous radio spectrum of J1509+6137. Symbols and colours explained in the figure. Filled symbols denote integrated flux densities and empty symbols mark peak flux densities, except empty red symbols with downward arrows that are used for upper limits. VLA 1 data from \citet{2020berton2} and VLA 2 data from this paper. \label{j1509spect}}
\end{center}
\end{figure}

\begin{table*}
\caption{Interferometric data for J1509+6137. Columns: (1) array; (2); band or central frequency used for the observation; (3) peak flux density, or an upper limit (3$\sigma$ for the JVLA and 6$\sigma$ for the VLBA); (4) rms level of the observation; (5) clean beam size; (5) beam position angle.}
\label{tab:j1509interf}
\centering
\footnotesize
\begin{tabular}{lccccr}
\hline\hline
array & band    & $S_{\mathrm{peak}}$ & rms                   & beam size         & beam PA \\
      &         & (mJy beam$^{-1}$)   & ($\mu$Jy beam$^{-1}$) & (" $\times$ ")    & (deg)\\  \hline\hline
JVLA  & X       & $<$ 0.021           &  7                    & 0.261 $\times$ 0.204 & 43.0  \\ 
JVLA  &  Ku     & $<$ 0.018           &  6                    & 0.171 $\times$ 0.133 & 33.4 \\
JVLA  &  K      & $<$ 0.027           &  9                    & 0.116 $\times$ 0.092 & 37.3 \\
JVLA  &  Ka     & $<$ 0.039           &  13                   & 0.080 $\times$ 0.061 & 44.6 \\
JVLA  &  Q      & $<$ 0.105           &  35                   & 0.061 $\times$ 0.052 & 61.0  \\
      &         &                     &                       &       &        \\
VLBA  &  Ku     &  $<$ 0.360          &  60                   & 0.00098 $\times$ 0.00056 & 0.1        \\ \hline 
\end{tabular}
\end{table*}

\begin{table}
\caption{Single-dish detections for J1509+6137. Columns: (1) telescope; (2) central frequency; (3) flux density; (4) date of the observation.}
\label{tab:j1509singledish}
\centering
\footnotesize
\begin{tabular}{lccc}
\hline\hline
telescope & frequency & $S_{\mathrm{int}}$  & date \\
          & (GHz)     & (mJy)               & (dec. year)\\  \hline\hline
MRO & 37.0 & 670 $\pm$ 130  & 2015.454862   \\
MRO & 37.0 & 840 $\pm$ 140  &  2015.457565   \\
MRO & 37.0 & 660 $\pm$ 70  &  2016.396338   \\
MRO & 37.0 & 480 $\pm$ 100  & 2016.412707   \\
MRO & 37.0 & 480 $\pm$ 100  & 2016.415428   \\
MRO & 37.0 & 810 $\pm$ 180  & 2016.418182   \\
MRO & 37.0 & 510 $\pm$ 120  & 2017.391984	   \\
MRO & 37.0 & 970 $\pm$ 140  & 2017.39745   \\
MRO & 37.0 & 610 $\pm$ 90  & 2017.413817   \\
MRO & 37.0 & 450 $\pm$ 90  & 2017.419258   \\
MRO & 37.0 & 660 $\pm$ 120  & 2017.454812    \\
MRO & 37.0 & 820 $\pm$ 120  & 2017.473956    \\
MRO & 37.0 & 820 $\pm$ 130  & 2017.520388	   \\
MRO & 37.0 & 520 $\pm$ 100  & 2018.54222   \\
MRO & 37.0 & 850 $\pm$ 120  & 2019.012217   \\
MRO & 37.0 & 1000 $\pm$ 160  & 2019.37014    \\
MRO & 37.0 & 1020 $\pm$ 160  & 2019.381063   \\
MRO & 37.0 & 610 $\pm$ 110  & 2019.564088   \\
MRO & 37.0 & 680 $\pm$ 120  & 2019.698009   \\
MRO & 37.0 & 700 $\pm$ 170  & 2020.399100   \\
MRO & 37.0 & 640 $\pm$ 130  & 2020.407293   \\
MRO & 37.0 & 790 $\pm$ 130  & 2021.725315   \\
MRO & 37.0 & 620 $\pm$ 130  & 2021.881091     \\ \hline 
\end{tabular}
\end{table}


\subsection{SDSS J151020.06+554722.0}
\label{sec:1510}

J1510+5547 has a high detection percentage of 17.6 per cent at 37~GHz (Table~\ref{tab:j1510singledish} and Figs.~\ref{j1510lc} and ~\ref{fig:J1510lc-ul}). It was last detected in 2019 even if the number of annual observations has stayed roughly the same. The mean luminosity at 37~GHz is $\nu L_{\nu}$ = 8.7 $\times$ 10$^{42}$ erg s$^{-1}$. It was detected in L, C, and X bands in our previous JVLA observations, but remained a non-detection in all bands, X through Q, in the recent observations (Table~\ref{tab:j1510interf}). The radio spectrum of J1510+5547 is shown in Fig.~\ref{j1510spect}. The X band upper limit is very close to the earlier X band detection flux density, and considering that the central frequencies of the two observations differ by 1~GHz, it is likely that the recent non-detection is due to the source being very close to the detection limit. 

This source is also detected by LOFAR and seems to be marginally resolved. There is another radio source north-east of it and within the MRO beam. This source is faint, has no optical counterpart, and is not seen in any JVLA band. The projected distance between J1510+5547 and the source is more than 40~kpc, thus it is unlikely that it is related to our source. The radio spectrum below 10~GHz is consistent with that of star forming galaxies with the characteristic spectral turnover seen toward lower frequencies.

\begin{figure}
\begin{center}
\includegraphics[width=\columnwidth]{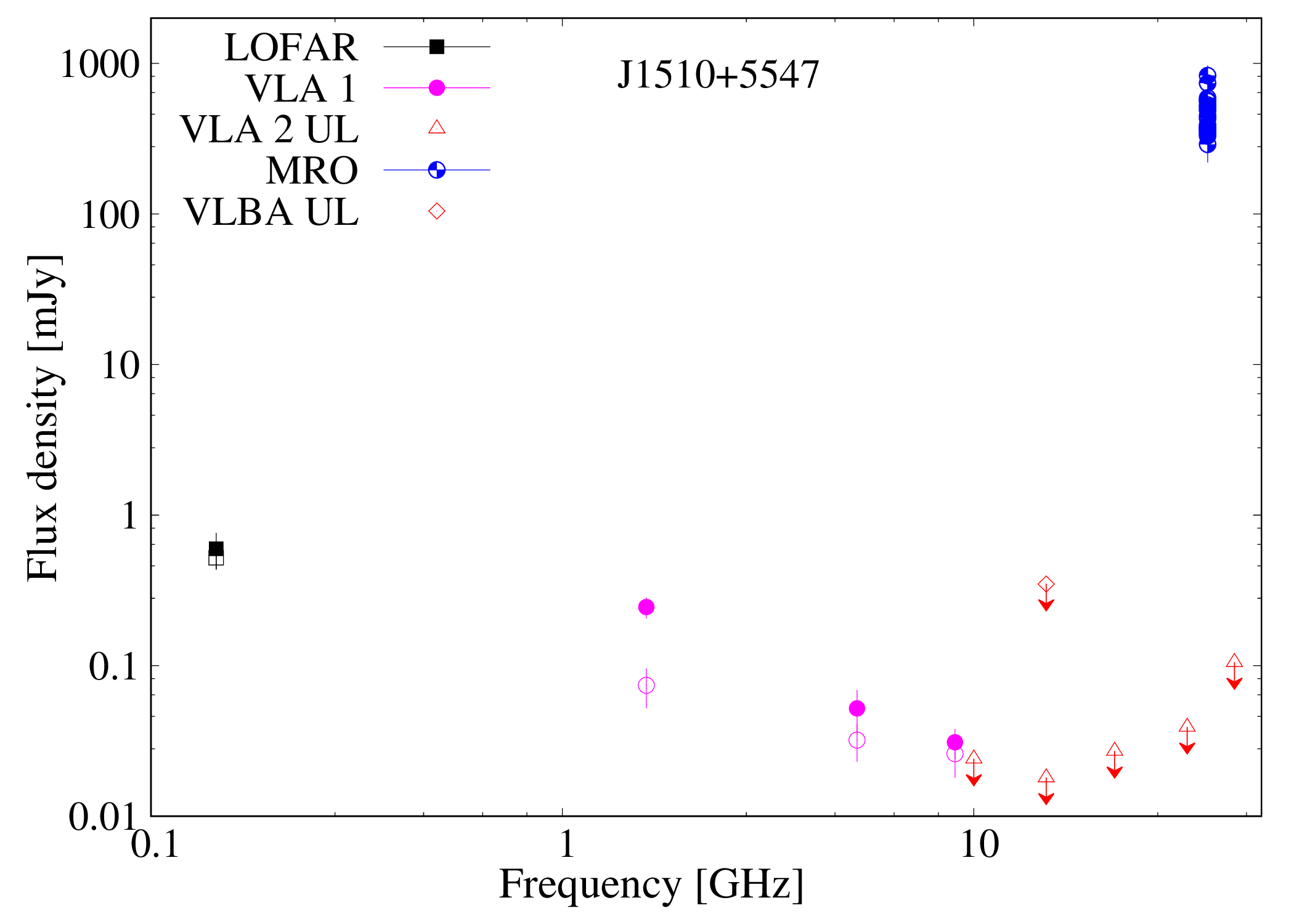}
\caption{Non-simultaneous radio spectrum of J1510+5547. Symbols and colours explained in the figure. Filled symbols denote integrated flux densities and empty symbols mark peak flux densities, except empty red symbols that are used for upper limits. VLA 1 data from \citet{2020berton2} and VLA 2 data from this paper. \label{j1510spect}}
\end{center}
\end{figure}

\begin{table*}
\caption{Interferometric data for J1510+5547. Columns: (1) array; (2); band or central frequency used for the observation; (3) peak flux density, or an upper limit (3$\sigma$ for the JVLA and 6$\sigma$ for the VLBA); (4) rms level of the observation; (5) clean beam size; (5) beam position angle.}
\label{tab:j1510interf}
\centering
\footnotesize
\begin{tabular}{lcccccr}
\hline\hline
array & band    & $S_{\mathrm{peak}}$ & $S_{\mathrm{int}}$ & rms                   & beam size         & beam PA \\
      &         & (mJy beam$^{-1}$)   & (mJy)              & ($\mu$Jy beam$^{-1}$) & (" $\times$ ")    & (deg)\\  \hline\hline
JVLA  & X       & $<$ 0.024           &                    & 8                     & 0.240 $\times$ 0.225 & 49.7 \\ 
JVLA  &  Ku     & $<$ 0.018           &                    & 6                     & 0.162 $\times$ 0.153 & 52.1 \\
JVLA  &  K      & $<$ 0.027           &                    & 9                     & 0.109 $\times$ 0.095 & 48.5 \\
JVLA  &  Ka     & $<$ 0.039           &                    & 13                    & 0.074 $\times$ 0.064 & 53.6  \\
JVLA  &  Q      & $<$ 0.105           &                    & 35                    & 0.059 $\times$ 0.051 & 74.0 \\
      &         &                     &                    &                       &                       &        \\
VLBA  &  Ku     &  $<$ 0.348          &                    & 58                    & 0.00094 $\times$ 0.00057 & 0.1        \\ 
      &         &                     &                    &                       &                        &        \\
LOFAR & 144~MHz & 0.521 $\pm$ 0.084   &	0.597 $\pm$ 0.164  & 85                    & 6 $\times$ 6           &       \\ \hline 
\end{tabular}
\end{table*}

\begin{table}
\caption{Single-dish detections for J1510+5547. Columns: (1) telescope; (2) central frequency of the observation; (3) flux density and its error; (4) date of the observation.}
\label{tab:j1510singledish}
\centering
\footnotesize
\begin{tabular}{lccc}
\hline\hline
telescope & frequency & $S_{\mathrm{int}}$ & date \\
          & (GHz)     & (mJy)              & (dec. year) \\  \hline\hline
MRO & 37.0 & 380 $\pm$ 80 & 2015.506706 \\
MRO & 37.0 & 370 $\pm$ 70 &  2015.784329 \\
MRO & 37.0 & 510 $\pm$ 60 &  2015.798000 \\
MRO & 37.0 & 450 $\pm$ 90 &  2015.801756 \\
MRO & 37.0 & 330 $\pm$ 70 &  2016.396263 \\
MRO & 37.0 & 490 $\pm$ 80 &  2016.412639 \\
MRO & 37.0 & 430 $\pm$ 80 &  2016.415360 \\
MRO & 37.0 & 570 $\pm$ 90 &  2016.418114	 \\
MRO & 37.0 & 290 $\pm$ 70 &  2016.426339 \\
MRO & 37.0 & 830 $\pm$ 140 &  2016.535468 \\
MRO & 37.0 & 560 $\pm$ 100 &  2017.056030 \\
MRO & 37.0 & 340 $\pm$ 80 &  2017.288073 \\
MRO & 37.0 & 360 $\pm$ 90 &  2017.413749 \\
MRO & 37.0 & 530 $\pm$ 90 &  2017.41666 \\
MRO & 37.0 & 350 $\pm$ 90 &  2018.165158 \\
MRO & 37.0 & 740 $\pm$ 110 & 2018.457715 \\
MRO & 37.0 & 390 $\pm$ 60 &  2018.531227 \\
MRO & 37.0 & 370 $\pm$ 80 &  2018.637850 \\
MRO & 37.0 & 590 $\pm$ 110 &  2019.569528  \\ \hline 
\end{tabular}
\end{table}


\subsection{SDSS J152205.41+393441.3}
\label{sec:1522}

J1522+3934 is a nearby source ($z$ = 0.077) that resides in a disk galaxy that is merging with a non-active galaxy \citep{2018jarvela1}. It shows almost symmetrical resolved emission on west/north-west and east/south-east sides of the nucleus from 144~MHz to Ku band, and is detected up to Ka band (Table~\ref{tab:j1522interf} and Figs.~\ref{j1522x}-\ref{j1522ka}). Interestingly, the extended radio emission is perpendicular to the host galaxy, indicating that it does not originate from the star formation activity in the host \citep{2021jarvela1}. To explain the 37~GHz flaring in J1522+3934 the jet emission needs to be relativistically boosted and thus the jet needs to point close to our line of sight. If this is the case, the extended emission would be a relic of past activity -- unless the jets are very bent, pointing at us close to the nucleus and turning perpendicular at larger distances. The spatially resolved spectral index map in the L band does show regions of steeper spectral index around $-1.0$, possibly indicative of synchrotron cooling.

The radio spectrum of J1522+3934, in Fig.~\ref{j1522spect}, has a very stable slope around $-0.7$ from 144~MHz all the way to the Ka band (Table~\ref{tab:j1522spind}). The VLASS points seem to deviate from this which is surprising considering that the Quick Look flux densities should underestimate the real flux densities. Overall the spectrum seems to be consistent with optically thin radio emission and we can assume its predominant origin to be the AGN.

J1522+3934 has the record 37~GHz flux density among our sources at 1430~mJy, whereas the other detections are much more modest. Its detection percentage at MRO is only 3.9 per cent, and the mean luminosity is $\nu L_{\nu}$ = 2.7 $\times$ 10$^{42}$ erg s$^{-1}$. In addition to these detections, it has also been detected at 15~GHz at OVRO on three different dates (Table~\ref{tab:j1522singledish}), with a maximum flux density of 45~mJy. The light curves of J1522+3934 are shown in Figs.~\ref{j1522lc} and ~\ref{fig:J1522lc-ul}.

\begin{figure}
\begin{center}
\includegraphics[width=\columnwidth]{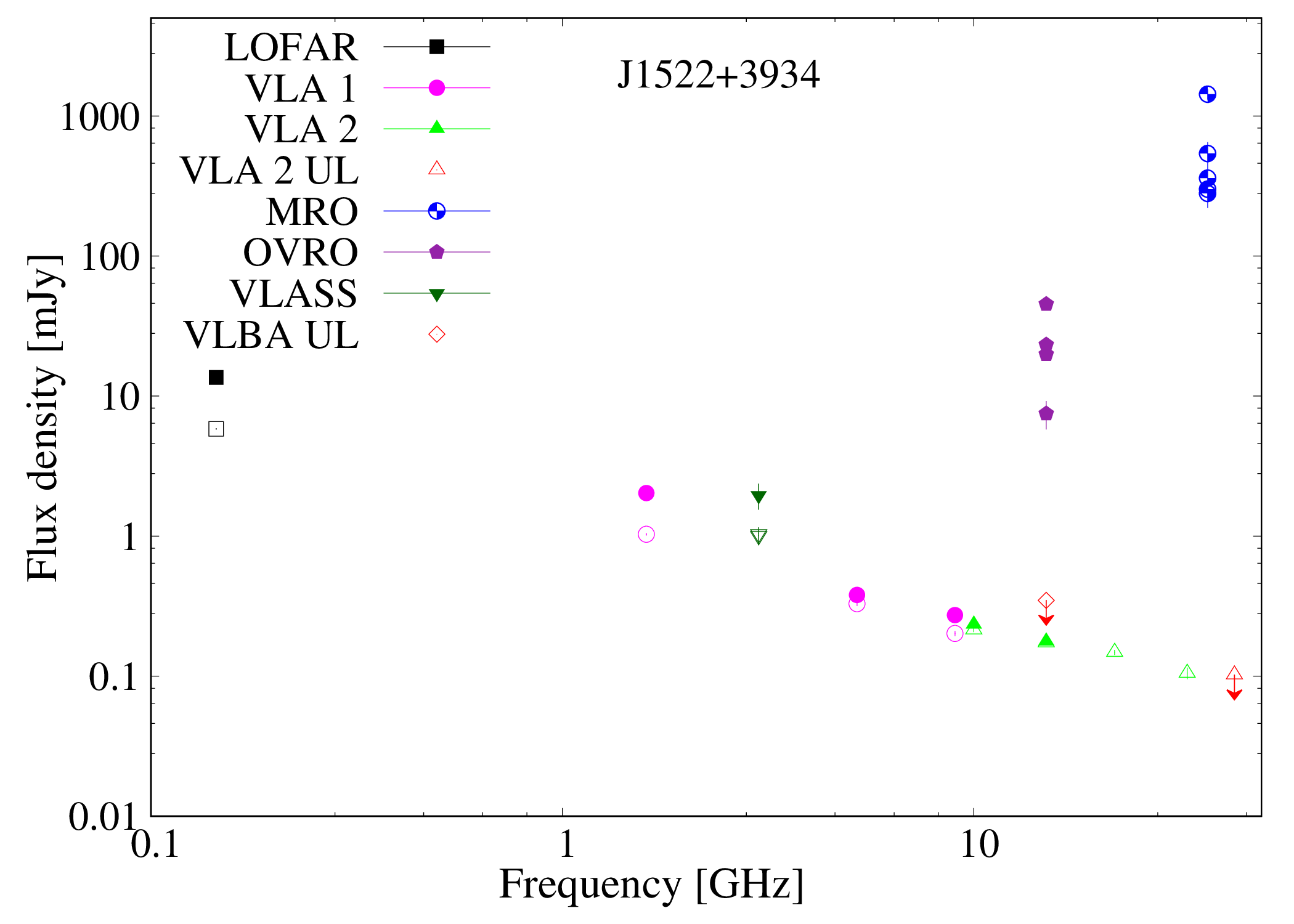}
\caption{Non-simultaneous radio spectrum of J1522+3934. Symbols and colours explained in the figure. Filled symbols denote integrated flux densities and empty symbols mark peak flux densities, except empty red symbols with downward arrows that are used for upper limits. VLA 1 data from \citet{2020berton2} and VLA 2 data from this paper. \label{j1522spect}}
\end{center}
\end{figure}

\begin{table*}
\caption{Interferometric data for J1522+3934. Columns: (1) array; (2); band or central frequency used for the observation; (3) peak flux density, or an upper limit (3$\sigma$ for the JVLA and 6$\sigma$ for the VLBA); (4) integrated flux density; (5) rms level of the observation; (6) peak radio luminosity; (7) integrated radio luminosity; (8) clean beam size; (9) beam position angle.}
\label{tab:j1522interf}
\centering
\footnotesize
\begin{tabular}{lcccccccr}
\hline\hline
array & band    & $S_{\mathrm{peak}}$ & $S_{\mathrm{int}}$ & rms                   & $\nu L_{\nu\textrm{, peak}}$ & $\nu L_{\nu\textrm{, int}}$ &   beam size         & beam PA \\
      &         & (mJy beam$^{-1}$)   & (mJy)              & ($\mu$Jy beam$^{-1}$) & (10$^{39}$ erg s$^{-1}$)     & (10$^{39}$ erg s$^{-1}$) &    (" $\times$ ")    & (deg)\\  \hline\hline
JVLA  & X       & 0.214 $\pm$ 0.008  & 0.234 $\pm$ 0.014   & 8                     & 0.274                        & 0.299   &    0.248 $\times$ 0.213 & $-88.7$  \\ 
JVLA  &  Ku     & 0.173 $\pm$ 0.007  & 0.177 $\pm$ 0.010   & 6                     & 0.332                        & 0.339   &    0.162 $\times$ 0.142 & 82.4 \\
JVLA  &  K      & 0.148 $\pm$ 0.006  &                     & 9                     & 0.416                        &         &    0.114 $\times$ 0.106 & 50.0 \\
JVLA  &  Ka     & 0.105 $\pm$ 0.010  &                     & 13                    & 0.443                        &         &    0.095 $\times$ 0.069 & 76.8 \\
JVLA  &  Q      & $<$ 0.102          &                     & 34                    &                              &         &    0.067 $\times$ 0.048 & $-83.7$ \\
      &         &                    &                     &                       &                              &         &                    &        \\
VLBA  &  Ku     &  $<$ 0.348         &                     & 58                    &                              &         &  0.00106 $\times$ 0.00056      & $-4.1$        \\ 
      &         &                    &                     &                       &                               &        &                 &        \\
LOFAR & 144~MHz & 5.833 $\pm$ 0.075  & 13.599 $\pm$ 0.240  & 71                    &                              &         & 6 $\times$ 6      &         \\
VLASS1 &  3~GHz & 1.040 $\pm$ 0.00   &                     &                       &                              &         & 2.5 $\times$ 2.5   &        \\
VLASS2 &  3~GHz & 1.008 $\pm$ 0.148  & 1.957 $\pm$ 0.411   &                       &                             &          &  2.5 $\times$ 2.5      &       \\ \hline 
\end{tabular}
\end{table*}

\begin{table}
\caption{Spectral indices for J1522+3934. Columns: (1) bands used; (2) spectral index using peak flux densities; (3) spectral index using integrated flux densities.}
\label{tab:j1522spind}
\centering
\footnotesize
\begin{tabular}{lcc}
\hline\hline
bands          & $\alpha_{\textrm{peak}}$ & $\alpha_{\textrm{int}}$  \\  \hline\hline
LOFAR - VLASS2 & $-0.58$ $\pm$ 0.05         & $-0.64$ $\pm$ 0.07              \\
VLASS2 - X     & $-1.29$ $\pm$ 0.15         & $-1.76$ $\pm$ 0.22              \\
X - Ku         & $-0.52$ $\pm$ 0.19         & $-0.69$ $\pm$ 0.29              \\ 
Ku - K         & $-0.41$ $\pm$ 0.21         &                             \\
K - Ka         & $-0.85$ $\pm$ 0.33         &                            \\ \hline 
\end{tabular}
\end{table}

\begin{table}
\caption{Single-dish detections for J1522+3934. Columns: (1) telescope; (2) central frequency of the observation; (3) flux density and its error; (4) date of the observation.}
\label{tab:j1522singledish}
\centering
\footnotesize
\begin{tabular}{lccc}
\hline\hline
telescope & frequency & $S_{\mathrm{int}}$ & date \\
          & (GHz)     & (mJy)              & (dec. year)\\  \hline\hline
MRO & 	37.0 & 360 $\pm$ 70 & 2014.397397  \\
MRO & 	37.0 & 300 $\pm$ 60 & 2017.071323  \\
MRO & 	37.0 & 1430 $\pm$ 120 & 2017.221661  \\
MRO & 	37.0 & 280 $\pm$ 60 & 2018.110510    \\
MRO & 	37.0 & 540 $\pm$ 110 & 2021.960079  \\
OVRO &  15.0 & 7.5 $\pm$ 1.7 & 2020.430300   \\ 
OVRO &  15.0 & 45.3 $\pm$ 3.0 & 2020.875700    \\
OVRO &  15.0 & 23.3 $\pm$ 1.9 & 2021.872600    \\ 
OVRO &  15.0 & 19.9 $\pm$ 2.0 & 2021.872600    \\ \hline 
\end{tabular}
\end{table}


\subsection{SDSS J164100.10+345452.7}
\label{sec:1641}

J1641+3454 is the only one of our sources with a statistically significant gamma-ray detection \citep{2018lahteenmaki1}, usually considered as proof of the presence of relativistic jets. Interestingly, its detection rate at 37~GHz is the lowest in the sample at 1.5 per cent. Its 37~GHz flux densities are modest, generally around 500~mJy and below, indicating that most of its flaring activity might not exceed the MRO detection threshold. Its average 37~GHz luminosity is $\nu L_{\nu}$ = 9.9 $\times$ 10$^{42}$ erg s$^{-1}$. J1641+3454 has also been detected at 15~GHz at OVRO with a flux density of $\sim$30~mJy (Table~\ref{tab:j1641singledish}). 

J1641+3454 was a target of an intense 20-month multiwavelength monitoring campaign in radio, optical, ultraviolet and, X-rays \citep{2023romano1}. During the campaign it flared twice at 37~GHz: the first radio flare was followed by brightening in X-rays, whereas the latter flare was not accompanied by any significant changes at other frequencies. Nevertheless, this was the first detection of a 37~GHz radio flare counterpart at another frequency.

J1641+3454 is detected in X, Ku, K, and Ka bands with the JVLA (Table~\ref{tab:j1641interf}). It is resolved in X and Ku bands, with extended emission seen on the north-west and the south-east sides of the nucleus. This emission is seen also at lower frequencies and it appears to be patchy, which points to star formation rather than the AGN as the origin \citep{2020berton2}. J1641+3454 is also detected at 144~MHz by LOFAR and at 3~GHz in VLASS. At 3~GHz it is not properly resolved but appears elongated in the north-west/south-east direction similarly to the JVLA maps. Interestingly, in the LOFAR map it seems to be elongated toward south-west. This emission has no optical counterpart, but it is clearly outside the host galaxy of J1641+3454, so it remains unclear whether it is related to J1641+3454.

The radio spectrum of J1641+3454, shown in Fig.~\ref{j1641spect}, clearly has a curvature, and it flattens towards lower frequencies and steepens toward higher frequencies, reaching a spectral index around $-1.0$. No AGN contribution is required to explain the properties of its high-resolution data radio spectrum and no signs of flattening in the spectrum or in the spatially resolved spectral index maps can be seen \citep{2021jarvela1}.

In addition to the 37~GHz detections, J1641+3454 has also been detected once at 15~GHz by OVRO with a flux density of $\sim$30~mJy. The OVRO detection is quite close to an MRO detection, within 23 days, but unfortunately in the case of these sources we cannot assume that these detections are necessarily from the same event. However, in the case they were, we can derive a quasi-simultaneous spectral index of 2.70 $\pm$ 0.63. Since these detections are not strictly simultaneous and we do not know which stage of the flare the detections represent, the spectral index is only a rough estimate. It agrees with SSA within the errors, but might imply that also another source of absorption is required. The light curves of J1641+3454 are shown in Figs.~\ref{j1641lc} and ~\ref{fig:J1641lc-ul}.

\begin{figure}
\begin{center}
\includegraphics[width=\columnwidth]{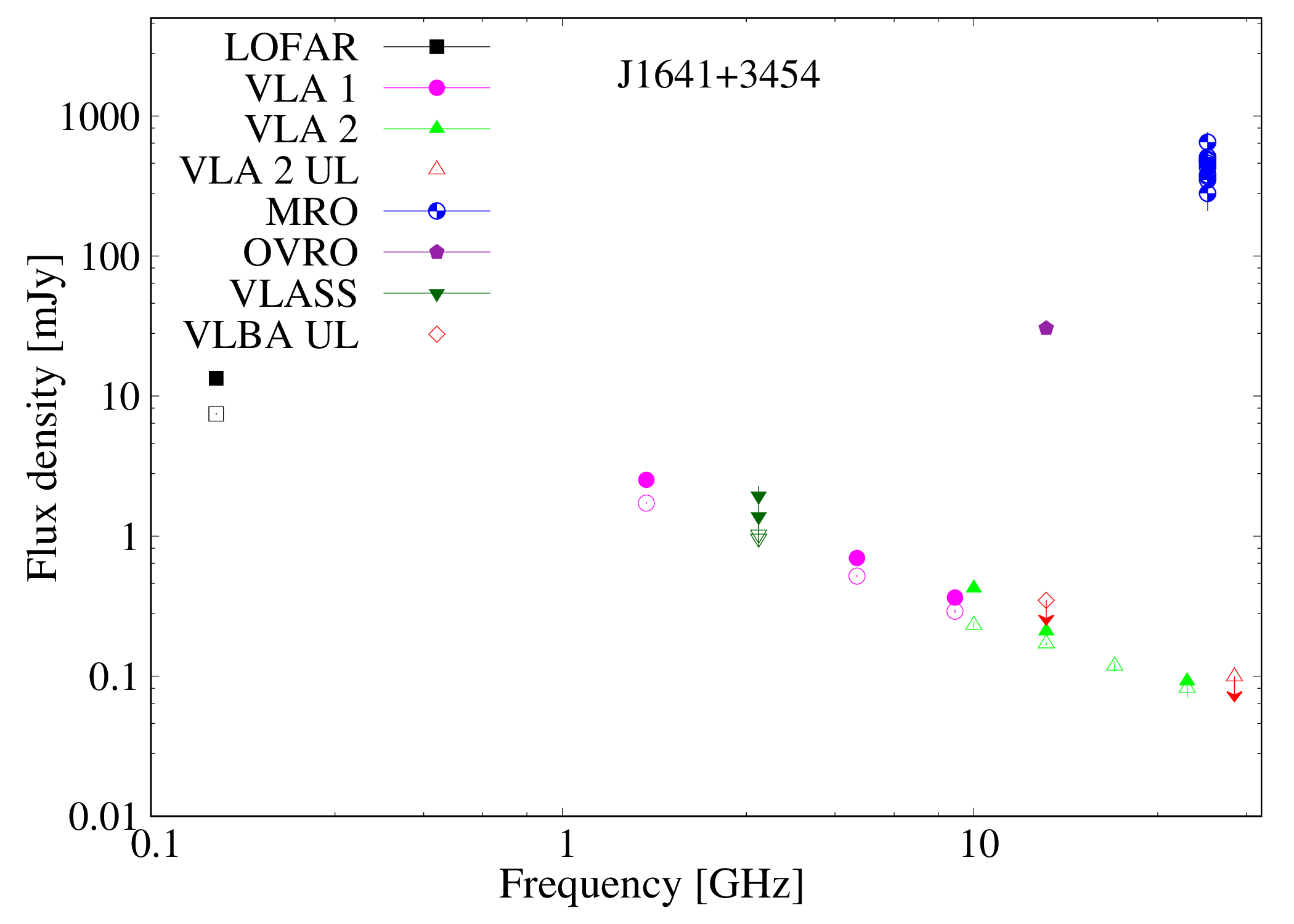}
\caption{Non-simultaneous radio spectrum of J1641+3454. Symbols and colours explained in the figure. Filled symbols denote integrated flux densities and empty symbols mark peak flux densities, except empty red symbols with downward arrows that are used for upper limits. VLA 1 data from \citet{2020berton2} and VLA 2 data from this paper. \label{j1641spect}}
\end{center}
\end{figure}

\begin{table*}
\caption{Interferometric data for J1641+3454. Columns: (1) array; (2); band or central frequency used for the observation; (3) peak flux density, or an upper limit (3$\sigma$ for the JVLA and 6$\sigma$ for the VLBA); (4) integrated flux density; (5) rms level of the observation; (6) peak radio luminosity; (7) integrated radio luminosity; (8) clean beam size; (9) beam position angle..}
\label{tab:j1641interf}
\centering
\footnotesize
\begin{tabular}{lcccccccr}
\hline\hline
array & band    & $S_{\mathrm{peak}}$ & $S_{\mathrm{int}}$ & rms                   & $\nu L_{\nu\textrm{, peak}}$ & $\nu L_{\nu\textrm{, int}}$ &  beam size       & beam PA \\
      &         & (mJy beam$^{-1}$)   & (mJy)              & ($\mu$Jy beam$^{-1}$) & (10$^{39}$ erg s$^{-1}$)      & (10$^{39}$ erg s$^{-1}$)   &  (" $\times$ ")  & (deg)\\  \hline\hline
JVLA  & X       & 0.231 $\pm$ 0.012   & 0.424 $\pm$ 0.019  & 7                     & 1.389  & 2.549  &   0.259 $\times$ 0.210 & $-80.5$  \\ 
JVLA  &  Ku     & 0.170 $\pm$ 0.006   & 0.209 $\pm$ 0.010  & 5                     & 1.533  & 1.885  &   0.174 $\times$ 0.146 & $-73.6$ \\
JVLA  &  K      & 0.118 $\pm$ 0.009   &                    & 8                     & 1.561  &        &   0.117 $\times$ 0.096 & $-75.9$ \\
JVLA  &  Ka     & 0.082 $\pm$ 0.012   & 0.092 $\pm$ 0.014  & 11                    & 1.627  & 1.825  &   0.103 $\times$ 0.065 & $-84.0$ \\
JVLA  &  Q      & $<$ 0.099           &                    & 33                    &        &        &   0.071 $\times$ 0.049 & $-77.9$ \\
      &         &                     &                    &                       &        &        &         &        \\
VLBA  &  Ku     &  $<$ 0.348          &                    & 58                    &        &       & 0.00109 $\times$ 0.00055  & 3.0        \\ 
      &         &                     &                    &                       &        &       &       &        \\
LOFAR & 144~MHz & 7.464 $\pm$ 0.100   & 13.415 $\pm$ 0.462 & 100                   &        &       & 6 $\times$ 6      &         \\
VLASS1 &  3~GHz & 0.965 $\pm$ 0.118   & 1.385 $\pm$ 0.266  &                       &        &       & 2.5 $\times$ 2.5       &        \\
VLASS2 &  3~GHz & 1.034 $\pm$ 0.145   & 1.942 $\pm$ 0.340  &                       &        &       & 2.5 $\times$ 2.5       &       \\ \hline 
\end{tabular}
\end{table*}

\begin{table}
\caption{Spectral indices for J1641+3454. Columns: (1) bands used; (2) spectral index using peak flux densities; (3) spectral index using integrated flux densities.}
\label{tab:j1641spind}
\centering
\footnotesize
\begin{tabular}{lcc}
\hline\hline
bands          & $\alpha_{\textrm{peak}}$ & $\alpha_{\textrm{int}}$  \\  \hline\hline
LOFAR - VLASS2 & $-0.65$ $\pm$ 0.05         & $-0.64$ $\pm$ 0.07          \\
VLASS2 - X     & $-1.24$ $\pm$ 0.16         & $-1.26$ $\pm$ 0.18           \\
X - Ku         & $-0.76$ $\pm$ 0.22         & $-1.74$ $\pm$ 0.23           \\ 
Ku - K         & $-0.95$ $\pm$ 0.29         &                           \\
K - Ka         & $-0.90$ $\pm$ 0.55         &                           \\ \hline 
\end{tabular}
\end{table}

\begin{table}
\caption{Single-dish detections for J1641+3454. Columns: (1) telescope; (2) central frequency of the observation; (3) flux density and its error; (4) date of the observation.}
\label{tab:j1641singledish}
\centering
\footnotesize
\begin{tabular}{lcccc}
\hline\hline
telescope & frequency & $S_{\mathrm{int}}$ & date \\
          & (GHz)     & (mJy)              & (dec. year)   \\  \hline\hline
MRO		& 37.0	& 460 $\pm$	80	&  2015.993015          \\         
MRO		& 37.0	& 280 $\pm$	70    &  2018.208142         \\         
MRO		& 37.0	& 370 $\pm$	90	&  2019.643459         \\        
MRO		& 37.0	& 650 $\pm$	120	&  2019.684388         \\         
MRO		& 37.0	& 380 $\pm$	90	&  2020.395579        \\         
MRO		& 37.0	& 510 $\pm$	110	&  2020.399177        \\         
MRO		& 37.0	& 490 $\pm$	120	&  2021.265503        \\        
MRO		& 37.0	& 480 $\pm$	110	&  2021.281906        \\          
MRO		& 37.0	& 480 $\pm$	90	&  2021.689678        \\       
MRO	    & 37.0	& 350 $\pm$	80	&  2021.779156      \\	        
MRO		& 37.0	& 430 $\pm$	100	&  2022.172608       \\ 
OVRO    & 15.0  & 30.5 $\pm$  3.3   &  2021.842500    \\ \hline 
\end{tabular}
\end{table}


\subsection{Flare characteristics using MRO data}
\label{sec:flarecharacteristics}

We can use the consecutive MRO detections to infer some properties of the radio emission in our sources. Following \citet{1999valtaoja1} and \citet{2009hovatta1} we can estimate the flare rise and decay $e$-folding timescales, variability brightness temperatures, and variability Doppler factors. We performed these calculations for all consecutive detections -- that is, there are no non-detections between them -- that were less than seven days apart and had different flux densities even when taking the errors into account. We cannot be sure if the two detections are from the same flare, but in case they are not, it means that the variability is even faster and more extreme. We also assume that the maximum amplitude of the flare is equal to the higher of the two flux densities. In case it is not, and the real amplitude of the flare is larger, the timescales would be shorter. Thus, these timescale estimates, and the parameters derived from them can be consider as lower limits. For simplicity, since our knowledge of these sources is so limited, we used the same equation for both rising and decaying flares:
\begin{equation}
\Delta S (t) = \Delta S_{\textrm{max}} \textrm{e}^{(t-t_{\textrm{max}}) / \tau} \quad [\mathrm{Jy}], \label{eq:tau}
\end{equation}
where $\Delta S_{\textrm{max}}$ is the maximum amplitude of the flare in Jy, after subtracting the baseline flux density level, $S_{\textrm{b}}$, $t_{\textrm{max}}$ is the epoch of the peak of the flare, and $\tau$ is the rise or decay time of the flare expressed in days ($e$-folding timescale). We do not know the exact quiescent flux density level, but based on the OVRO observations it cannot be much higher than $\sim$10~mJy (see Sect.~\ref{sec:obseffects}), so we chose this number as the baseline flux density level. The results are shown in Table~\ref{tab:flaretau}.

In order to estimate the variability Doppler factors of our sources, we calculated the variability brightness temperature, $T_{\textrm{b, var}}$, (in the source proper frame) with:
\begin{equation}
T_{\textrm{b, var}} = 1.548 \times 10^{-32} \frac{\Delta S_{\textrm{max}} d_L^2}{\nu^2 \tau^2 (1+z)} \quad [\mathrm{K}],
\end{equation}
where $\nu$ is the observed frequency in GHz, $d_{L}$ is the luminosity distance in metres, and $\Delta S_{\textrm{max}}$ and $\tau$ are defined in Eq.~\ref{eq:tau}. The numerical factor corresponds to using $H_0 = 72$~km s$^{-1}$ Mpc$^{-1}$, and $\Omega_\Lambda = 0.73$, and to assuming that the source is a homogeneous sphere. Since estimating the brightness temperature from the flux density variability is based on a causality argument, these values are in fact lower limits. We calculated the variability brightness temperatures for all flares with $\tau$ values. It should be kept in mind that the brightness temperatures derived from variability are systematically larger by a factor of $\delta^2$, where $\delta$ is the Doppler factor, than those obtained directly from VLBI measurements due to the different dependence on the Doppler factor. 

Once we know the variability brightness temperature we can use it to estimate the variability Doppler factor, assuming we know the intrinsic brightness temperature, $T_{\textrm{b, int}}$:

\begin{equation}
\delta_{\textrm{var}} = \left( \frac{T_{\textrm{b, var}}}{T_{\textrm{b, int}}} \right) ^{1/3} .
\end{equation}

For the intrinsic brightness temperature, we use 5\,$\times$\,10$^{10}$\,K \citep{1994readhead1, 1999lahteenmaki1}, which assumes equipartition between the energy densities of the magnetic field and the radiating particles. However, we do not know if these sources really are in equipartition and therefore cannot say how accurate the Doppler factor estimates are. Indeed, the rapid variability suggests that this may not be the case, thus these estimates should be taken with a grain of salt.

Keeping these caveats in mind, the results are reported in Table~\ref{tab:flaretau}. There are three sources with consecutive MRO detections within one week: J1228+5017, J1509+6137, and 1510+5547, but after excluding all the detections that can be the same within the error bars, only one source, J1509+6137, remains. It has shown two rising and one decaying flare that meet our criteria. In all cases the \textit{e}-folding timescales are of the order of days, or a maximum of a few weeks, the variability brightness temperatures around $10^{14}$-$10^{15}$\,K, and the variability Doppler factors between 5 and 50. These parameters, except the timescale, are comparable to what is seen in flat-spectrum radio quasars \citep{2009hovatta1}.

We can use a simple light travel time argument to infer an approximate size of the radio emitting region. The size needs to be $r < c \tau \delta / (1 + z)$. Assuming $z \sim$ 0, for $\tau$ of five days this gives 0.0042~pc $\times~\delta$ and for ten days 0.0084~pc $\times~\delta$. Taking into account the Doppler factor the size of the emitting region can increase by about an order of magnitude. These sizes are rough estimates since we cannot properly estimate the timescales with the current data, but it is probably safe to assume that the order of magnitude is correct and that the emitting region needs to be milliparsec in size. This indicates that the emission originates close to the black hole, well within the BLR, or from spatially limited regions inside the jet.

\begin{table*}
\caption{Flare properties from the MRO data. Columns: (1) source; (2, 3) start and stop times of the flare; (4, 5) flux density in the beginning and in the end of the flare; (6) timescaley; (7) variability brightness temperature; (8) variability Doppler factor.}
\label{tab:flaretau}
\centering
\footnotesize
\renewcommand{\arraystretch}{1.5}
\begin{tabular}{lccccccc}
\hline\hline
source     & start time  & stop time    & $S_{\textrm{begin}}$ & $S_{\textrm{end}}$ & $\tau$                            & $T_{\textrm{b, var}}$                & $\delta_{\textrm{var}}$ \\ 
           & (dec. year) & (dec. year)  & (Jy)                 &  (Jy)              & (day)                             & (10$^{14}$ K)                        & \\ \hline
rising     &             &              &                      &                    &                                   &                                      & \\
J1509+6137 & 2016.415428 &	2016.418182 & 0.48 $\pm$ 0.10      & 0.81 $\pm$ 0.18    & 2.0 $\substack{+13.1 \\ -0.9}$ & 17.2 $\substack{+55.1 \\ -16.9}$  & 32.5 $\substack{+20.0 \\ -24.4}$  \\
J1509+6137 & 2017.391984 &	2017.397450 & 0.51 $\pm$ 0.12      & 0.97 $\pm$ 0.14    & 3.2 $\substack{+4.4 \\ -2.8}$  & 8.0 $\substack{+16.7 \\ -6.8}$    & 25.2 $\substack{+11.5 \\ -11.9}$ \\
decaying   &             &              &                      &                    &                                   &                                      &     \\	
J1509+6137 & 2017.397450 &	2017.413817 & 0.97 $\pm$ 0.14      & 0.61 $\pm$ 0.09    & 13.2 $\substack{+24.6 \\ -5.2}$ & 0.5 $\substack{+1.0 \\ -0.4}$     & 9.7 $\substack{+4.5 \\ -5.2}$   \\ \hline 

\end{tabular}
\end{table*}

\section{Discussion}
\label{sec:discussion}

All of their variability properties considered, these seven sources exhibit flux density variations at a level never observed in AGN before at high radio frequencies. The short variability timescales they show are rare, but not unheard of, even in the radio regime \citep{2013rani1}, whereas the amplitude of the variability -- three to four orders of magnitude -- coupled with the short timescales, is unprecedented to the best of our knowledge.

Based on the 37~GHz light curves (Figs.~\ref{fig:J1029lc-ul}-\ref{fig:J1641lc-ul}), including both detections and upper limits (see App.~\ref{app:mrodetails} for details), most of the sources are usually detected very close to the detection threshold of MRO. J1509+6137 -- that has not been detected in interferometric observations at all -- is an exception, and consistently shows activity that is clearly above the detection limit. In general there do not seem to be notable trends in the detections, other than that the sources are detected more when they are observed more, which is not surprising. In some sources (for example, J1228+5017 and J1232+4957) there seem to be higher upper limits crowding around detections, possibly indicating an increased level of activity during that particular epoch (but see App.~\ref{app:mrodetails} for caveats). In others, such as J1641+3454, the detections are embedded amongst upper limits that show no apparent trends of activity. On the other hand, many detections are not accompanied by other nearby observations at all.

At OVRO all detections, except the first detection of J1522+3934, are clearly above the detection threshold. However, the detectability at 15~GHz compared to 37~GHz is significantly different. Only three sources have been detected at 15~GHz and the highest detection rate is only 4.5 per cent. The sources with the highest detection rates at 37~GHz have not been detected at 15~GHz at all despite the comparable number of observations. This might indicate that the flaring behaviour is stronger, in terms of the amplitudes, towards higher frequencies. Though, it should be noted that for many sources most MRO detections are from the time before OVRO started monitoring them, so it is also possible that these sources have been less active throughout the OVRO observations.

For some sources (J1509+6137, J1522+3934, and J1641+3454) there are a few MRO detections with OVRO observations within $\sim$1-5 days before or after the MRO detection. Using these detections and the OVRO upper limits, these quasi-simultaneous observations can be used to estimate a lower limit for the 15-37~GHz spectral index. The spectral index lower limits are around 4 to 5, strongly suggesting that there are external factors in play resulting in the observed phenomenon.

Despite frequent detections at 37~GHz, and some at 15~GHz, all sources were in the low state in the JVLA observations. However, considering the low-to-moderate detectabilities (1.5-25 per cent) and the short timescales of the sources at 37~GHz, it is not infeasible that none of them were flaring at the time of the two epochs of the JVLA observations.

In the following we discuss different phenomena able to cause variability in AGN. It should be kept in mind that the physical explanation for the observed variability might not be the same in all sources or that it can be a combination of more than one mechanism. For completeness we include a number of explanations that we have been able to reject or that are unlikely to be responsible for the extreme behaviour. Since not much can be said regarding the sources that have very few detections only in some of the bands, the discussion mostly considers the sources with the most complete data.

\subsection{Rejected explanations}

More data, especially multifrequency monitoring of the flares, are absolutely necessary to narrow down the possible explanations, however, based on the current data some scenarios can already be ruled out. These alternatives alone cannot explain the observed properties of our sources, but we cannot totally discard their presence in them. 

\subsubsection{Normal relativistic jets}

Based on the results in this paper and in \citet{2020berton2} it is obvious that the sources in our sample do not host persistent, continuously visible relativistic jets similar to those seen in other jetted NLS1s or any other class of jetted AGN. Several jetted NLS1s exhibit 37~GHz behaviour similar to the sources studied in this paper \citep{2017lahteenmaki1}, and all of them also show core or core-jet structures in mas-scale VLBI observations \citep[e.g., ][]{2011doi1, 2015richards2, 2016lister1}. In general, the VLBI flux densities of the previously studied jetted NLS1s vary from a few mJy to hundreds of mJy, and thus are at a level that should have been easily detectable in our VLBA observations. However, the non-detections of these sources either imply that the radio core is very faint, $<$ 0.5~mJy, or possibly absorbed (see Sect~\ref{sec:pureffa} and ~\ref{sec:ffaclouds}). 

We did not expect to be able to resolve the possible jet with the JVLA -- except perhaps in the highest-frequency bands -- since the flaring behaviour implies that we are seeing these sources at quite small angles. However, our initial assumption, again based on the observations of other jetted NLS1s, was that these sources would show flat or inverted spectra toward higher frequencies. Only one of our sources, J1228+5017, shows a radio spectrum that can be deemed flat, and none of the detected sources show any hints of an inverted spectrum in the JVLA observations. Regarding the non-detected sources, from these results we can only infer that their spectra do not turn inverted toward higher frequencies.

With these combined results we are able to reliably rule out the possibility that the variability in our sources is due to flares in a relativistic jet similar to those in other jetted NLS1s or AGN. This does not necessarily mean that the jet is absent, but in the low state it seems to be undetectable, implying that there must be also other contributors to the observed behaviour.

\subsubsection{Kinematically young jets}
\label{sec:kineyoungjets}

These results also rule out one of our early hypotheses, that these sources would be kinematically young and have considerably high radio spectrum turnover frequencies due to that \citep{2021odea1}. The 37~GHz behaviour could be explained as radio flares superimposed on a convex radio spectrum of a peaked source \citep{2001tornikoski1, 2005torniainen1, 2009tornikoski1}. Obviously this is not the case, as we do not see any signs of spectra resembling those of peaked sources. Also the long-term temporal behaviour disagrees with this scenario, since several of these sources have been detectable at 37~GHz at the same flux density level for the past $\sim$ten years, ever since the observations first started. In case of a kinematically young source, the turnover frequency is expected to decrease very fast during the early stages of its life, staying above $>$ 40~GHz only for 6-20 years \citep{2020berton2} -- the kind of evolution we should be able to recognise at 37~GHz, and also at 15~GHz, as increasing or decreasing detectability, or as long-term permanent changes in the flux density levels. There are a few sources that have not been detected during the past few years even when they have been observed regularly (J1232+4957 and J1510+5547), which indicates temporal changes in these sources. Even in these cases kinematically young jets seem improbable since the evolution is not so fast that we would not have been able to detect a convex spectrum at lower frequencies with the JVLA. It should be noted that whereas kinematically young jets with SSA cannot explain the behaviour of our sources, it does not mean that the jets in these sources could not be young.

\subsubsection{Fast radio bursts}
\label{sec:frb}

The seemingly sporadic detectability, implying very short timescales, raised the question of whether this phenomenon could be related to fast radio bursts (FRB). FRBs are short, sub-second duration broad-band Jy-level pulses of extragalactic origin \citep[for a recent review, see][]{2022petroff1}. Several repeating FRBs have been found, and in principle they could fall into the MRO beam during an observation. In practise, it is very unlikely that such an event could account for the detections of these sources: first, the moderately long 1600-1800~s integration time used at MRO would average out even a Jy-level, sub-second pulse to an undetactable level, and second, FRBs have very steep spectra with an average spectral index of -1.5 \citep{2019macquart1}, making them fainter and even harder to detect at high radio frequencies.

\subsubsection{Tidal disruption events}
\label{sec:tde}

Tidal disruption events (TDE) occur when a star passes by too close to a supermassive black hole and gets disintegrated. In some extreme cases these events can result in launching of (mildly) relativistic jets, reaching luminosities around 10$^{42}$ erg s$^{-1}$ , and therefore possibly bright enough to explain our 37~GHz detections \citep[][and references therein]{2020alexander1}. However, the timescales of TDEs are in the range of tens to hundreds of days and thus not compatible with the behaviour of our sources. Furthermore, so far a TDE has never been observed twice in the same source, and thus it seems extremely improbable that repeated detections over ten years could be due to TDEs. There are some records of partial TDEs \citep{2015campana1} when the whole star does not get destroyed but continues to orbit the black hole, causing small TDEs once per orbit. Whereas partial TDEs could be responsible for repeated radio flares, they are unlikely to produce variability at a timescale of days.

\subsection{Unlikely explanations}

In the following we discuss some alternatives that are unlikely, but cannot be totally ruled out yet, or are not able to explain our sources on their own, but might contribute to the observed properties.

\subsubsection{Observational effects}
\label{sec:obseffects}


Interestingly, it seems that in all cases an inverted spectrum or a high state is seen only in single-dish observations, whereas interferometric observations show a barely flat or a steep spectrum, if the source is detected at all. This raises the question of whether the difference could be explained by contamination by nearby compact sources that the larger beams of the single-dish telescopes pick up, or by emission resolved-out with radio arrays. The first explanation -- different beam sizes -- can be ruled out since based on the JVLA images mapping the OVRO beam there are no other strong radio sources close to any of our targets, and thus even the largest beams (MRO and OVRO) should not suffer from confusion. 

On the other hand, resolved-out emission can contribute to the discrepancy, but not explain all of it. In A-configuration the largest angular scales that the JVLA can see are approximately 5.3, 3.6, 2.4, 1.6, and 1.2~arcsec in X, Ku, K, Ka, and Q bands, respectively. In the worst case scenario, the lowest-$z$ source in Q band, this translates to 1.70~kpc. It is obvious that emission at these scales cannot explain the variability timescales seen in our sources. There can be a contribution from the resolved-out emission, but, for example, at 37~GHz based on the MRO detection threshold, it cannot exceed $\sim$200-300~mJy, otherwise we would be able to detect these sources much more frequently. Similarly, OVRO, with a beam of the same size as MRO, gives an upper limit of $\sim$10~mJy for the 15~GHz resolved-out emission. Since there are no emission sources at kpc-scale that can produce such an inverted spectrum between 15 and 37~GHz, it is reasonable to assume that the real 37~GHz flux density is at a similar or lower level than the 15~GHz flux density, suggesting that extreme variability is still present.

In addition, based on the preliminary results of our JVLA monitoring campaign of J1522+3934 using the B-configuration in X and K bands (VLA/23A-061, PI Berton), the beam size does not have a significant impact on the flux density. In the B-configuration the beam is about three times larger than in the A-configuration in both bands, and also the largest detectable angular scales -- 17~arcsec in X and 7.9~arcsec in K, 24.1 and 11.2~kpc at the redshift of J1522+3934, respectively -- are significantly more extended than in A-configuration. However, the observed flux densities in A- and B-configurations are the same within the errors, further supporting that any resolved-out emission is not able to explain the difference.

\subsubsection{Precessing jet}
\label{sec:precession}

One alternative to explain variability in AGN is the precession in the jets \citep[e.g.][]{2011kudryavtseva1}, leading to changes in the viewing angle and thus in the strength of relativistic boosting. Precession can be caused by a tilted accretion disk via different mechanisms, such as the radiation-driven warping instability \citep{1996pringle1} or the Bardeen-Petterson effect \citep{1975bardeen1} due to Lense-Thirring precession \citep{1918thirring1}. Precession can also be observed in binary supermassive black hole systems \citep{1980begelman1}. However, in all these cases the expected, and so far observed, precession period is of the order of years \citep[e.g.][]{2011kudryavtseva1, 2018liska1, 2020horton1}, rather than days as in our case. It is therefore unlikely that precession on its own could explain the properties of these sources.

\subsubsection{Intermittent activity}
\label{sec:intermittentjet}

The lack of detectable jets in these NLS1s might indicate a kinematically young age -- that was already discussed in Sect.~\ref{sec:kineyoungjets} -- or intermittent activity. Intermittent activity due to radiation pressure instabilities in the accretion disk was evoked to explain the excessive number of kinematically young radio AGN, such as GPS sources, and especially their subclass of compact symmetric objects \citep[CSOs][]{2009czerny1}. For a black hole with a mass of $10^8 M_{\odot}$ the duration of the activity phases is estimated to be $10^3$- $10^4$ years, and the breaks between them $10^4$- $10^6$ years. For lower black hole mass sources, such as NLS1s, these timescales are shorter, but certainly not short enough to explain the variability we are observing.

Also 3D general-relativistic magnetohydrodynamic (GRMHD) simulations have yielded similar results; \citet{2022lalakos1} find that before establishing stable, powerful relativistic jets an AGN can go through several cycles of intermittent activity, with the jets turning on and off and drastically changing direction. This leads to an X-shaped radio morphology seen in 5-10 per cent of radio galaxies, and, naturally, considerable variability. Using the results in \citet{2022lalakos1} we can estimate that the launch-to-quench timescale for a black hole with a mass of $10^7 M_{\odot}$ is 10-100 years, and the jets re-emerge after 100-1000 years. The timescale is too long for our sources, but it suggests that in lower black hole mass AGN we could be able to follow, at human timescales, the chain of events from the initial launch of the jets until they are quenched by the infalling gas. As low black hole mass jetted sources NLS1s could be an optimal target for these kind of studies. 

Shorter timescale intermittency can manifest itself as a result of changing injection rate of plasma into the jet base / jet \citep[e.g.][]{2013lohfink1, 2023fedorova1}. Between these events the jet can be totally absent or very weak, possibly explaining the low state of our sources. What remains unclear is whether these kind of events can account for the required short timescales and high variability amplitudes, and how these events manifest themselves in the radio regime. The classical viscous and thermal timescales associated with an accretion disk around a black hole with a mass of $\sim$10$^7 M_{\odot}$ are too long to explain the variability, whereas the magnetic timescale dominating the inner parts of the disk can be considerably shorter \citep{2003livio1, 2004king1}. The magnetic timescale is the time on which the poloidal magnetic fields in different parts of the disk can spontaneously align, possibly changing the dissipation in the disk and its coupling to the jet. Local changes in the magnetic field alignment can cause small-amplitude flickering at very short timescales, whereas large-amplitude events, where the magnetic field is aligned in a considerable fraction of the disk, are more rare. Thus this kind of intermittency could possibly explain either the short timescales or the high amplitudes, but not both. 

It is worth noting that even if intermittent activity would not be the culprit in this case, we do see signs of that among these sources. Assuming that we are now observing the jets in our sources at small angles as indicated by the variability, it is evident from the misalignment between the radio emission and the host galaxy in Fig.~5, panel c) in \citet{2021jarvela1} that J1522+3934 has experienced an earlier activity period. However, the projected size of the structure is almost 20~kpc, well beyond the host galaxy, implying that the activity period has been longer than what would be expected in the aforementioned scenarios. Based on the current data we also cannot determine whether the jets turned off or just changed direction.

\subsubsection{Pure FFA}
\label{sec:pureffa}

A possible way to explain the flares is to assume that the underlying radio emission of the relativistic jet is totally free-free absorbed by ionised gas in the low state, and would only occasionally break through the absorbing screen due to intrinsic flaring, or due to very fast drops in the absorption (see Sect.~\ref{sec:ffaclouds}). By solving the transfer equation, it is possible to prove that such a scenario is not impossible, as it does not require an unreasonable amount of gas. Let us assume that the radiation produced by the jet is free-free absorbed as follows: 
\begin{equation}
    I_\nu = I_{\nu,0} e^{-\tau_\nu}  \quad [\mathrm{erg\,s^{-1}}],
\end{equation}
where $\tau_\nu$ is the optical depth, $I_{\nu,0}$ is the radiation produced by the jet, and $I_\nu$ is the radiation we observe after it has crossed the ionised gas. For simplicity, let us do our calculation at 10~GHz, and assume that the jet emission is not detected. The detection threshold of the JVLA for our observations in the X band is 10~$\mu$Jy, so we can assume an upper limit for the observed flux density of 30~$\mu$Jy. Let us also assume that the jet has an underlying flat spectrum, and that the unabsorbed flux density at 10~GHz is 1~Jy. Using the previous equation, we can obtain an optical depth $\tau_\nu \sim 10$. The optical depth of the ionised gas cloud depends on the absorption coefficient $k_\nu^{\mathrm{ff}}$, following
\begin{equation}
    \tau_\nu = \int_0^l k_\nu^{\mathrm{ff}}dr ,
\end{equation}
where $l$ is the size of the absorbing cloud. The free-free absorption coefficient is
\begin{equation}
    k_\nu^{\mathrm{ff}} \simeq 3.69\times10^8 Z^2 \frac{N_eN_i}{\sqrt{T_e}} \frac{1}{\nu^3} g_{\mathrm{ff}}  \quad [\mathrm{cm^{-1}}],
\end{equation}
where $N_e$ is the electron number density, $N_i$ is the number density of the ions, $T_e$ the electron temperature, $Z$ the atomic number, and $g_{\mathrm{ff}}$ is the Gaunt factor. Assuming hydrogen gas ($N_e = N_i$), and using the approximation of the Gaunt factor between 0.3 and 30~GHz, the coefficient becomes
\begin{equation}
    k_\nu^{\mathrm{ff}} \simeq 0.21 N_e^2 T_e^{-1.35} \nu^{-2.1}  \quad [\mathrm{cm^{-1}}].
\end{equation}

If we integrate this assuming that the cloud has a uniform density and temperature, the optical depth becomes
\begin{equation}
    \tau_\nu^{\mathrm{ff}} \simeq 0.0824 T_e^{-1.35} \nu^{-2.1} N_e^2 l .
\end{equation}
Inverting this equation, we can derive
\begin{equation}
    l \simeq \frac{\tau_\nu^{\mathrm{ff}}}{0.0824 N_e^2} T_e^{1.35} \nu^{2.1}  \quad [\mathrm{pc}].
\end{equation}

Since we now know that $\tau_\nu^{\mathrm{ff}} \sim 10$, we can try to calculate the size of the absorbing clouds by assuming different values of electron density and temperature, at the frequency of 10~GHz. For $N_e = 10^4$\,cm$^{-3}$ and $T_e = 10^4$\,K, which are rather typical values, we obtain $l = 38$\,pc. For a higher density, possibly similar to the conditions of a shock, of $N_e = 10^5$\,cm$^{-3}$ and $T_e = 10^5$\,K, the size decreases to $l = 8.6$\,pc. Such size is comparable to that of the Orion Nebula. Finally, if $N_e = 10^5$\,cm$^{-3}$ and $T_e = 10^4$\,K, the resulting $l = 0.38$\,pc, which is too small for a star forming region, but may be closer to the expectations of a region of gas ionised via shock by the jet itself. Due to the $\nu^{2.1}$ dependence, the required size of the ionised cloud increases at higher frequencies. For example, at 50~GHz it would need to be $\sim$30 times larger to effectively absorb all the emission. This would imply sizes of hundreds of parsecs, unlikely ionised by the AGN, but of a characteristic size for a star forming region \citep[e.g.,][]{2023congiu1}. Lower densities and temperatures instead require unreasonably large sizes. For instance, $N_e = 10^3$\,cm$^{-3}$ and $T_e = 10^4$\,K lead to $l = 3.8$~kpc, which is not realistic since this requires a uniform distribution of ionised gas as large as a small galaxy. 

Even if the previous considerations show that this scenario is feasible, there are some issues that we cannot ignore. First of all, in this scenario in the low state the jet emission needs to be totally absorbed -- otherwise we would see an inverted spectrum -- thus the JVLA radio emission needs to originate outside the absorbed region. Were the absorption due to a star forming region, it could as well be the source of the faint low-state radio emission. As the star forming region cannot explain the variability, it would have to be intrinsic to the jet that would occasionally get bright enough to break through the FFA screen. However, assuming that the underlying relativistic jet is similar to those in other jetted NLS1s, we would assume the timescales to be comparable too, which is not the case.  

Another way of producing the observed flares is by means of a variable optical depth, which in turn requires either fast moving clouds (see Sect.~\ref{sec:ffaclouds}) or a rapid propagation of the jet throughout an interstellar medium with variable density and temperature. 

\subsubsection{Geometrical effects}

The changes in the Doppler factor due to circumstances internal or external to the jet have been evoked to explain large-amplitude flares in AGN. Such circumstances could be result of changes in the orientation of the jet, or parts of it, or due to the jet substructure, such as a helical magnetic field \citep{1999villata1, 2010mignone1, 2017raiteri1, 2021raiteri1}. This variability is characterised by achromatic frequency behaviour in the affected bands. 

For example, an FSRQ CTA 102 has shown in the optical a somewhat similar behaviour to what we see in our sources in radio \citep{2017raiteri1}. The source increased its optical magnitude by six magnitudes, but in comparison the other frequencies were almost unaffected by the flare. In our case the flare seems to predominantly affect the radio emission and not other wavelengths \citep{2023romano1}. \citet{2017raiteri1} suggested that the variability in CTA 102 was caused by changes in the viewing angle due to peculiar jet geometry. If this is the case, we are observing different regions of the jet at different angles. In our sources only the radio emission would be seen at small viewing angle, experiencing stronger relativistic boosting due to the higher Doppler factor. This scenario could be consistent with what is seen in several simulations. Jets propagating in dense ISM cannot proceed in a straight line but tend to wiggle around the least resistance path \citep{2012wagner1}.

To estimate the feasibility of this scenario, we can estimate the level of change in the Doppler factor required to explain the extreme variability we observe in our sources. We assume the unbeamed flux density, $S_0$, to be at the level of the JVLA values, and the beamed, $S_{\mathrm{obs}}$, to be close to the MRO detections. The emission is boosted as:
\begin{equation}
    S_{\mathrm{obs}} = S_{0} \delta ^p ,  \label{eq:boost}
\end{equation} 
where $p$ = $2 - \alpha$ for a continuous jet stream, and  $p$ = $3 - \alpha$ for a transient emission region, such as a blob or a knot in the jet. We assume the jet spectral index to be $\alpha$ = 0. A few different cases of the unbeamed and beamed flux densities are shown in Table~\ref{tab:doppler}. The Doppler factors in case of a continuous stream are very high, but more reasonable in the case of a transient emission region in the jet. We calculated the required change in the viewing angle resulting in the estimated changes in the Doppler factor (Table~\ref{tab:doppler}). We did the calculation with two different Lorentz factors ($\Gamma)$ characteristic for jetted NLS1s: 10 and 20 \citep{2009abdo2}. In case of the continuous stream, when $p$ = 2, $\Gamma$ = 10 is not high enough to reach the Doppler factors shown in Table~\ref{tab:doppler}, and even $\Gamma$ = 20 yields results only in case of $\delta$ = 50 ($\Delta \theta$ = 18$\degree$), thus we list the viewing angle changes only for the $p$ = 3 case. The required changes are not unreasonable, for example, in \citet{2017raiteri1} the viewing angle change is $\sim$9~$\degree$. However, in their case the timescale of the change is of the order of several weeks, whereas in our case it is of the order of days. Also other issues remain, as discussed below.

This hypothesis requires a relativistic jet to be present, but we do not see any clear signs of this in any of our sources. In the first order approximation, in this scenario either the jet needs to change direction and consequently its Doppler factor, or new components, possibly with higher Lorentz factors, would need to be ejected. Also other factors, for example, temporal variability in the physical conditions of the jet -- such as the magnetic field, and the density and energy distribution of the relativistic particles -- may contribute, but their impact can be expected to be less significant.

If the changes are due to the re-orientation of some parts of the jet it is hard to explain why we observe the flaring behaviour only in radio. This might require the same part of the jet to consistently change its orientation, which does not seem likely. In this case the variability should be achromatic, which is something we cannot yet study with the current data. If the flares are due to new components ejected, we would expect to see the underlying jet also when it is not flaring, since it should be relativistically boosted also between flares unless the blobs have considerably higher Doppler factors than the continuous stream. In both these scenarios the emission comes from the whole jet and therefore require the emitting region to be very close, within the innermost parsec, to the black hole, to be able to match the estimated timescales. An alternative way of producing drastic changes in the Doppler factor only in some parts of the jet is magnetic reconnection, which will be discussed in Sec.~\ref{sec:magrec}.

\begin{table}[h]
\caption{Required Doppler factors and changes in the viewing angle of the jet. Columns: (1) unbeamed flux density; (2) beamed flux density; (3) required Doppler factor assuming a continuous jet stream; (4) required Doppler factor assuming a moving component in the jet; (5) required change in the viewing angle assuming $p$ = 3 and  $\Gamma$ = 10; (6) required change in the viewing angle assuming $p$ = 3 and  $\Gamma$ = 20.}
\label{tab:doppler}
\centering
\footnotesize
\begin{tabular}{lccccc}
\hline\hline
S$_0 $ & S$_{\mathrm{obs}}$ & $\delta_{\mathrm{stream}}$ & $\delta_{\mathrm{blob}}$ & $\Delta \theta_{\Gamma = 10}$ & $\Delta \theta_{\Gamma = 20}$ \\ 
(mJy)  & (mJy)              &  $p$ = 2                   & $p$ = 3                  & ($\degree$) & ($\degree$)\\ \hline\hline
0.05   & 500                &  100.0                     & 21.5                     & --          & 15.1 \\
0.05   & 1000               &  141.4                     & 27.1                     & --          & 15.7 \\
0.1    & 500                &  70.7                      & 17.1                     & 22.9        & 14.5 \\
0.1    & 1000               &  100.0                     & 21.5                     & --          & 15.1 \\ 
0.2    & 500                &  50.0                      & 13.6                     & 21.3        & 13.8 \\
0.2    & 1000               &  70.7                      & 17.1                     & 22.9        & 14.5 \\ \hline 
\end{tabular}
\end{table}

Another geometrical effect in relativistic jets that causes changes in the observed flux density is due to large-scale, ordered helical magnetic fields. If the jet is magnetically dominated, the magnetic field can drive helical streams within the jet. These streams can experience differential Doppler boosting along the jet when on one side of the helix the radiation gets relativistically boosted and on the other side it gets diminished \citep{1997steffen1, 2011clausenbrown1, 2018gabuzda1}. 

In case of a continuous stream we should be able to see the jet at all times, which is not the case, so we can assume that in this scenario the flares are caused by a blob moving in the jet, thus $p$ = 3 in Eq.~\ref{eq:boost}. For simplicity and to maximise the strength of the effect, let us assume a helical magnetic field seen exactly at the helix angle. Assuming constant $\beta$ the changes in the flux density only depend on the Doppler factor whose value depends on the angle between the helical stream within the jet and  the line of sight as:
\begin{equation}
    \delta = \frac{ \sqrt{1-\beta ^2} }{1 - \beta ~\mathrm{cos} \theta},  \label{eq:doppler}
\end{equation}
where $\beta$ = $v/c$, and $\theta$ is the angle compared to the line of sight. In our scenario $\theta$ has a minimum of 0$\degree$. Let us estimate the radius of the jet in case of the longest $e$-folding timescale in a MRO-detected flare from Table~\ref{tab:flaretau}; the 2017 decaying flare of J1509+6137. The flux density decreased from 970~mJy to 610~mJy in 5.97 days in our reference frame, thus in 5.97 days $\times$ $\delta$ in the source frame. Based on Table~\ref{tab:j1509interf} let us assume that $S_0$ = 0.1~mJy, and that $S_{\mathrm{obs,max}}$ = 970~mJy, which happens when $\theta$ = 0$\degree$. Using Eq.~\ref{eq:boost} we can estimate that the required Doppler factor at the maximum flux density is $\delta_{\mathrm{max}}$ = 21.3, and at $S$ = 610~mJy it is $\delta_{\mathrm{610~mJy}}$ = 18.3. Using Eq.~\ref{eq:doppler} with $\delta_{\mathrm{max}}$ = 21.3 and $\theta$ = 0$\degree$ we get $\beta$ = 0.996. Assuming $\beta$ stays constant, and using $\delta_{\mathrm{610~mJy}}$ = 18.3, we can solve for $\theta_\mathrm{610~mJy}$ = 2.41$\degree$. The radius of the jet can be solved from
\begin{equation}
    R = \frac{b}{2 \pi} \frac{360\degree}{2.41\degree} \mathrm{[m]},
\end{equation}
where $R$ is the radius of the jet, and $b$ is the distance the blob has travelled along the arc of the outer edge of the jet. In this case $b$ = 5.97 days $\times$ $\delta_{\mathrm{max}}$ $\times$ 0.996~$c$ = 3.28 $\times$ 10$^{15}$~m, and $R$ = 7.80 $\times$ 10$^{16}$~m = 2.53~pc. This is the least extreme case, and in other flares where the changes were faster also the radius of the jet would have to be smaller to account for the variability. In cases when $\theta_{\mathrm{min}} \neq$ 0, $\beta$ would have to be larger to result in the same $\delta$, and $R$ would have to be smaller than in the $\theta_{\mathrm{min}}$ = 0 case.

Based on other AGN, jet diameters of a few parsecs are measured at projected distances from $\sim$1 to 10~pc from the AGN core \citep{2020kovalev1}, and thus most likely outside the BLR. This brings us to the same question again: where is the jet when it is not flaring? Though it should be noted that we estimated the radius assuming the most favourable conditions for Doppler factor changes, thus it is likely that in reality the radii should be smaller, but by how much is unclear.


\subsection{Viable explanations}

\subsubsection{Jet - cloud/star interaction}
\label{sec:jetisminteraction}

Shocks in the interaction region of a jet and ISM can efficiently accelerate the electrons and thus increase the observed flux density of the jet \citep[e.g.][]{1992fraixburnet1}. In the case that the ISM consists of clumpy clouds only, parts of the jet might come in contact with them resulting in regions of enhanced emission that are smaller than the radius of the jet \citep{2000gomez1}. Particularly relevant in our case is the possible interaction between the jet base and BLR clouds or stars \citep{2010araudo1, 2019delpalacio1, 2012boschramon1}. According to simulations the timescales of these events are of the same order as the estimated timescales of our sources, that is, from less than a day to a few days, and they can considerably increase the luminosity of the source. Whereas the timescales fit our observations, a BLR cloud or a massive star entering the jet is expected to produce a flare that should be observable over the whole electromagnetic spectrum, which is behaviour not consistently observed in our sources. On the other hand, based on \textit{Fermi} data, there does not seem to be strong evidence pointing at the BLR photons interacting with the jet since most blazars do not show the expected high-energy cut-off \citep{2018costamante1}. However, this result can be explained if the main gamma-ray-emitting region in AGN is outside the BLR and swamps the gamma-ray emission originating inside the BLR. As a result, jet - cloud/star interaction can still cause flares observable in lower energies, for example, in the optical and radio regimes \citep{2000romero1, 2002romero1}.

The issue of the missing jet also remains with this explanation. Although if the jet is small and embedded in the BLR clouds, also FFA could play a role in this scenario. Furthermore, since no dedicated simulations exist, it is unclear what the temporal evolution of these flares in radio is. More detailed simulations will be required to estimate if this hypothesis could provide a feasible explanation for the extreme variability of our sources.




\subsubsection{Relativistic jet and FFA with moving clouds}
\label{sec:ffaclouds}

In this scenario the starting point is similar to that in Sect.~\ref{sec:pureffa}, but the region of ionised gas is not uniform and stationary, but consists of moving ionised gas clouds. The AGN would be totally free-free absorbed most of the time and the flares take place when the nucleus is temporarily revealed. In other words, the behaviour we observe would be caused by a combination of obscuration and geometry, and not by an intrinsic change in the jet activity. Some support for this hypothesis was found in J1641+3454 in which no absorption was detected in X-rays just after a flare when the nucleus probably was exposed, but a possible warm absorber is seen in the X-ray spectrum when the source is in a low state (Lähteenmäki et al. in prep.) In this scenario the timescale only depends on the size of the gap in the clouds, its distance from the radio emitting source, and its orbital velocity, so the timescales can be arbitrarily short.

In this hypothesis the covering medium would most likely be ionised BLR clouds that are considerably denser and smaller than ISM clouds. The BLR clouds can be as dense as $N_e$ = 10$^{11}$~cm$^{-3}$ \citep{1992ferland1} with sizes around 100-400 solar radii and thus easily able to absorb bright radio emission even at high frequencies. The covering factor of the BLR in optical is believed to be around 10-50 per cent, but reaching $\sim$100 per cent towards certain directions \citep{2009gaskell1}.

However, open questions remain also in this case. This scenario requires that the jets of these sources are kinematically very young and still within the BLR, and also that their advancement is hindered enough by the BLR so that they have stayed within the BLR our whole observing period, about 10 years. Assuming a BLR outer radius of 0.1~pc, the jet propagation speed would have to be $\lesssim$ 0.03~\textit{c} for this hypothesis to be viable. \citet{2012wagner1} report a jet propagation speed of 0.003-0.16~\textit{c} in the presence of clouds impeding its progress. Thus a slow jet could stay within the BLR up to a hundred years, easily enough for our case \citep{2021kino1, 2023savolainen1}.

\subsubsection{Magnetic reconnection}
\label{sec:magrec}

Magnetic reconnection in the jet or in the black hole magnetosphere has been evoked to explain fast variability in AGN, especially at GeV and TeV energies \citep[e.g.,][]{2010degouveiadalpino1, 2013giannios1, 2015kadowaki1, 2020shukla1}. It can account for high-amplitude variability at timescales from minutes to days. If magnetic reconnection were to take place in the jet in the form of so-called jets-in-jets or minijets \citep[e.g.,][]{2008ghisellini1, 2009giannios1, 2011nalewajko1}, the jet would need to be heavily absorbed, since it remains undetected, and it would likely still be within the BLR. Proof of classic gamma-ray flares happening inside the BLR does exist \citep{2013vovk1, 2015liao1}, and also signs of gamma-ray pair attenuation have been found \citep{2010poutanen1}, further suggesting that flares can happen inside the BLR. However, the research so far has concentrated on the high-energy characteristics of minijets, and the production of radio emission and flares in the context of magnetic reconnection in the jet has not been studied, thus it is unclear whether this scenario could result in the behaviour we see in our NLS1s.

An alternative for the magnetic reconnection in the jet is the magnetic reconnection in the black hole magnetosphere \citep[e.g.,][]{2010degouveiadalpino1, 2015kadowaki1, 2022ripperda1, 2022kimura1}. The advantage of this explanation is that it does not require the presence of a permanent relativistic jet. The emission characteristics of these kind of events have been studied utilising general-relativistic magnetohydrodynamic simulations \citep{2022ripperda1} and also development of the theoretical framework has been started \citep{2022kimura1}, but we still lack any direct evidence of this. \citet{2010degouveiadalpino1} and \citet{2015kadowaki1} argue that the radio and gamma-ray emission in low-luminosity AGN can be explained with magnetic reconnection in the black hole magnetosphere, whereas blazars also require a significant contribution from the relativistic jet. Based on their model, an effectively accreting black hole with a mass of 10$^{7}$ M$_{\odot}$ and turbulence-induced fast reconnection can show magnetic reconnection power spanning from 10$^{39}$ to 10$^{43}$\,erg\,s$^{-1}$ and thus likely enough to explain the flares in our sources. 


\subsection{Implications}

It is evident that more data, especially simultaneous multifrequency observations of the flaring state, are required to determine the origin of the extreme variability seen in these NLS1s. Already based on the current data, the most common causes of radio variability in AGN can be ruled out, or they would require considerable fine-tuning. The strictest requirements come from the variability timescales, especially coupled with the extremely high, three to four orders of magnitude, amplitude of the variability. The timescales are extraordinarily short and therefore require a compact, milliparsec-scale, emitting region, or, alternatively, a peculiar interplay between the jet and the BLR clouds. Whereas intrinsic variability mechanisms allowing very short timescales and high amplitudes exist, most of them are still very little studied or only based on simulations or theoretical work, and lack observational evidence. To determine if any of them could explain the behaviour of our sources, more detailed theoretical framework, possibly dedicated simulations, and especially targeted observations will be needed. It should be kept in mind that we cannot exclude the possibility that we are seeing a new type of variability either. Either way, catching flares in these sources will be challenging due to the short timescales and sporadic activity, but considering that these NLS1s exhibit one of the most extreme radio variability seen in AGN so far, they do deserve our full attention.

One of the most interesting aspects of the discovery of these sources is that they were found among two very differently selected samples whose final detection percentage at MRO turned out to be very high at 12 per cent -- eight sources out of 66 were detected. Whether our selection criteria actually helped us select NLS1s exhibiting this behaviour or if it was pure coincidence remains unclear. Observations of other NLS1 samples selected using similar and different selection criteria will be needed to estimate the impact of the selection effects.

Either way, detecting $>$10 per cent of a presumably mostly radio-silent NLS1 sample is extraordinary and raises the question whether this variability phenomenon is characteristic of NLS1s or possibly early-stage AGN, or if similar sources are hiding also among radio-weak AGN of other classes. For obvious reasons radio-weak AGN have not been a target of extensive high radio frequency monitoring campaigns and we therefore know very little about their behaviour in that regime. It is possible that also strong radio AGN exhibit this kind of behaviour, but that it is swamped by other sources of radio emission and thus has remained undetected. Investigating in which kind of sources this phenomenon can be seen can help us to determine the cause of the variability. Being able to identify any common properties these sources have will also help us to find more of them.

Whether this kind of variability is limited to early-stage AGN or if it is a more common phenomenon has implications on our current understanding of AGN. These sources clearly represent an unknown population of AGN, that has gone unnoticed so far. If they are a new type of jetted AGN or something else entirely, is unclear, as is their evolution and relation to other classes of AGN. Furthermore, we do not know how common they are or if they are characteristic to the local Universe, or also exist at higher redshifts. Further studies are also required to estimate which kind of a role they play in AGN feedback over the cosmic time.

\section{Conclusions} 
\label{sec:summary}

In this paper we investigated the origin of the extreme radio variability seen in seven NLS1s using the JVLA, the VLBA, MRO, and OVRO observations. These extraordinary sources defy an easy explanation, but the new data presented in this paper allowed us to rule out some alternatives and set additional constraints on the possible explanations. Our main conclusions are:

\begin{itemize}
    \item The behaviour of these sources is hard to explain with the usual variability mechanisms in AGN -- instead a more complex scenario or possibly a new type of physical mechanism to produce variability is required.
    \item The amplitude of the variability -- three to four orders of magnitude -- seen in these sources is unprecedented, but it remains unclear whether it is intrinsic to the source, or caused by some external circumstances.
    \item The variability timescales indicate that if the variability is intrinsic the emitting region needs to be milliparsec in size. This implies that the emission originates close to the black hole, clearly inside the BLR, or from limited, confined regions in the jet.
    \item The high detection percentage among the original sample, that were not expected to be strong radio emitters, implies that this kind of sources could be quite common, but so far our understanding of this new population of AGN is very limited.
\end{itemize}

Revealing the nature of these peculiar sources is of utmost importance as they might be the first representatives of a new type of AGN variability, and/or a new class of jetted AGN entirely. In the future, an increase in the sample size will be essential to explore this new population. Their short timescales and sporadic activity pose an observational challenge, also given how diverse their behaviour is at different frequencies. High-cadence multifrequency radio monitoring with an instrument sensitive enough to detect also the rising and decaying parts of the flares will be essential to better characterise their variability, and set additional constraints to the different hypotheses concerning these sources. Furthermore, given the small spatial scales implied by the variability timescales, many of the upcoming telescopes and instruments currently under development, such as the next generation VLA (ngVLA) in radio, the Multi-Conjugated Adaptive Optics (MCAO) Assisted Visible Imager and Spectrograph (MAVIS) and the High Angular Resolution Monolithic Optical and Near-infrared Integral field spectrograph (HARMONI) in the optical/near-infrared, and \textit{Athena} in X-rays, will be crucial to study the spatial properties and evolution of these remarkable sources.

\section*{Acknowledgements}

The University of Oklahoma Land Acknowledgement Statement:

Long before the University of Oklahoma was established, the land on which the university now resides was the traditional home of the "Hasinais" Caddo Nation and "Kirikiris" Wichita \& Affiliated Tribes. This land was also once part of the Muscogee Creek and Seminole nations.

We acknowledge this territory once also served as a hunting ground, trade exchange point, and migration route for the Apache, Comanche, Kiowa, and Osage nations. Today, 39 federally recognized tribal nations dwell in what is now the State of Oklahoma as a result of settler colonial policies designed to assimilate Indigenous peoples.

The University of Oklahoma recognizes the historical connection our university has with its Indigenous community. We acknowledge, honor, and respect the diverse Indigenous peoples connected to this land. We fully recognize, support, and advocate for the sovereign rights of all of Oklahoma’s 39 tribal nations.

This acknowledgment is aligned with our university’s core value of creating a diverse and inclusive community. It is our institutional responsibility to recognize and acknowledge the people, culture, and history that make up our entire OU Community.

This publication makes use of data obtained at Metsähovi Radio Observatory, operated by Aalto University in Finland.

The National Radio Astronomy Observatory is a facility of the National Science Foundation operated under cooperative agreement by Associated Universities, Inc. CIRADA is funded by a grant from the Canada Foundation for Innovation 2017 Innovation Fund (Project 35999), as well as by the Provinces of Ontario, British Columbia, Alberta, Manitoba and Quebec.

This research has made use of data from the OVRO 40-m monitoring program \citep{2011richards1}, supported by private funding from the California Institute of Technology and the Max Planck Institute for Radio Astronomy, and by NASA grants NNX08AW31G, NNX11A043G, and NNX14AQ89G and NSF grants AST-0808050 and AST- 1109911.

This research has made use of the NASA/IPAC Extragalactic Database (NED), which is operated by the Jet Propulsion Laboratory, California Institute of Technology, under contract with the National Aeronautics and Space Administration. This research has made use of the SIMBAD database, operated at CDS, Strasbourg, France.

T.S. was partly supported by the Academy of Finland project 315721. 

T.H. was supported by the Academy of Finland projects 317383, 320085, 322535, and 345899.

S.K. acknowledges support from the European Research Council (ERC) under the European Unions Horizon 2020 research and innovation programme under grant agreement No.~771282.

I.V. would like to thank the Magnus Ehrnrooth Foundation for their continuing support.

R.R. and W.M. are supported by the ANID BASAL project FB210003.

\section*{Data Availability}

The JVLA (Legacy ID: AJ442) and the VLBA (Legacy ID: BJ109) data are publicly available in the NRAO Data Archive: https://data.nrao.edu. The MRO and OVRO data will be made available via CDS.



\bibliographystyle{mnras}
\bibliography{artikkeli} 



\newpage
\appendix

\section{Details of observations}
\label{app:obsdetails}

Here we provide some further details regarding the observations and the data reduction procedures.

\subsection{JVLA}

Table~\ref{tab:jvlaobssummary} summarises the JVLA observations, including the date of the observation for each source, and the integration times in all bands.

\begin{table}
\caption{Summary of the JVLA observations. Columns: (1) source name; (2) date of observations; (3) integration time in each band.}
\label{tab:jvlaobssummary}
\centering
\footnotesize
\begin{tabular}{lcc}
\hline\hline
Source     & date          & $T_{\mathrm{int}}$ (X / Ku / K / Ka / Q) \\
           &  (yyyy-mm-dd) & (s)                \\  \hline\hline
J1029+5556 &  2022-04-22   & 594 / 594 / 534 / 706 / 754  \\
J1228+5018 &  2022-04-24   & 596 / 596 / 532 / 706 / 754  \\ 
J1232+4957 &  2022-04-25   & 594 / 594 / 532 / 708 / 754 \\ 
J1509+6137 &  2022-03-26   & 594 / 594 / 532 / 590 / 754  \\ 
J1510+5547 &  2022-03-13   & 594 / 594 / 534 / 706 / 868  \\ 
J1522+3934 &  2022-03-16   & 594 / 594 / 532 / 704 / 810  \\ 
J1641+3454 &  2022-03-15   & 594 / 594 / 543 / 708 / 696  \\ \hline 
\end{tabular}
\end{table}

\subsection{VLBA}

Table~\ref{tab:vlbaobssummary} summarises the VLBA observations, including the name of the calibrator source, its distance from the target as well as its VLBI scale flux density at 15~GHz, and the phase referencing cycle time of the observation.

\begin{table}
\caption{Summary of the VLBA observations. Columns: (1) target source name; (2) phase reference calibrator; (3) distance between the target and the calibrator; (4) calibrator's VLBI scale flux density at 15\,GHz; (5) phase referencing cycle time.}
\label{tab:vlbaobssummary}
\centering
\footnotesize
\begin{tabular}{lcccc}
\hline\hline
Source     & Calibrator    & Distance & $S_{\nu,\mathrm{cal}}$ & $T_\mathrm{cycle}$ \\
           &               &  (deg)   &  (mJy)                 &  (s)  \\ \hline\hline
J1029+5556 & J1035+5628    &  0.99    &  300                   & 180  \\
J1228+5018 & J1223+5037    &  0.87    &  62                    & 160  \\ 
J1232+4957 & J1223+5037    &  1.53    &  62                    & 160  \\ 
J1509+6137 & J1526+6110    &  2.05    &  38                    & 220  \\ 
J1510+5547 & J1510+5702    &  1.26    &  200                   & 160  \\ 
J1522+3934 & J1528+3816    &  1.82    &  68                    & 200  \\ 
J1641+3454 & J1635+3458    &  1.21    &  190                   & 180  \\ \hline 
\end{tabular}
\end{table}

\subsection{MRO}
\label{app:mrodetails}

Measurements that are considered to be of poor quality, for example, due to unfavourable weather conditions, are discarded semi-automatically. Additionally faint detections are checked manually in the final data reduction stage. Bad weather conditions or other environmental effects are taken into account and also, for example, conspicuous but rare flux density increases caused by aircraft in the telescope beam. The general flux density levels are checked to be consistent with adjacent measurements (i.e. other sources observed before and after the target source). In addition, we checked if the observations of these sources could be contaminated by a bright radio source falling into the reference beam of the MRO telescope. Using LOFAR, FIRST, and VLASS data we concluded that whereas there are a few moderately bright sources with flux densities around a few hundred mJy at low radio frequencies that could be in the reference beam, all of them have steep spectra, and it is thus very unlikely that any of them could affect these observations.

Due to the fairly high detection limit of the telescope (i.e. approx. 200~mJy in optimal conditions, which is more than adequate for the bright AGN monitoring programmes conducted at MRO), we typically only see the highest tips of the flares in faint sources, whereas most of the lower level activity remains below the detection threshold \citep[e.g.,][]{2014acciari1}. This is also seen in the upper limits, the level of which can drastically change in even a short time due to compromised weather conditions that can also significantly raise the detection limit. The undetected source could be actually fainter due to variability or the observing conditions could be worse (or both), and it is therefore not detected. The upper limits describe the largest flux density the source could have in the current conditions, but still remain below the 4$\sigma$ detection limit and cannot therefore be used for data analysis. Occasionally there are several high upper limit values clustered around detections, which indicate that the source could indeed be active, but for some reason it does not exceed the current detection limit, for example, due to weather. However, it has been shown that the high activity periods of NLS1 sources detected at MRO correspond to flare peaks in OVRO data \citep{2017lahteenmaki1}, confirming that at least most of the time the two telescopes are catching the same events.

\subsection{OVRO}
\label{app:ovrodetails}

The AGN monitoring sample at OVRO mostly consists of bright blazar-type objects, with the majority having a mean flux density $>60$~mJy \citep{richards2014}. Therefore in the schedules, each observation consists of four on-on integrations, each 8~s long, resulting in a total integration time of 32~s. Given the small number of on-on integrations, it is possible that atmospheric fluctuations or pointing errors result in outliers in the light curves \citep{richards2014}. Moreover, the number of observations performed each day is large (up to 500) so that it is possible that some, apparently high signal-to-noise ratio observations, occur purely due to random fluctuations. The data are processed with an automated pipeline, where manual editing is done to flag obviously bad periods of data, and data are automatically flagged based on large changes within the four on-on integrations and other diagnostics \citep[see][for details]{2011richards1}. However, individual data points are not typically manually inspected, as for the variability analysis of bright blazars, the effect of outliers in the data is small \citep{richards2014}.

Because of the faintness of the NLS1 targets, we have done additional manual checks to inspect the quality of the detections, which we describe here. We note that in all the cases described below, the flux density of the spurious detection has been less than 20~mJy and mostly $<10$~mJy, or the uncertainty has been larger than usual so that similar observations in our blazar light curves would not be as problematic. 

The receiver records both right- and left-hand circular polarisation separately with the final observation being a weighted average of the two. We can thus inspect the two values separately to verify that the source has been detected in both polarisations (here we assume that the circular polarisation of the objects is negligible, as is the case for most blazars at 15~GHz \citep[e.g.,][]{homan2006}. This made us reject two spurious detections in J1232+4957. Additionally, we have inspected observations of other nearby sources to see whether there are data that have been automatically flagged in the pipeline close to the observation of the NLS1, indicative of potentially poor observing conditions. This resulted in the rejection of one spurious detection in J1641+3454.

In October 2021 we also changed the observing strategy of these NLS1 targets so that they are observed twice in a row in the schedules. This way we can see whether short-term atmospheric effects or bad conditions have resulted in spurious detections if the two consecutive observations show a large difference, as we would not expect large changes on a time scale of $\sim 1$~min. This resulted in the rejection of single spurious detections in J1029+5556, J1509+6137, J1510+5547, and J1522+3934, all of which had moderate S/N values of $\sim$4-9. 

The remaining detections in the paper either show detections in two consecutive observations (J1522+3934) or consistent flux densities in the right- and left-hand circular polarisation and no apparent problems with nearby targets (J1029+5556, J1522+3934, J1641+3454). However, we cannot fully exclude additional, unknown effects in the observations before October 2021 when the sources were observed only once in a row, especially in J1029+5556 and J1641+3454 that do not show any other detections in the OVRO light curves. J1522+3934 on the other hand seems more reliable because of its multiple detections.

\section{Radio maps}
\label{app:radiomaps}

The JVLA radio maps with likely detections are shown here. This includes X, Ku, K, Ka, and Q band maps of J1228+5017 (Figs.~\ref{j1228x}-\ref{j1228q}), X, and Ku band maps of J1232+4957 (Figs.~\ref{j1232x}-\ref{j1232ku}), X, Ku, K, and Ka band maps of J1522+3934 (Figs.~\ref{j1522x}-\ref{j1522ka}), and X, Ku, K, Ka, and Q band maps of J1641+3454 (Figs.~\ref{j1641x}-\ref{j1641q}).


\begin{figure}
\begin{center}
\includegraphics[width=\columnwidth]{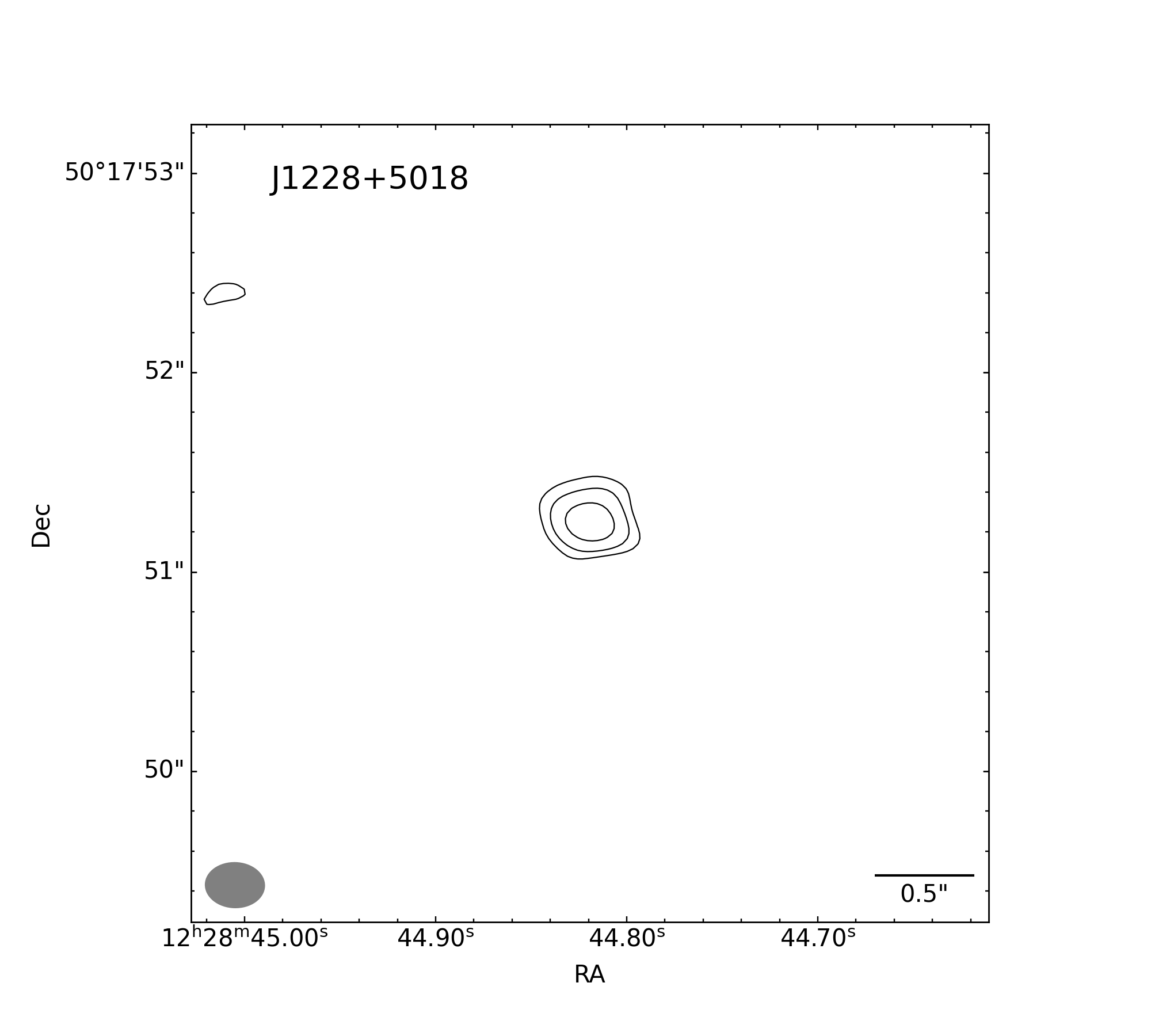}
\caption{JVLA X band radio map of J1228+5017, rms = 7$\mu$Jy beam$^{-1}$, contour levels at -3, 3, 6, 12 rms, beam size 1.20 $\times$ 0.91~kpc. \label{j1228x}}
\end{center}
\end{figure}

\begin{figure}
\begin{center}
\includegraphics[width=\columnwidth]{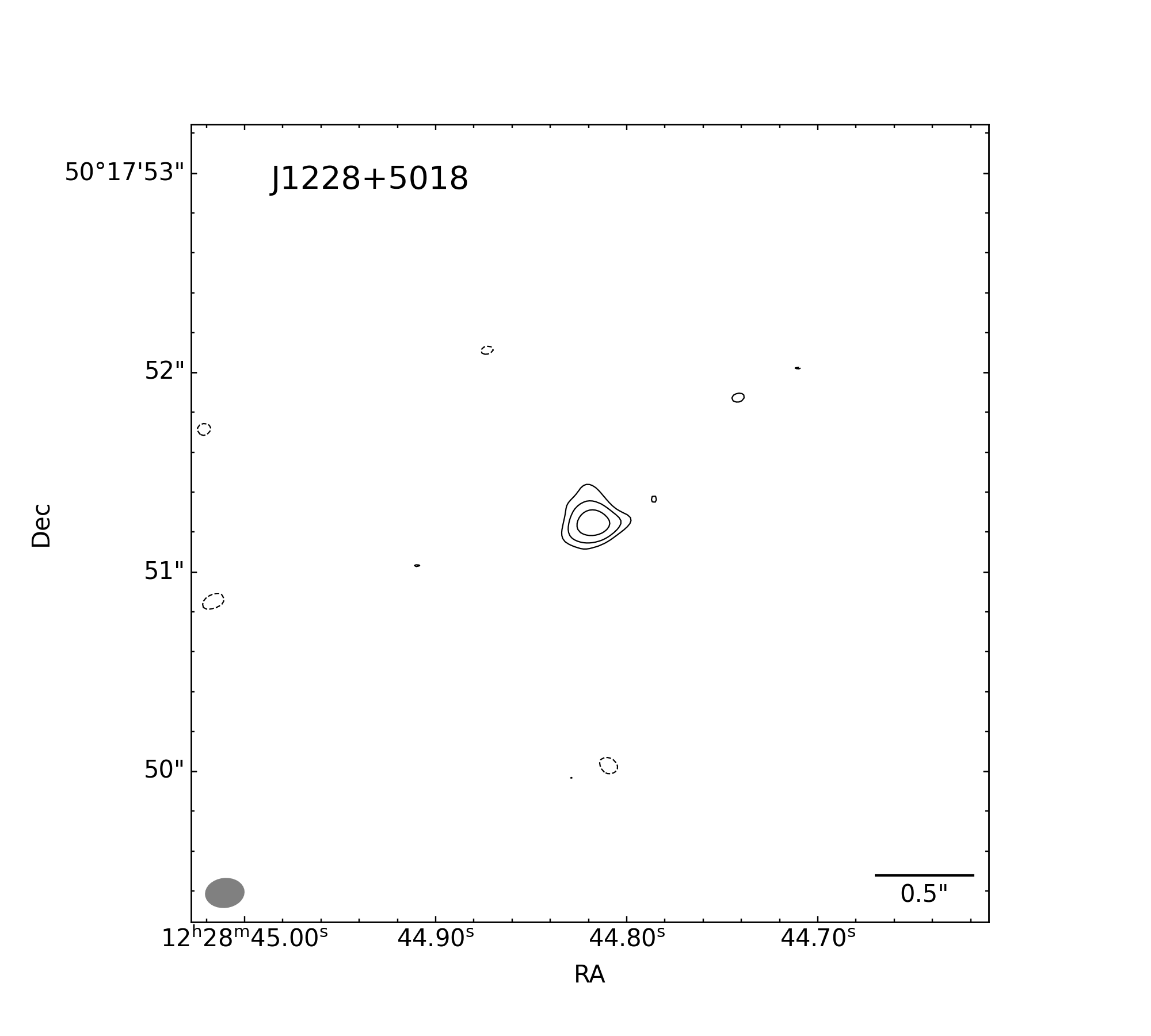}
\caption{JVLA Ku band radio map of J1228+5017, rms = 6$\mu$Jy beam$^{-1}$, contour levels at -3, 3, 6, 12 rms, beam size 0.78 $\times$ 0.58~kpc. \label{j1228ku}}
\end{center}
\end{figure}

\begin{figure}
\begin{center}
\includegraphics[width=\columnwidth]{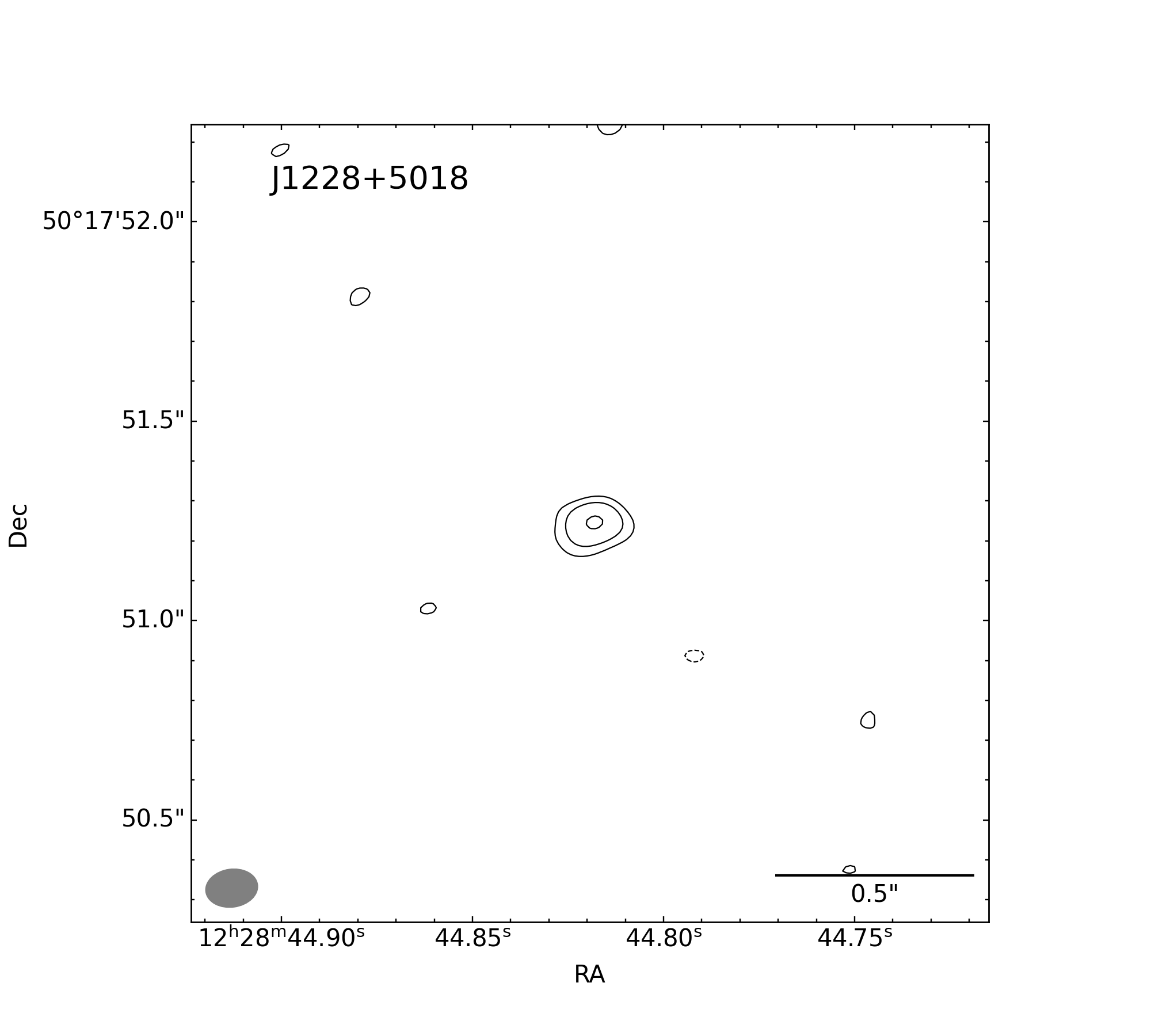}
\caption{JVLA K band radio map of J1228+5017, rms = 9$\mu$Jy beam$^{-1}$, contour levels at -3, 3, 6, 12 rms, beam size 0.52 $\times$ 0.38~kpc. \label{j1228k}}
\end{center}
\end{figure}

\begin{figure}
\begin{center}
\includegraphics[width=\columnwidth]{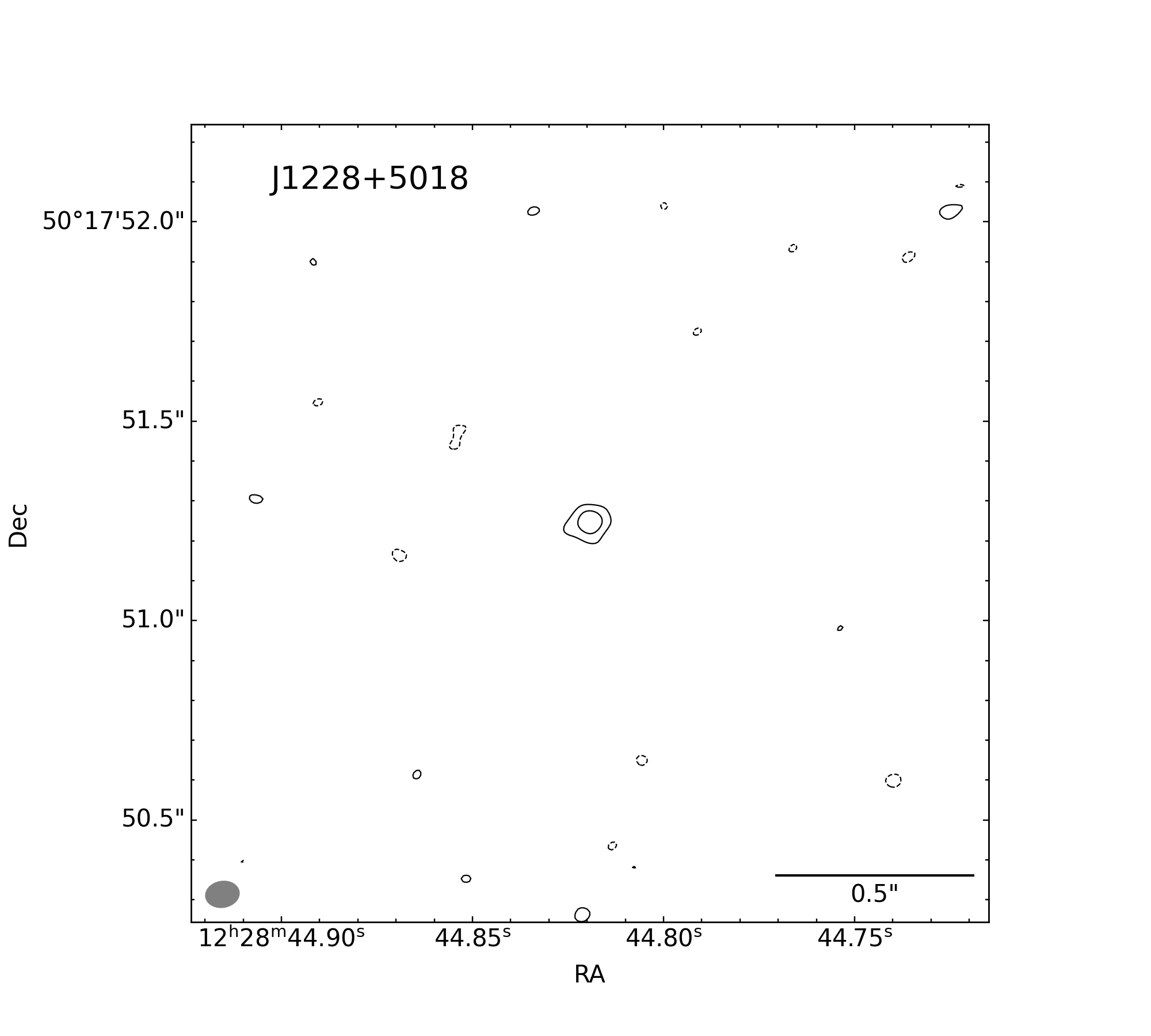}
\caption{JVLA Ka band radio map of J1228+5017, rms = 12$\mu$Jy beam$^{-1}$, contour levels at -3, 3, 6 rms, beam size 0.34 $\times$ 0.26~kpc. \label{j1228ka}}
\end{center}
\end{figure}

\begin{figure}
\begin{center}
\includegraphics[width=\columnwidth]{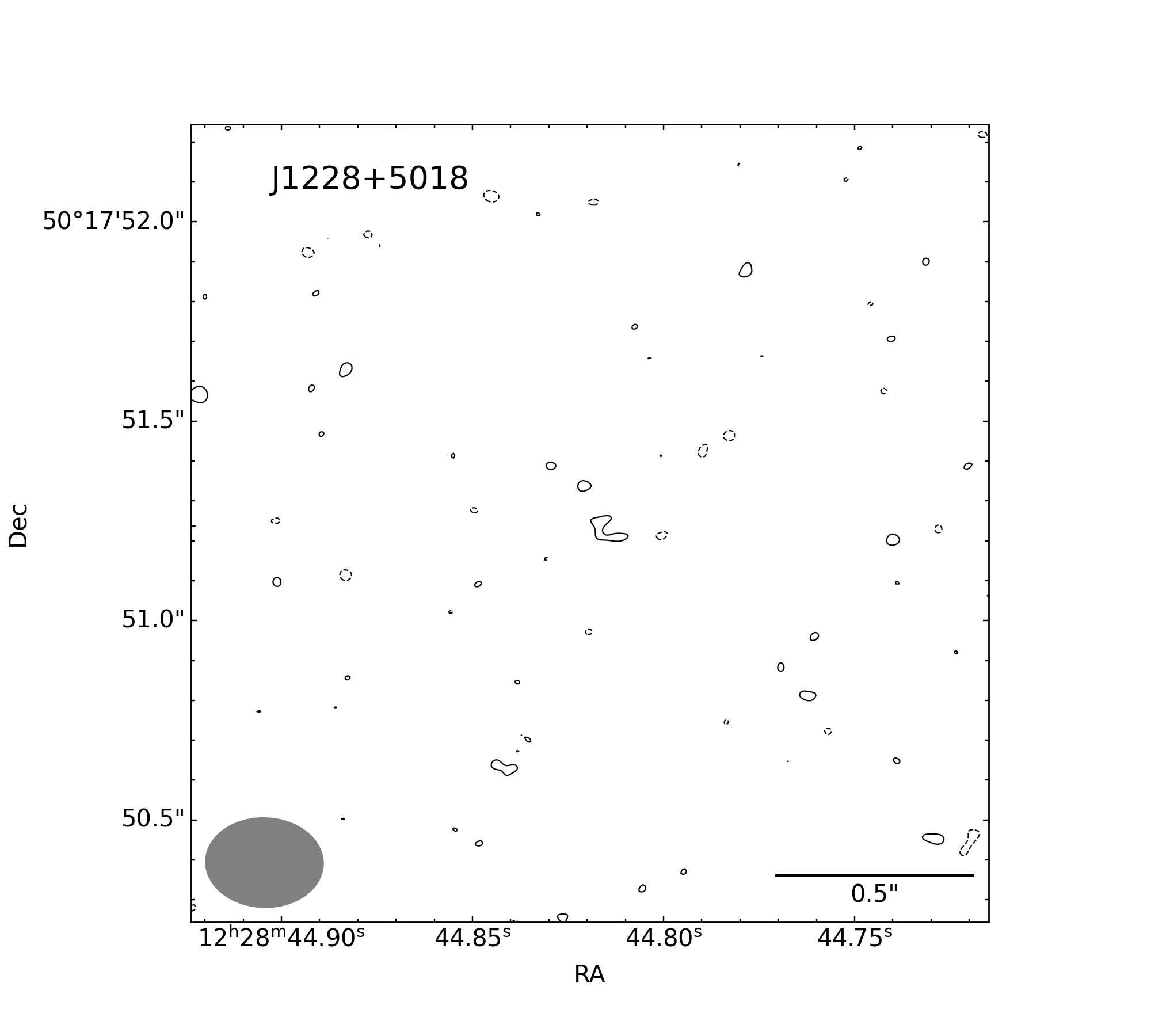}
\caption{JVLA Q band radio map of J1228+5017, rms = 31$\mu$Jy beam$^{-1}$, contour levels at -3, 3 rms, beam size 0.24 $\times$ 0.20~kpc. \label{j1228q}}
\end{center}
\end{figure}


\begin{figure}
\begin{center}
\includegraphics[width=\columnwidth]{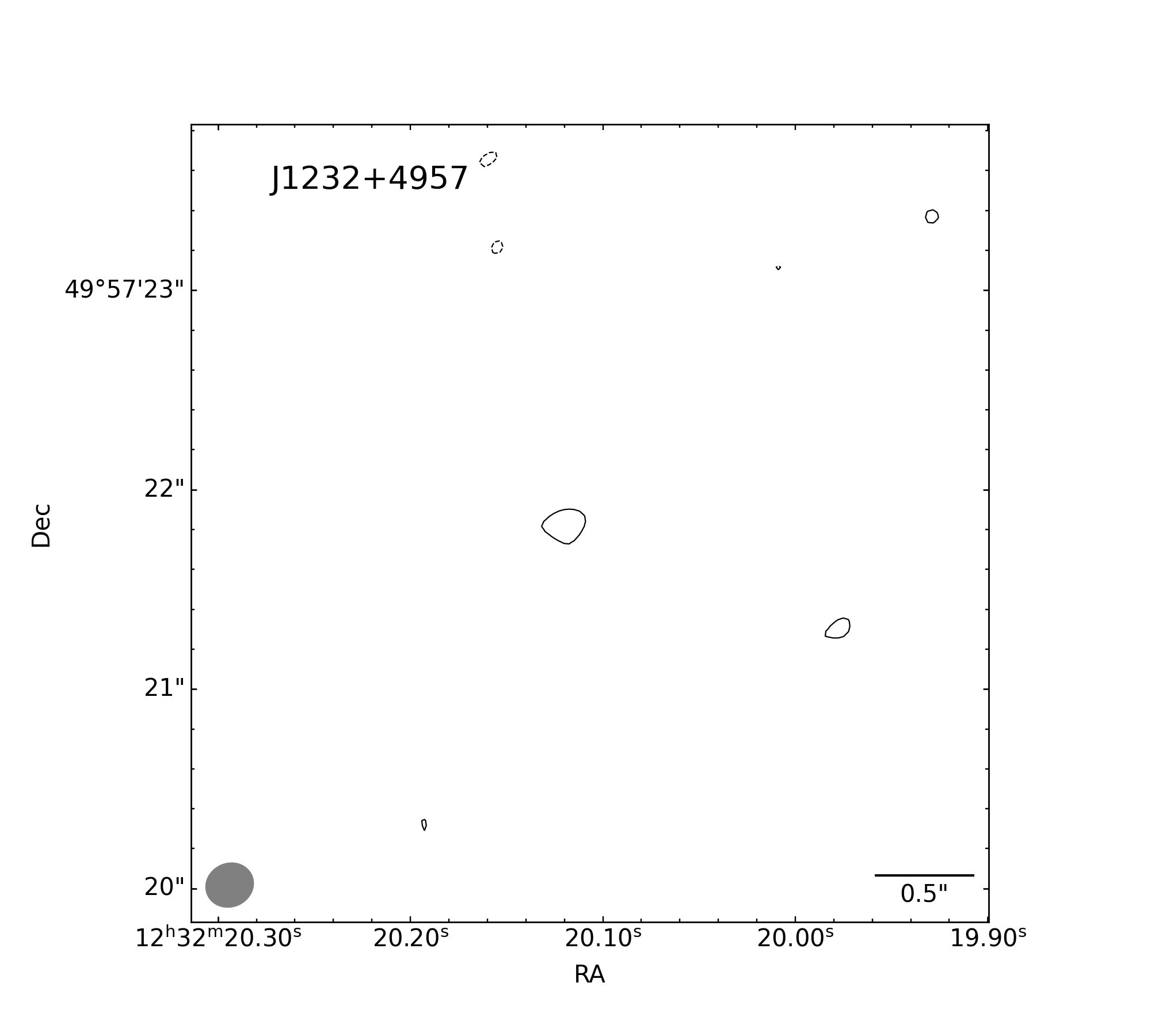}
\caption{JVLA X band radio map of J1232+4957, rms = 7$\mu$Jy beam$^{-1}$, contour levels at -3, 3 rms, beam size 0.98 $\times$ 0.88~kpc. \label{j1232x}}
\end{center}
\end{figure}

\begin{figure}
\begin{center}
\includegraphics[width=\columnwidth]{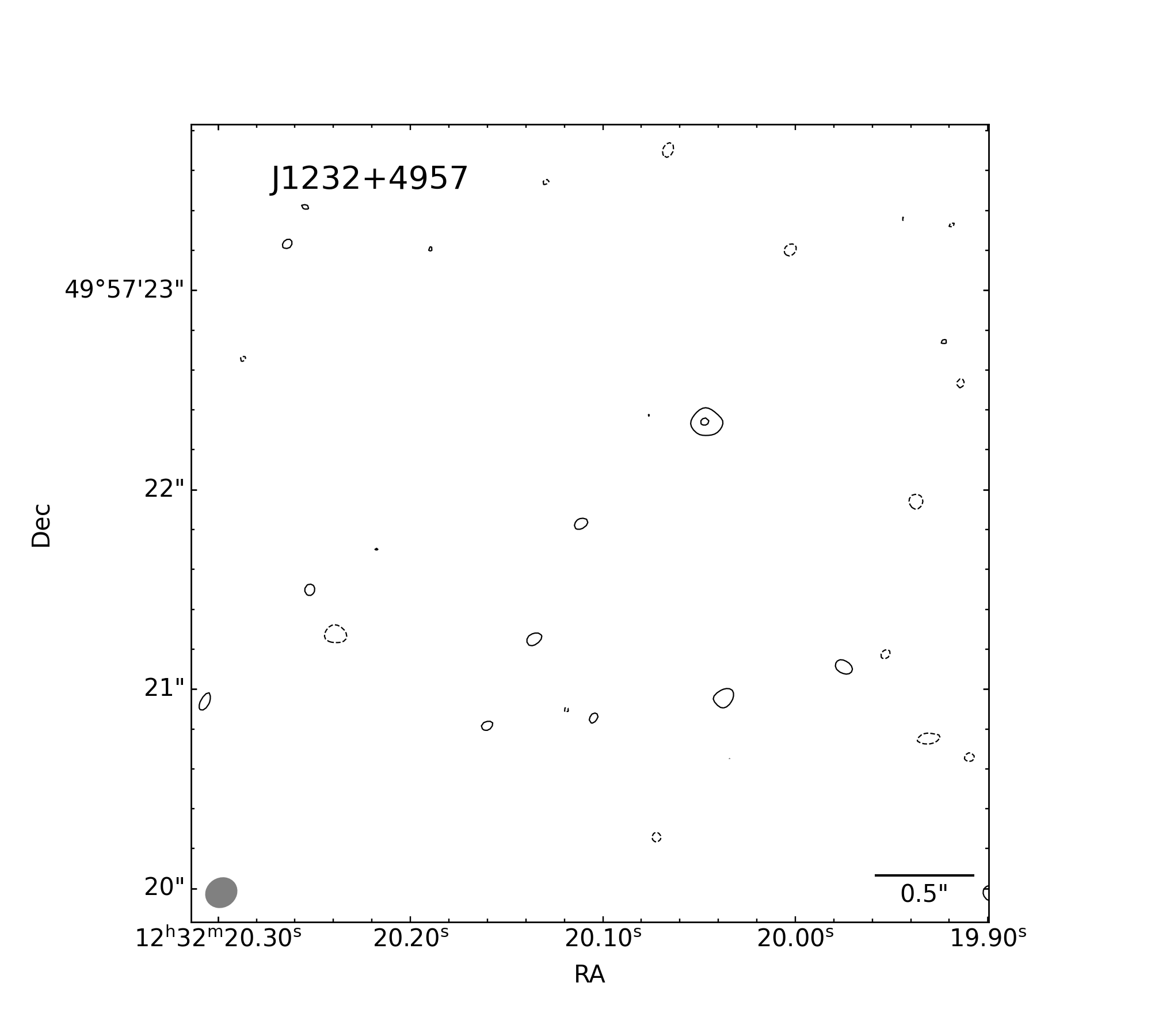}
\caption{JVLA Ku band radio map of J1232+4957, rms = 5$\mu$Jy beam$^{-1}$, contour levels at -3, 3, 6 rms, beam size 0.65 $\times$ 0.57~kpc. \label{j1232ku}}
\end{center}
\end{figure}



\begin{figure}
\begin{center}
\includegraphics[width=\columnwidth]{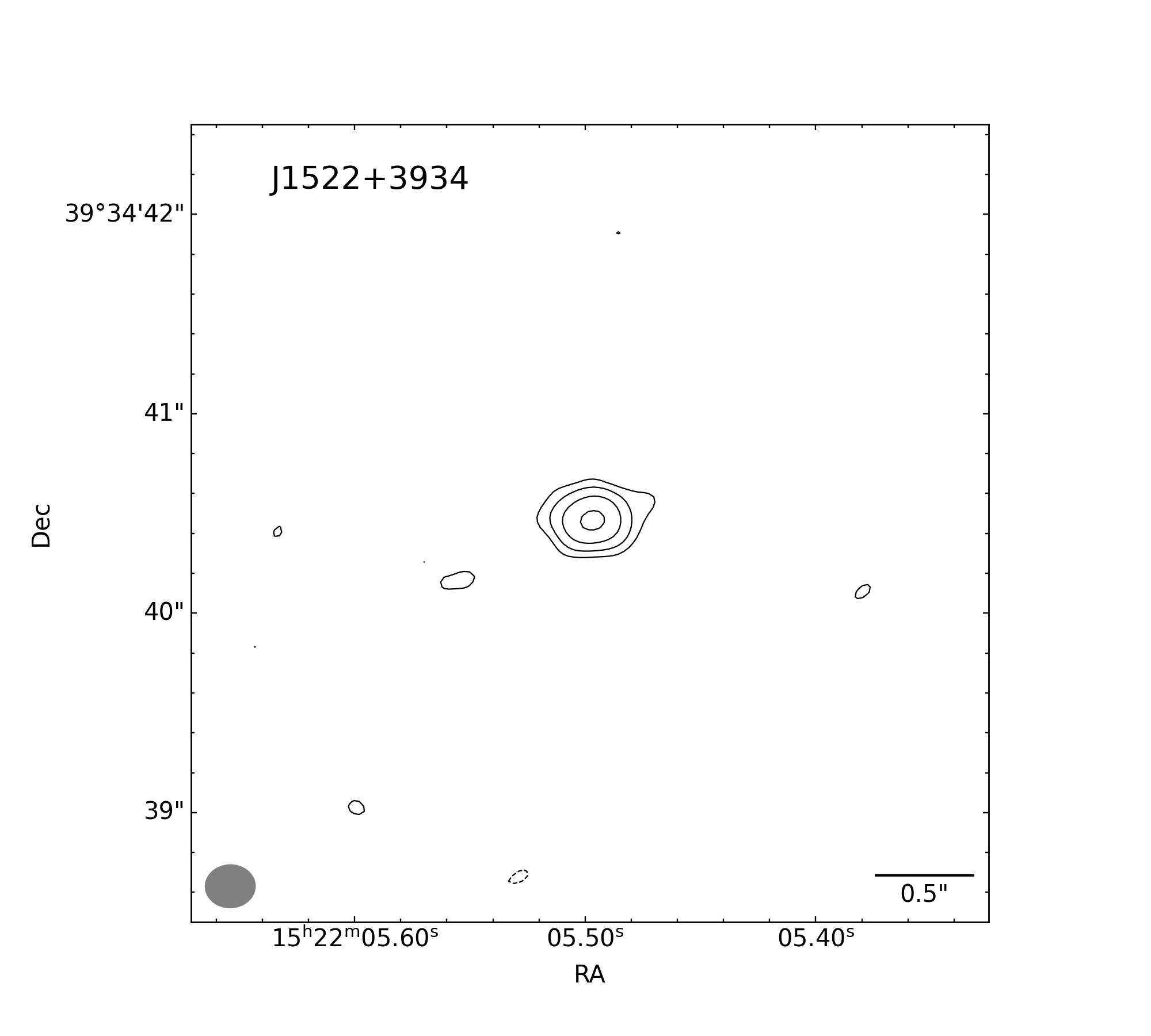}
\caption{JVLA X band radio map of J1522+3934, rms = 8$\mu$Jy beam$^{-1}$, contour levels at -3, 3, 6, 12, 24 rms, beam size 0.36 $\times$ 0.31~kpc. \label{j1522x}}
\end{center}
\end{figure}

\begin{figure}
\begin{center}
\includegraphics[width=\columnwidth]{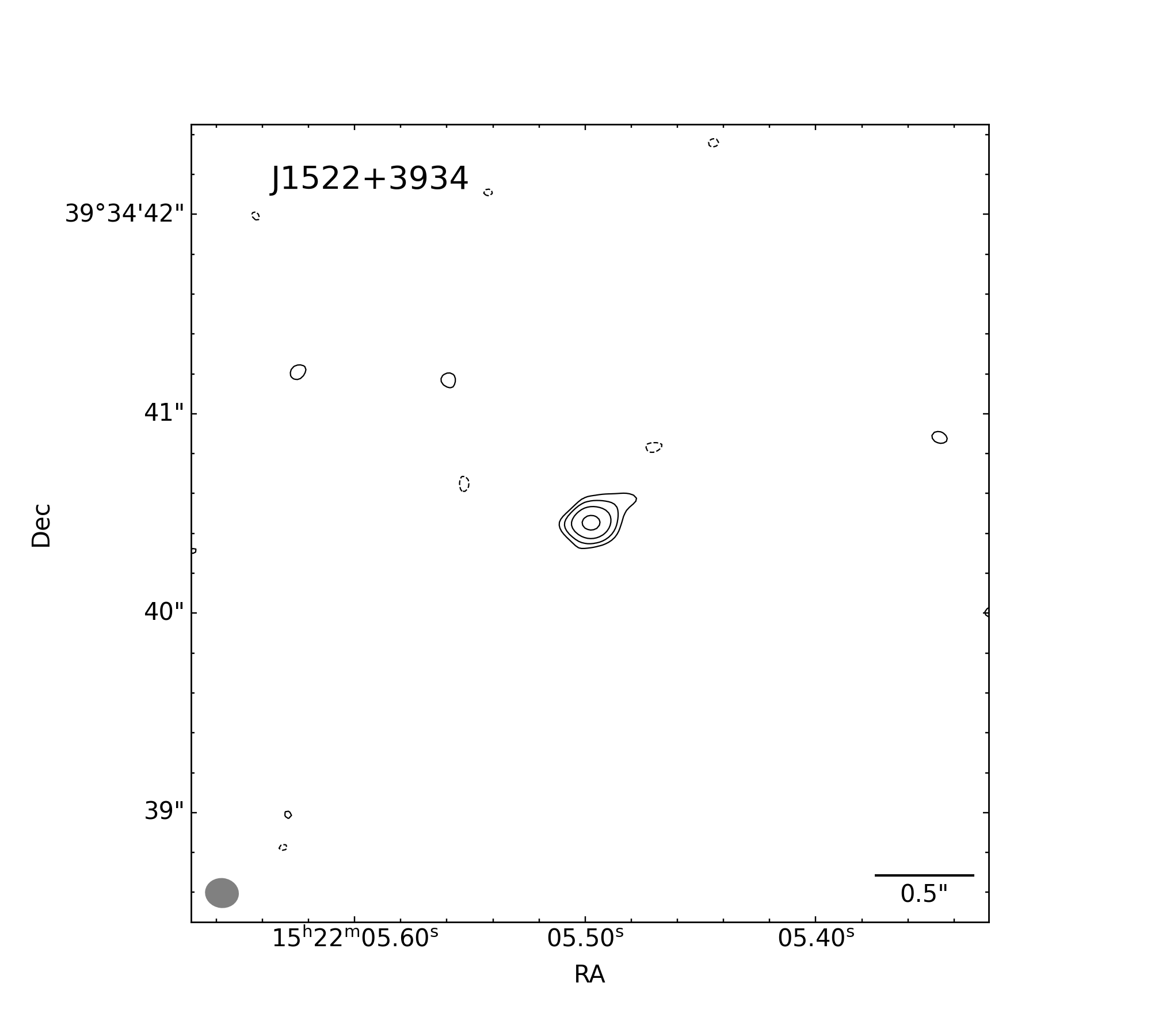}
\caption{JVLA Ku band radio map of J1522+3934, rms = 6$\mu$Jy beam$^{-1}$, contour levels at -3, 3, 6, 12, 24 rms, beam size 0.24 $\times$ 0.21~kpc. \label{j1522ku}}
\end{center}
\end{figure}

\begin{figure}
\begin{center}
\includegraphics[width=\columnwidth]{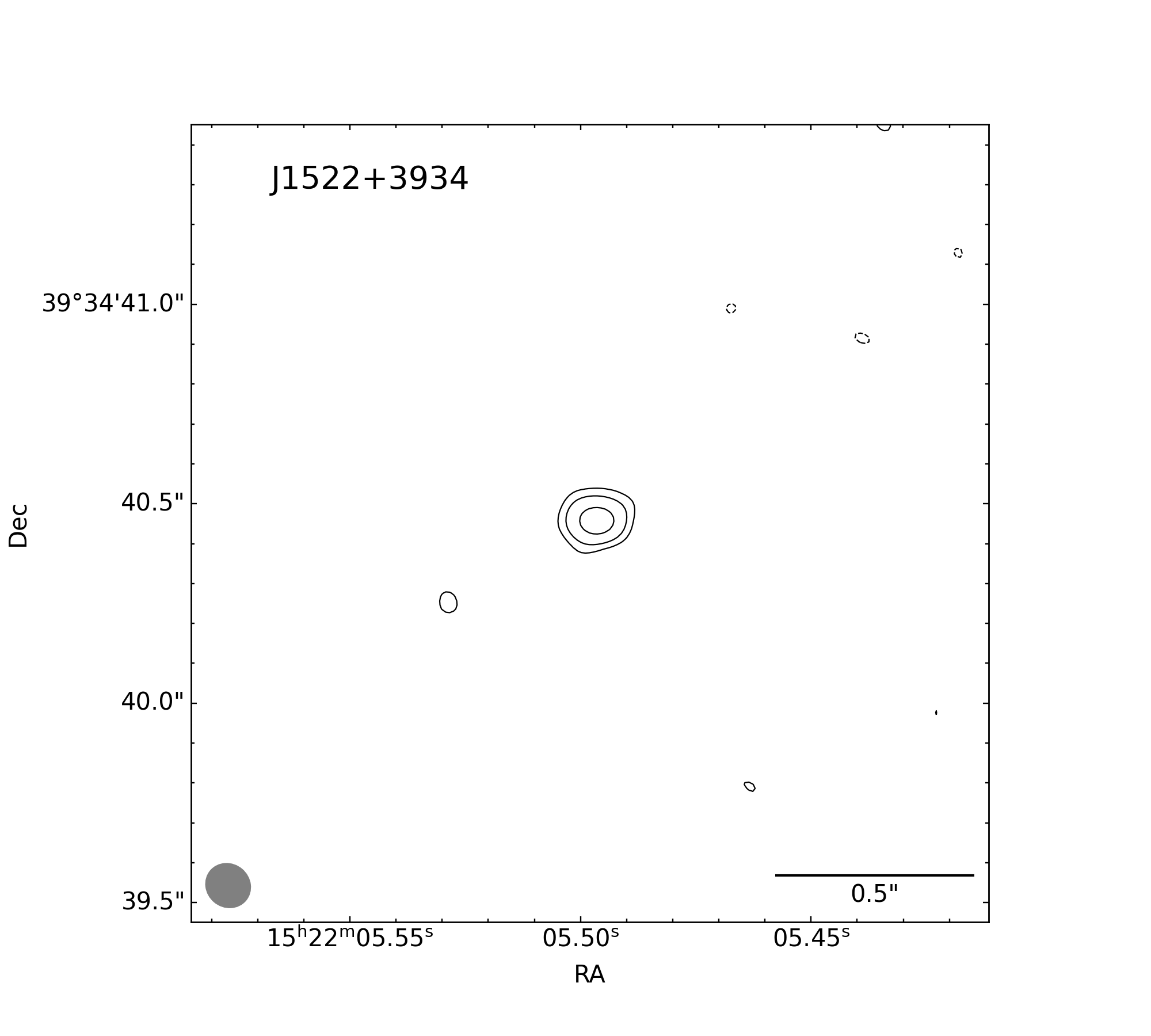}
\caption{JVLA K band radio map of J1522+3934, rms = 9$\mu$Jy beam$^{-1}$, contour levels at -3, 3, 6, 12 rms, beam size 0.17 $\times$ 0.16~kpc. \label{j1522k}}
\end{center}
\end{figure}

\begin{figure}
\begin{center}
\includegraphics[width=\columnwidth]{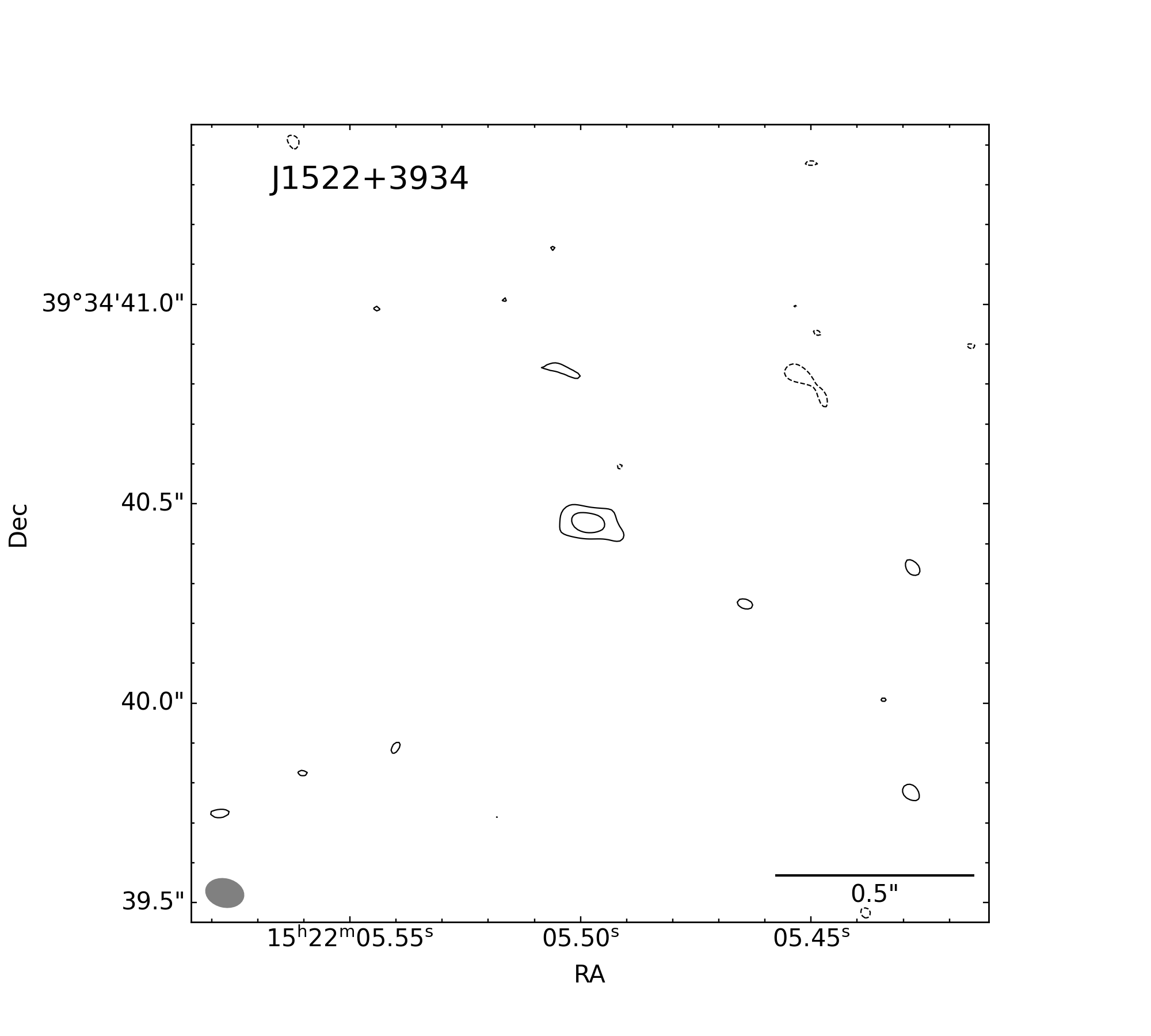}
\caption{JVLA Ka band radio map of J1522+3934, rms = 13$\mu$Jy beam$^{-1}$, contour levels at -3, 3, 6 rms, beam size 0.10 $\times$ 0.07~kpc. \label{j1522ka}}
\end{center}
\end{figure}


\begin{figure}
\begin{center}
\includegraphics[width=\columnwidth]{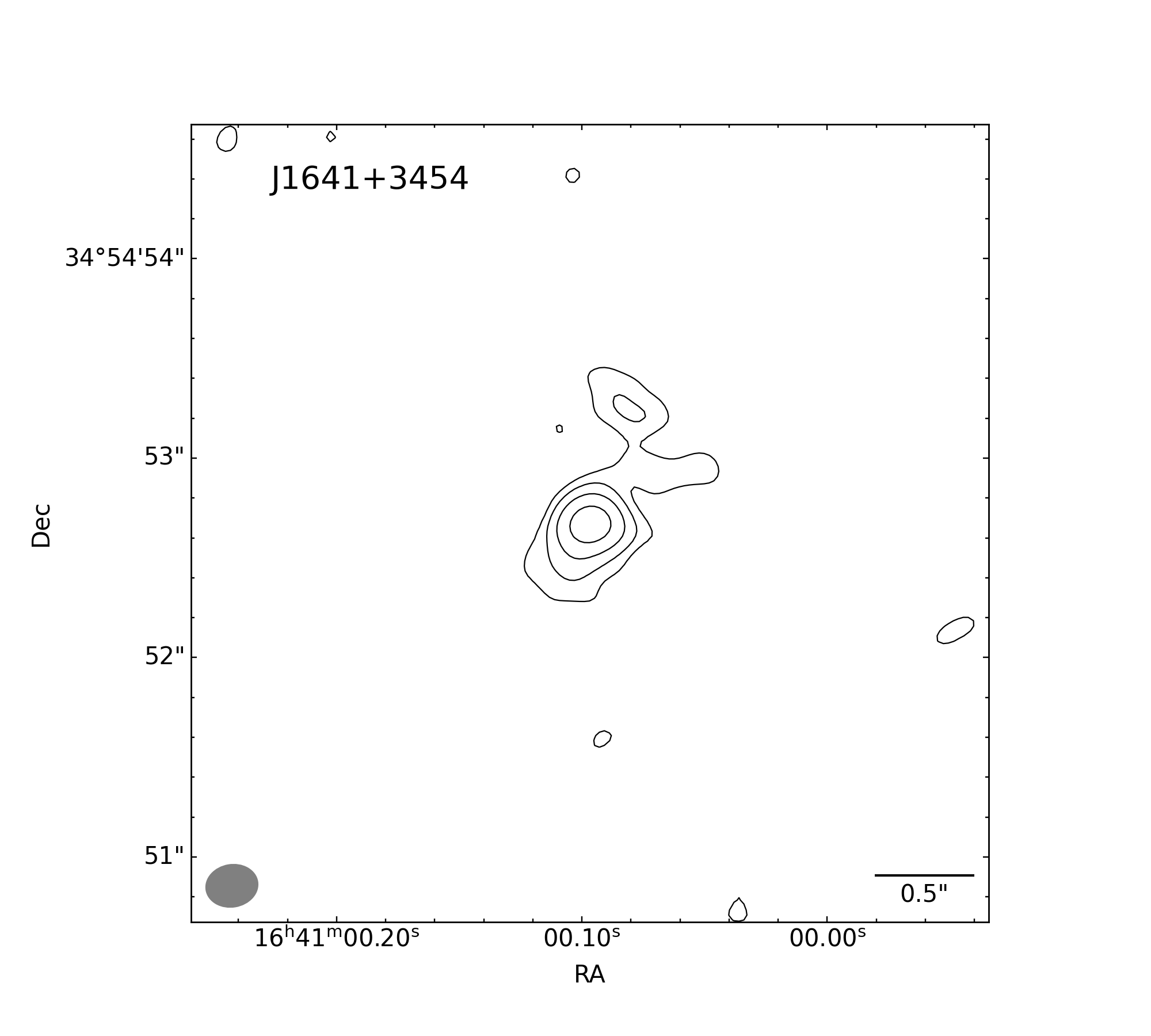}
\caption{JVLA X band radio map of J1641+3454, rms = 7$\mu$Jy beam$^{-1}$, contour levels at -3, 3, 6, 12, 24 rms, beam size 0.73 $\times$ 0.59~kpc. \label{j1641x}}
\end{center}
\end{figure}

\begin{figure}
\begin{center}
\includegraphics[width=\columnwidth]{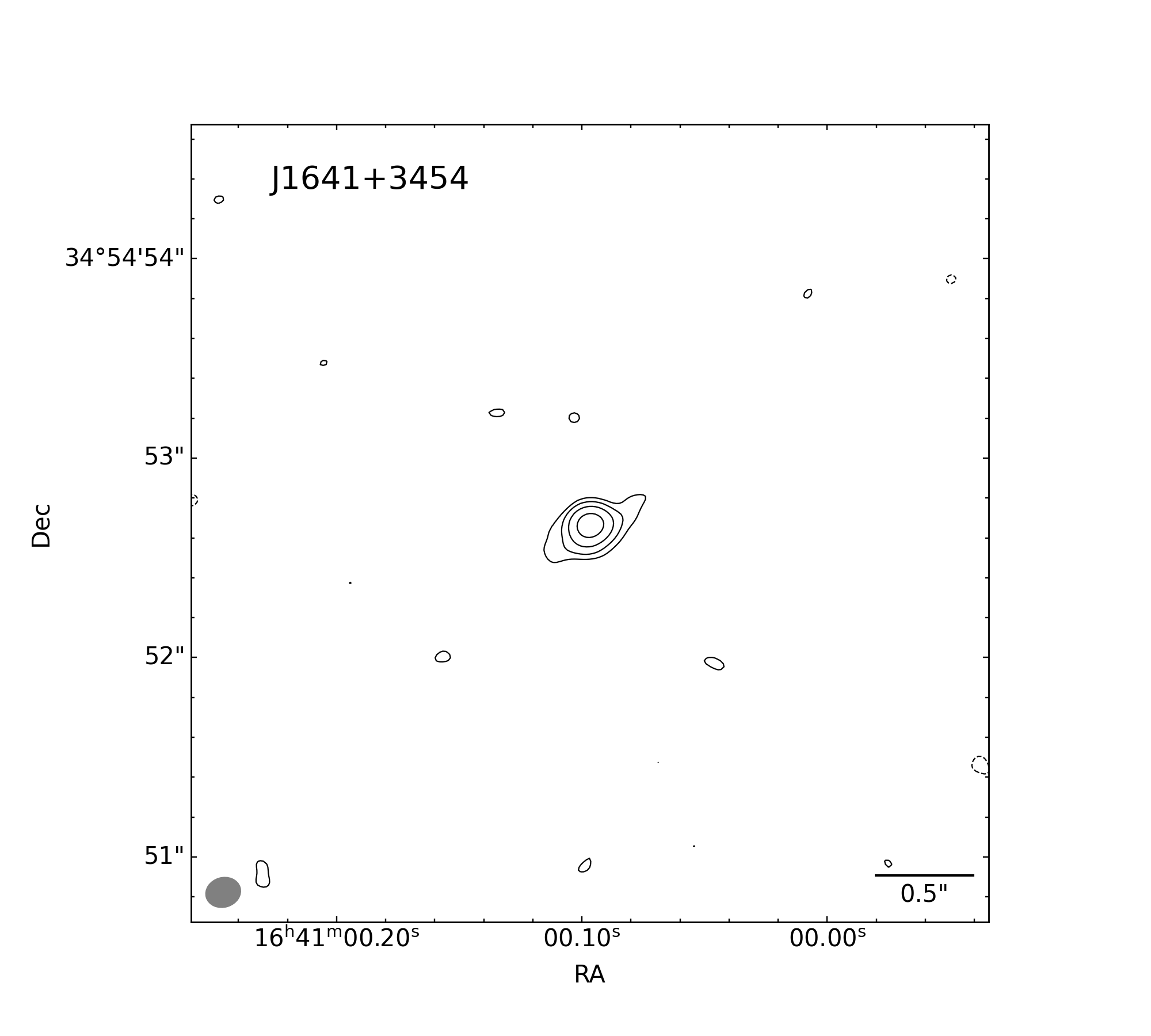}
\caption{JVLA Ku band radio map of J1641+3454, rms = 5$\mu$Jy beam$^{-1}$, contour levels at -3, 3, 6, 12, 24 rms, beam size 0.49 $\times$ 0.41~kpc. \label{j1641ku}}
\end{center}
\end{figure}

\begin{figure}
\begin{center}
\includegraphics[width=\columnwidth]{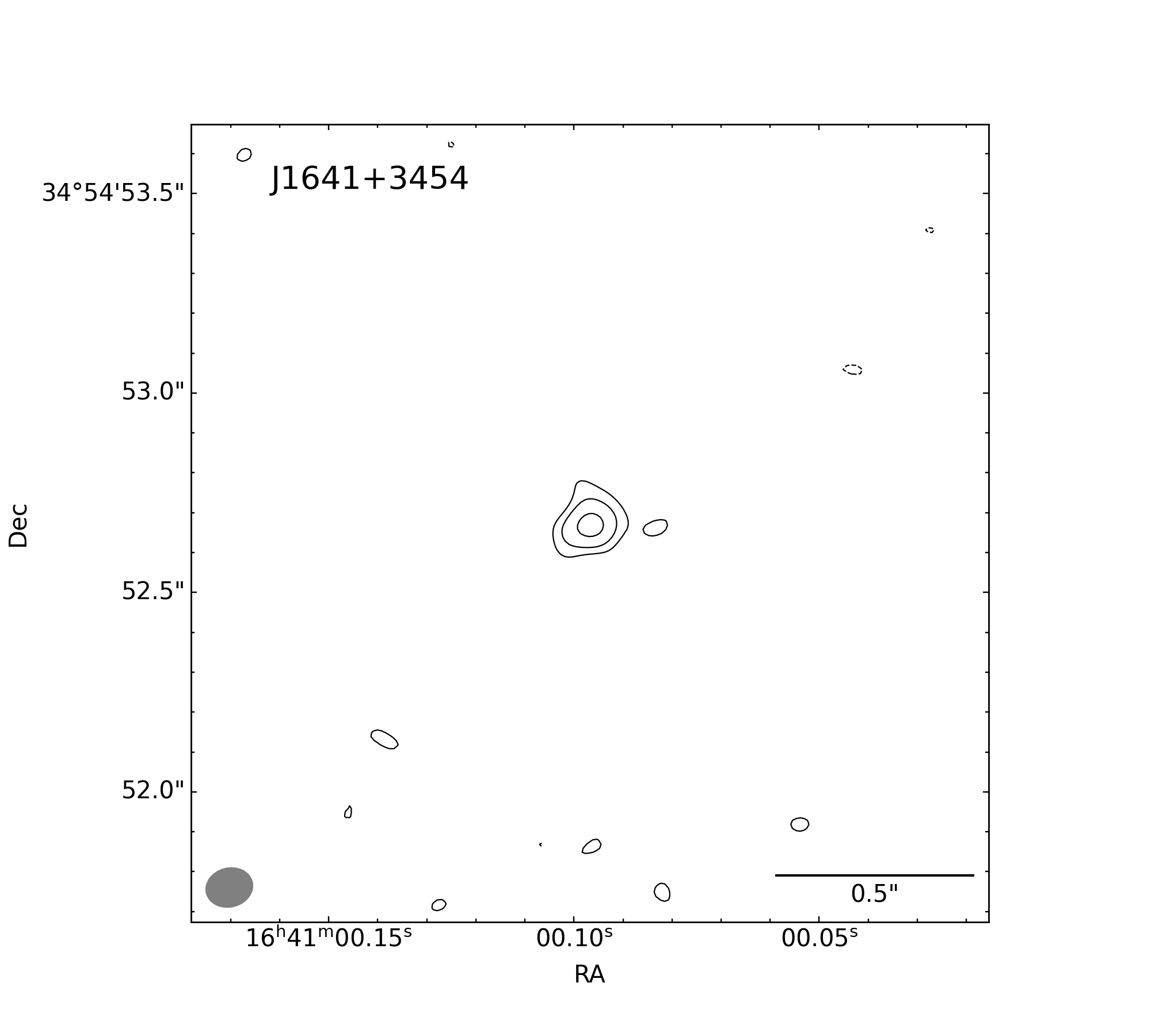}
\caption{JVLA K band radio map of J1641+3454, rms = 8$\mu$Jy beam$^{-1}$, contour levels at -3, 3, 6, 12 rms, beam size 0.33 $\times$ 0.27~kpc. \label{j1641k}}
\end{center}
\end{figure}

\begin{figure}
\begin{center}
\includegraphics[width=\columnwidth]{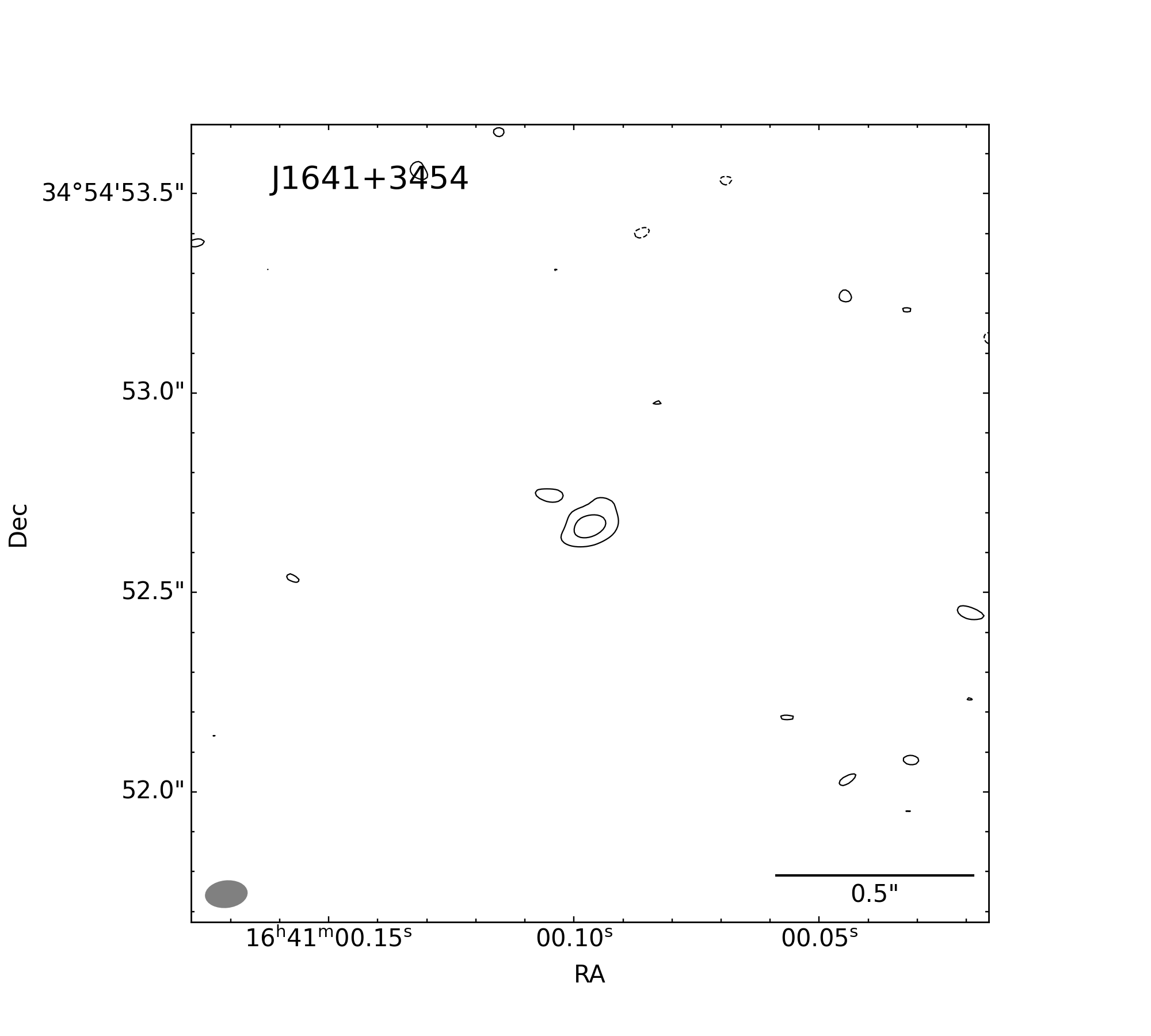}
\caption{JVLA Ka band radio map of J1641+3454, rms = 11$\mu$Jy beam$^{-1}$, contour levels at -3, 3, 6 rms, beam size 0.29 $\times$ 0.18~kpc. \label{j1641ka}}
\end{center}
\end{figure}

\begin{figure}
\begin{center}
\includegraphics[width=\columnwidth]{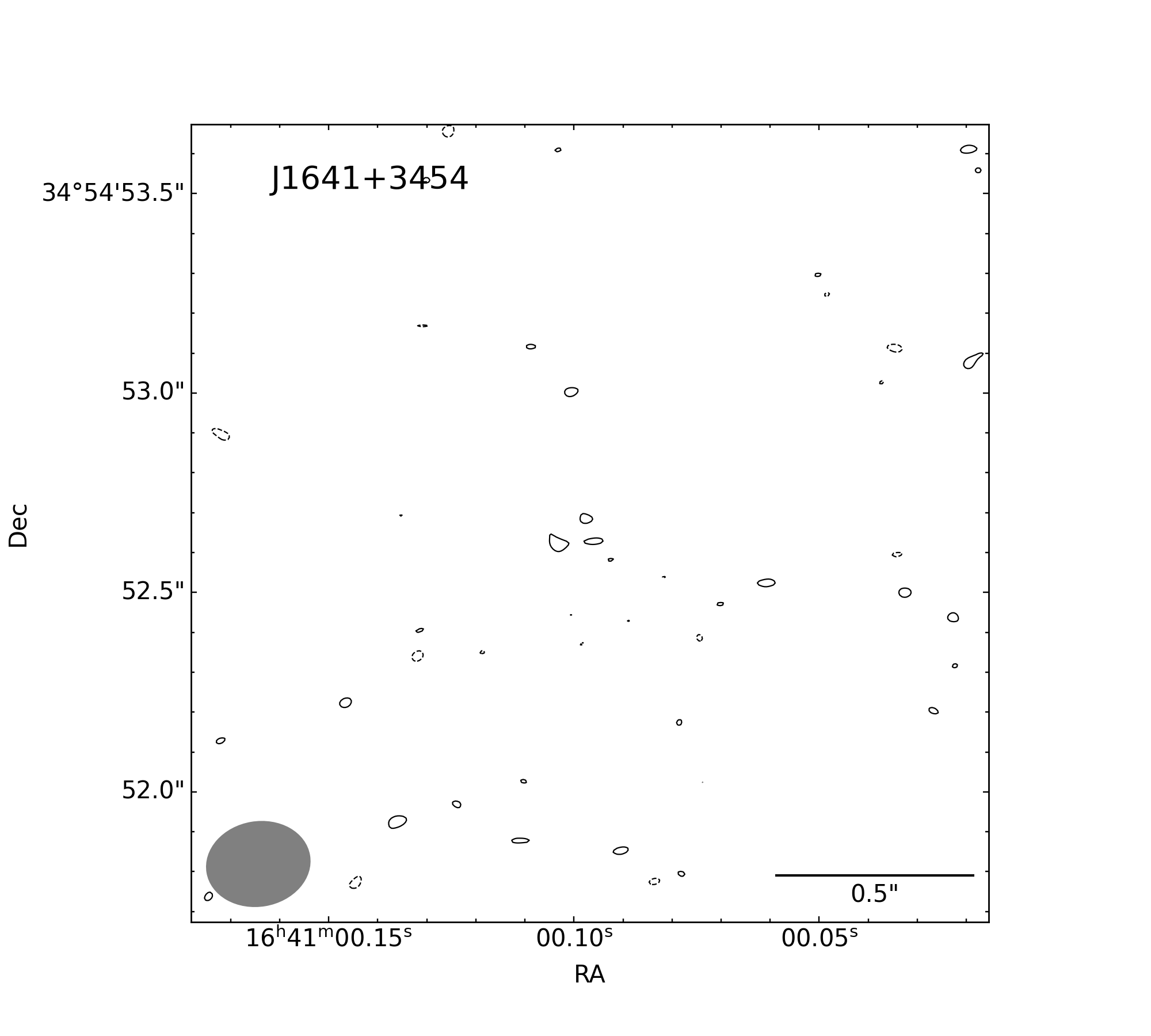}
\caption{JVLA Q band radio map of J1641+3454, rms = 33$\mu$Jy beam$^{-1}$, contour levels at -3, 3 rms, beam size 0.20 $\times$ 0.14~kpc. \label{j1641q}}
\end{center}
\end{figure}

\section{Light curves}
\label{app:lcs}

The light curves of our sources from the beginning of 2014 to mid-2022 are shown here. Figs.~\ref{j1029lc}-\ref{j1641lc} show light curves including low-resolution (MRO and OVRO) and high-resolution (JVLA, VLBA, and VLASS) data. Due to the strongly varying flux densities these plots are in logarithmic scale. The light curves in Figs.~\ref{fig:J1029lc-ul}-\ref{fig:J1641lc-ul} show only the MRO and OVRO data in linear scale, and include also the 4$\sigma$ upper limits for both observatories.

\begin{figure}
\begin{center}
\includegraphics[width=\columnwidth]{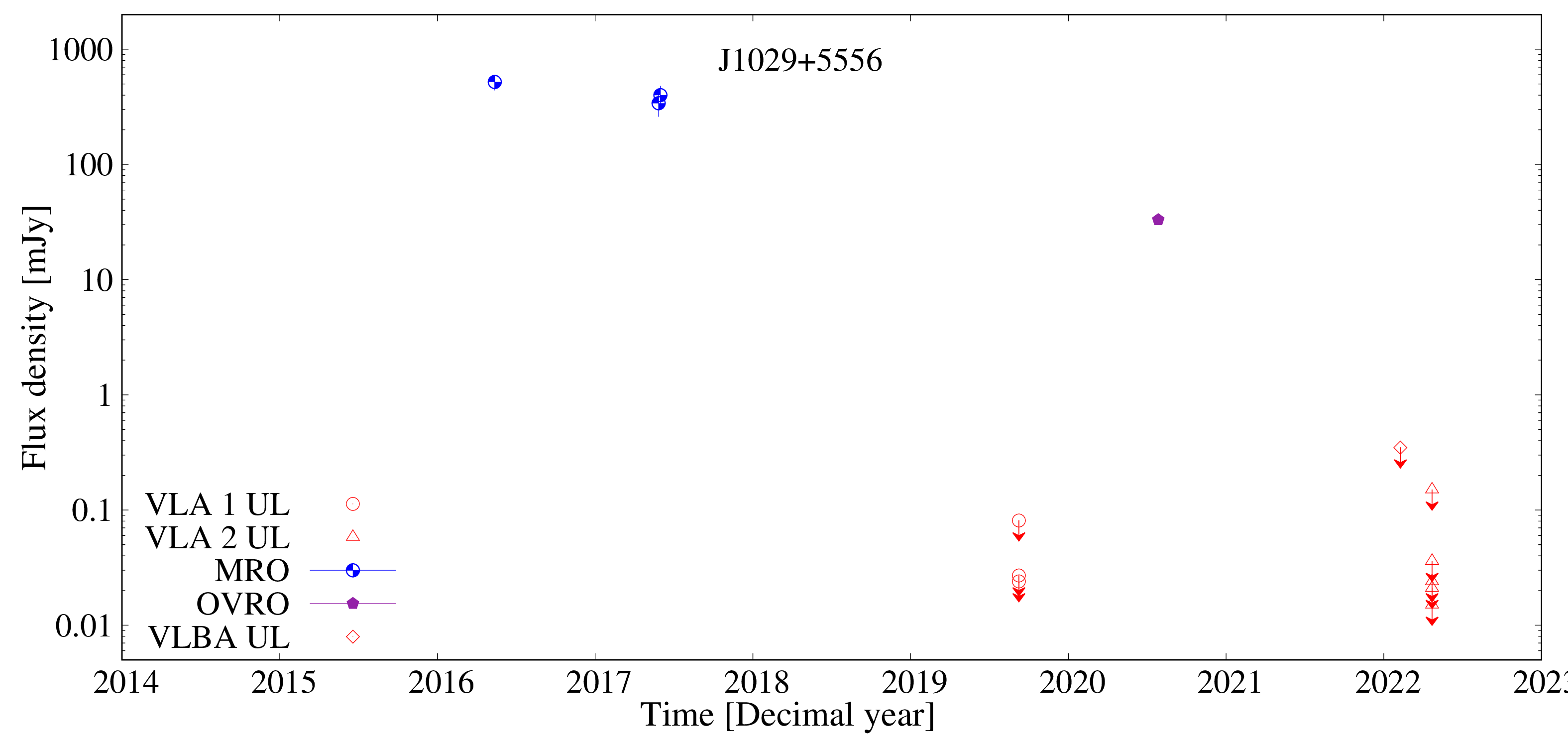}
\caption{Light curve of J1029+5556. Symbols and colours explained in the figure. Filled symbols denote integrated flux densities and empty symbols mark peak flux densities, except empty red symbols with downward arrows that are used for upper limits. VLA 1 data from \citet{2020berton2} and VLA 2 data from this paper. \label{j1029lc}}
\end{center}
\end{figure}

\begin{figure}
\begin{center}
\includegraphics[width=\columnwidth]{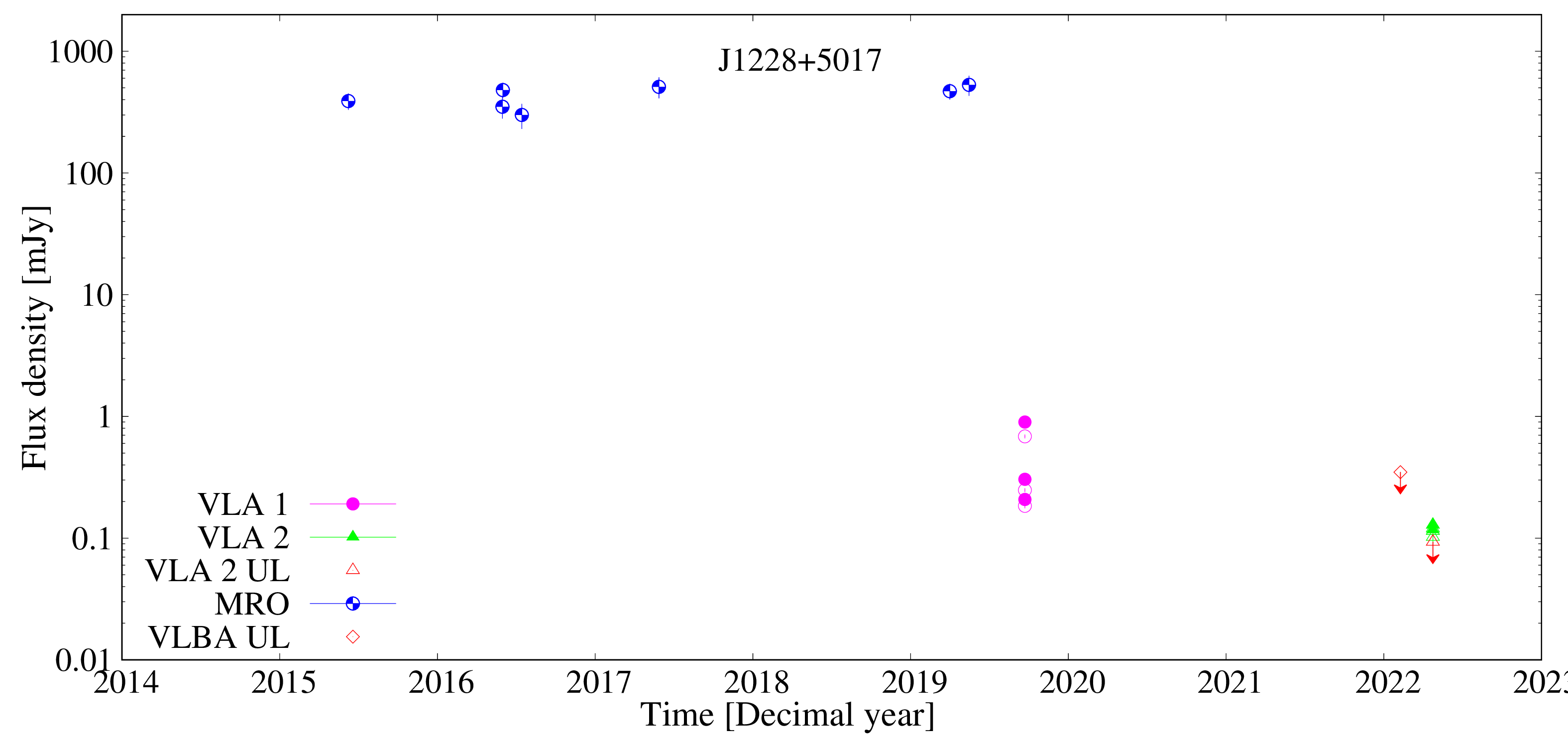}
\caption{Light curve of J1228+5017. Symbols and colours as in Fig.~\ref{j1029lc}. \label{j1228lc}}
\end{center}
\end{figure}

\begin{figure}
\begin{center}
\includegraphics[width=\columnwidth]{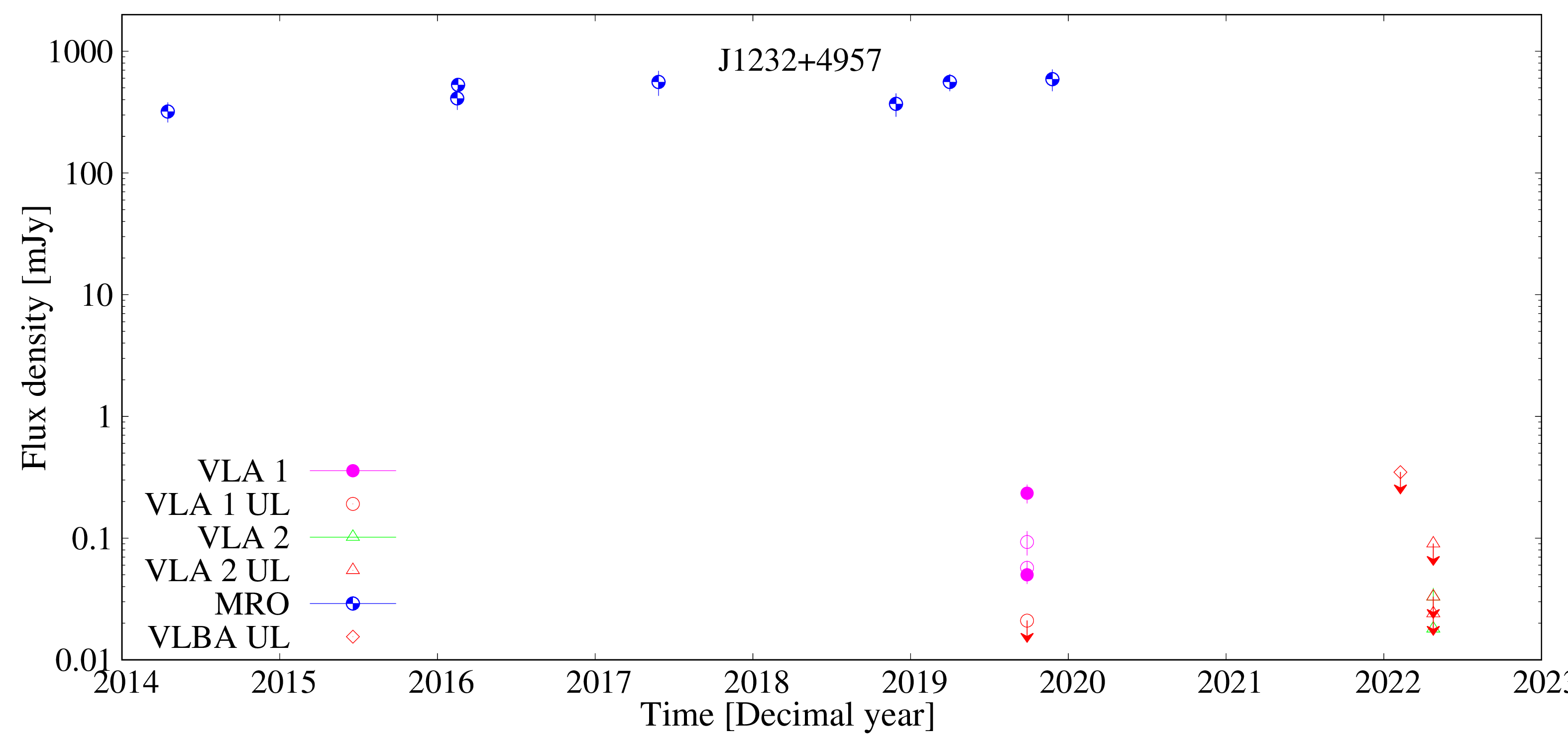}
\caption{Light curve of J1232+4957. Symbols and colours as in Fig.~\ref{j1029lc}. \label{j1232lc}}
\end{center}
\end{figure}

\begin{figure}
\begin{center}
\includegraphics[width=\columnwidth]{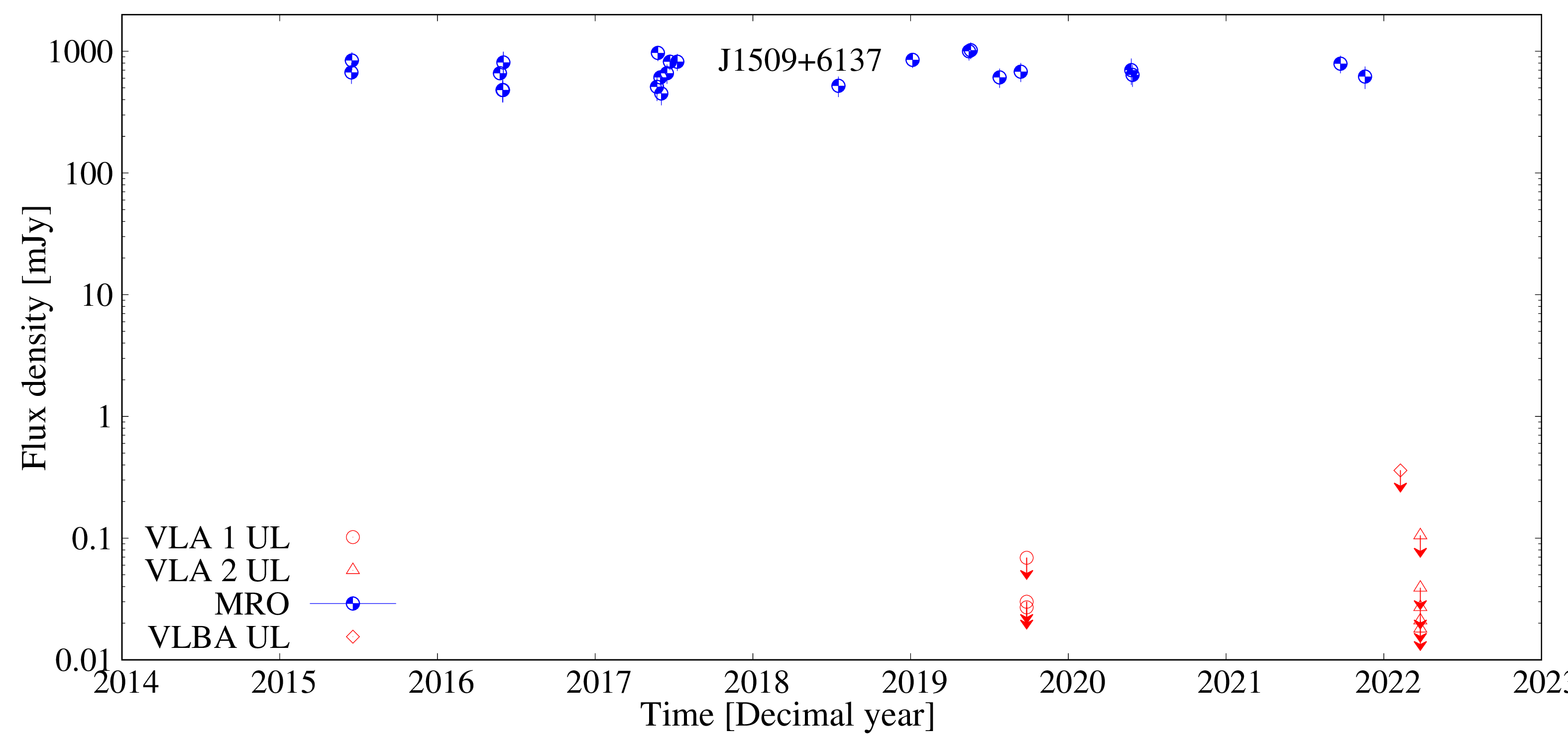}
\caption{Light curve of J1509+6137. Symbols and colours as in Fig.~\ref{j1029lc}. \label{j1509lc}}
\end{center}
\end{figure}

\begin{figure}
\begin{center}
\includegraphics[width=\columnwidth]{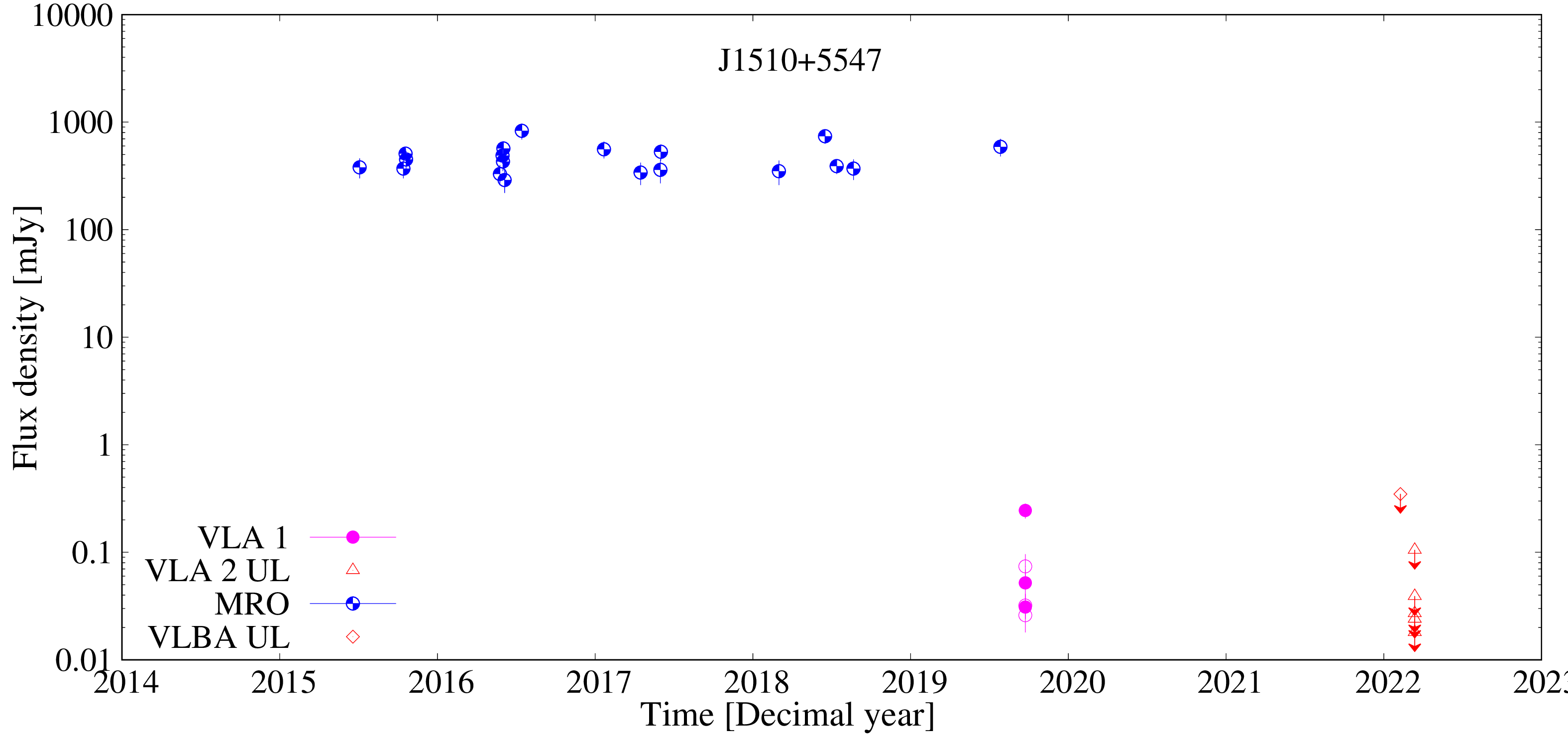}
\caption{Light curve of J1510+5547. Symbols and colours as in Fig.~\ref{j1029lc}. \label{j1510lc}}
\end{center}
\end{figure}

\begin{figure}
\begin{center}
\includegraphics[width=\columnwidth]{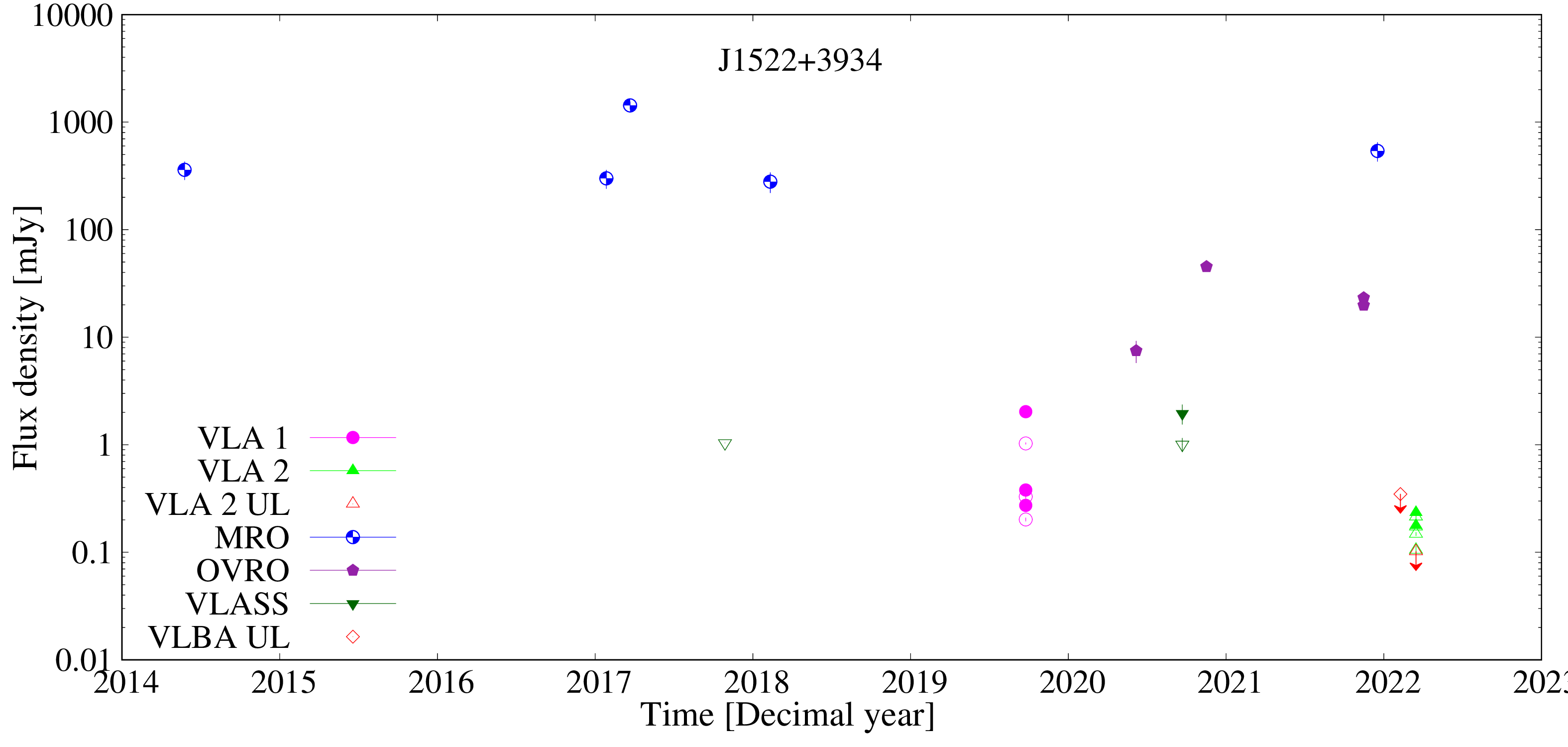}
\caption{Light curve of J1522+3934. Symbols and colours as in Fig.~\ref{j1029lc}. \label{j1522lc}}
\end{center}
\end{figure}

\begin{figure}
\begin{center}
\includegraphics[width=\columnwidth]{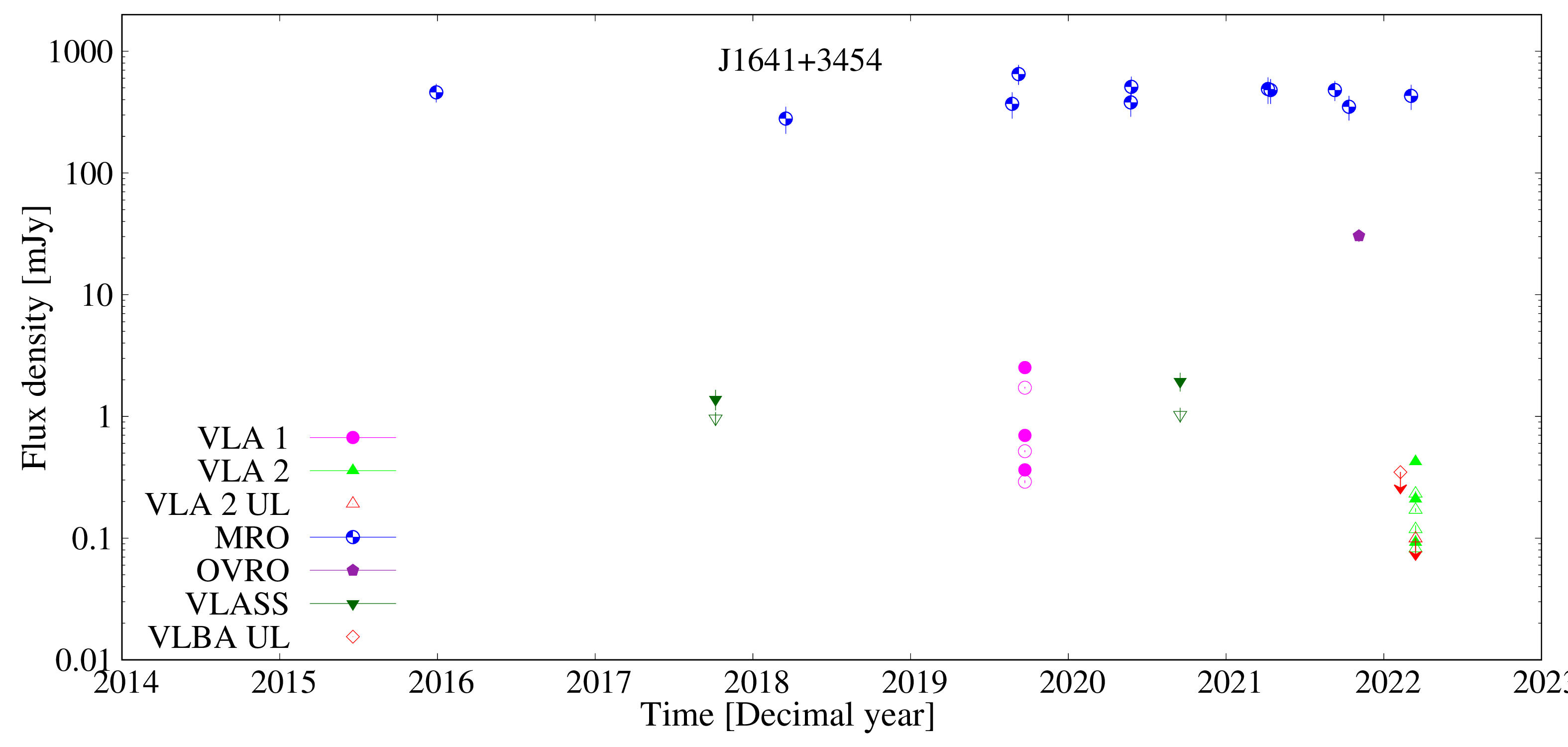}
\caption{Light curve of J1641+3454. Symbols and colours as in Fig.~\ref{j1029lc}. \label{j1641lc}}
\end{center}
\end{figure}

\begin{figure*}
    \centering
    \includegraphics[trim={0 1.6cm 0 0}, clip,width=16cm]{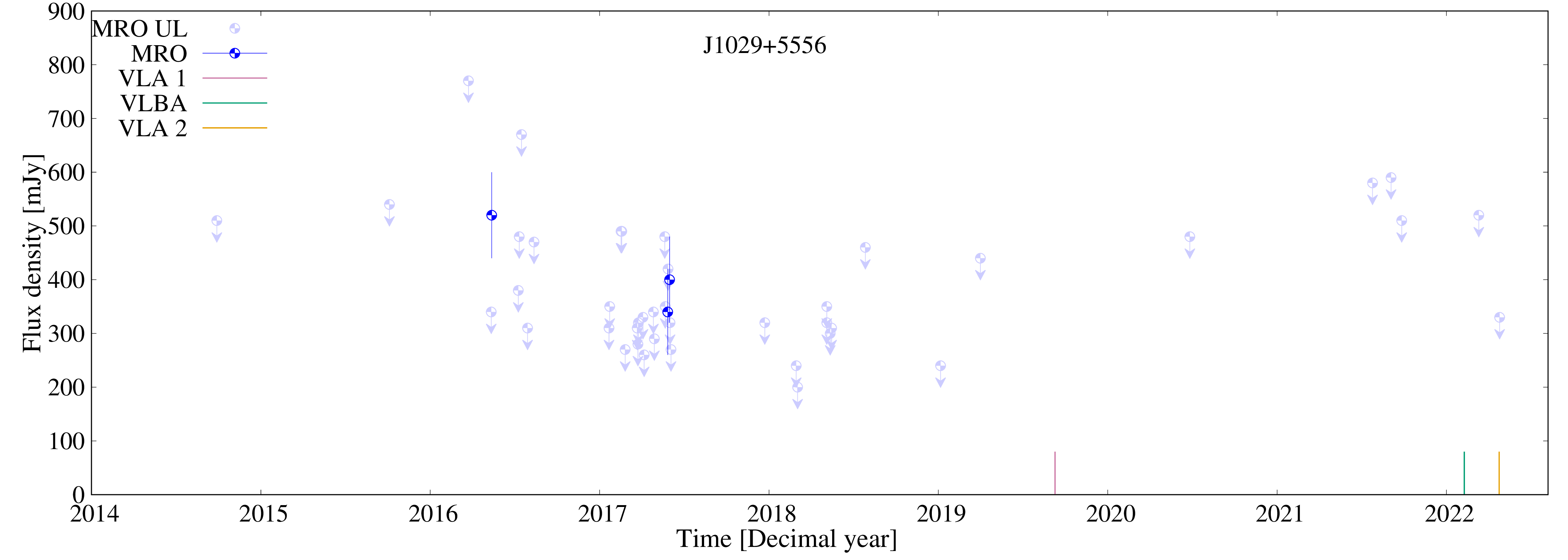}
\end{figure*}

\begin{figure*}
    \centering
    \includegraphics[width=16cm]{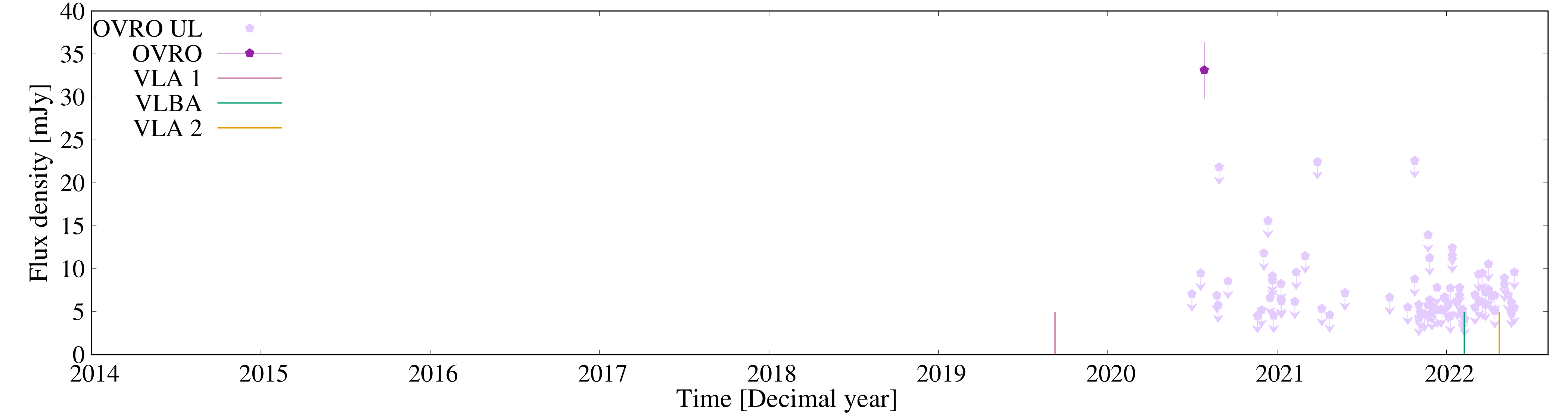}
    \caption{MRO and OVRO light curves of J1029+5556. Symbols and colours explained in the figure. Symbols with downward arrows denote upper limits, for the JVLA and the VLBA only the epochs of the observations are marked.}
    \label{fig:J1029lc-ul}
\end{figure*}


\begin{figure*}
    \centering
    \includegraphics[trim={0 1.6cm 0 0}, clip,width=16cm]{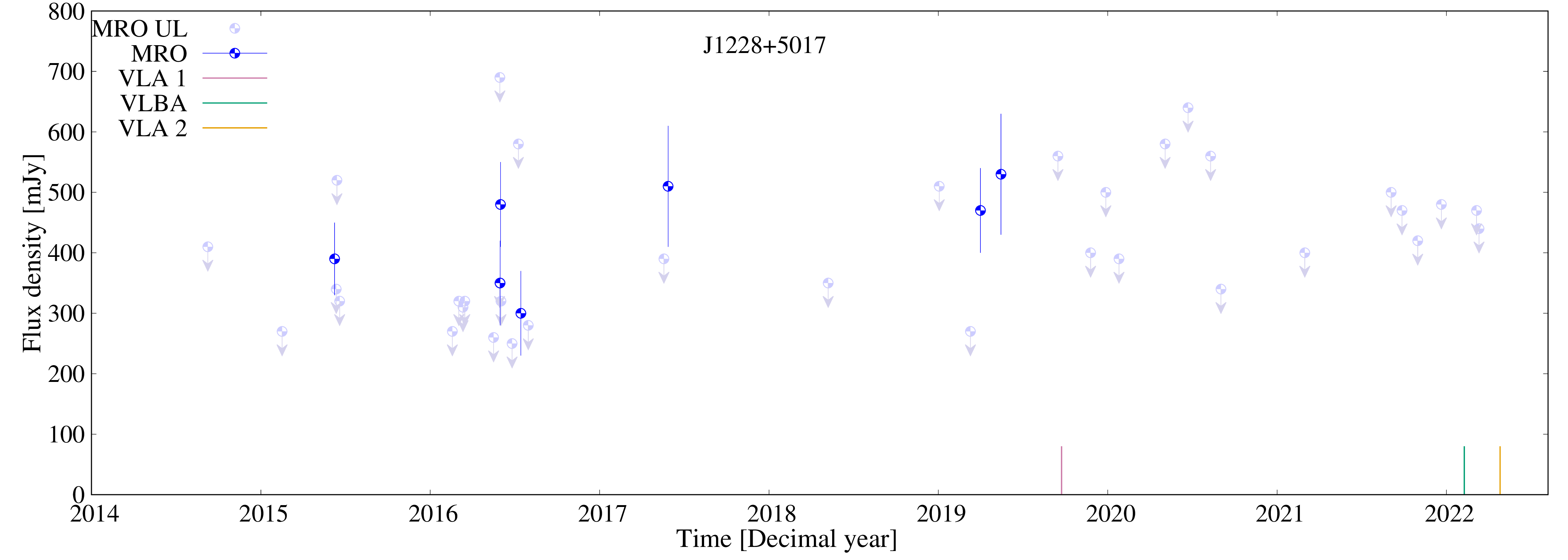}
\end{figure*}

\begin{figure*}
    \centering
    \includegraphics[width=16cm]{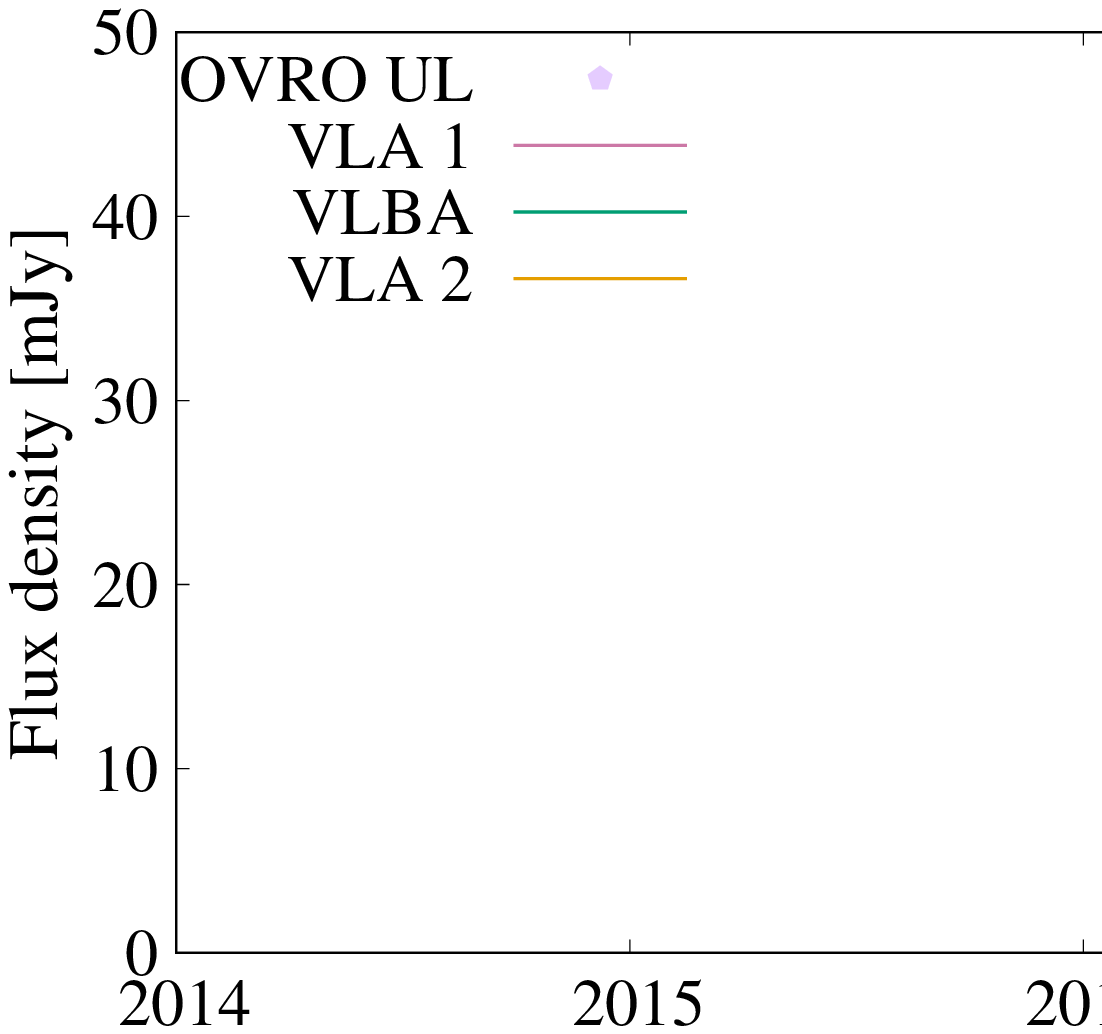}
    \caption{MRO and OVRO light curves of J1228+5017. Symbols and colours explained in the figure. Symbols with downward arrows denote upper limits, for the JVLA and the VLBA only the epochs of the observations are marked.}
    \label{fig:J1228lc-ul}
\end{figure*}


\begin{figure*}
    \centering
    \includegraphics[trim={0 1.6cm 0 0}, clip,width=16cm]{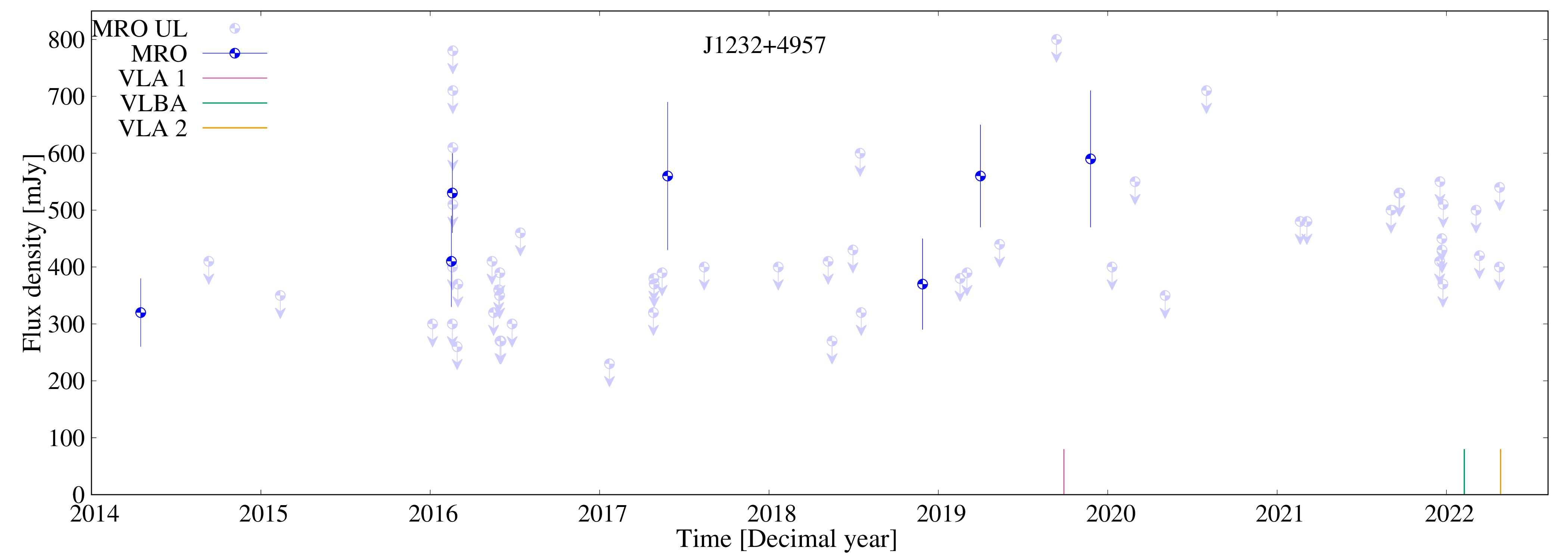}
\end{figure*}

\begin{figure*}
    \centering
    \includegraphics[width=16cm]{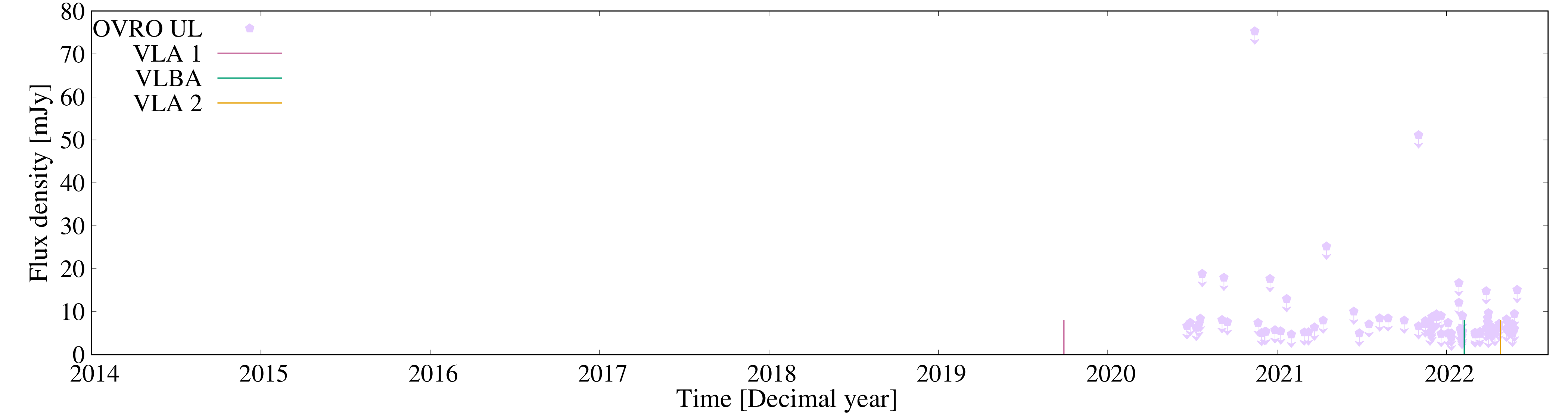}
    \caption{MRO and OVRO light curves of J1232+4957. Symbols and colours explained in the figure. Symbols with downward arrows denote upper limits, for the JVLA and the VLBA only the epochs of the observations are marked.}
    \label{fig:J1232lc-ul}
\end{figure*}


\begin{figure*}
    \centering
    \includegraphics[trim={0 1.6cm 0 0}, clip,width=16cm]{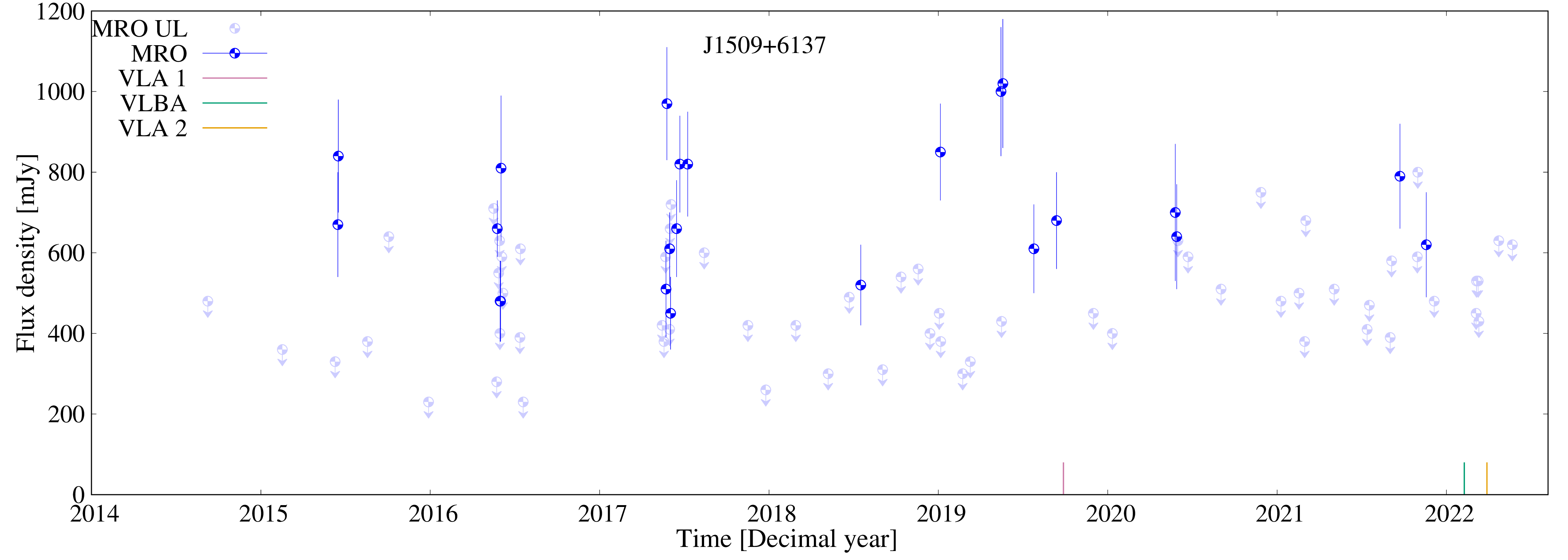}
\end{figure*}

\begin{figure*}
    \centering
    \includegraphics[width=16cm]{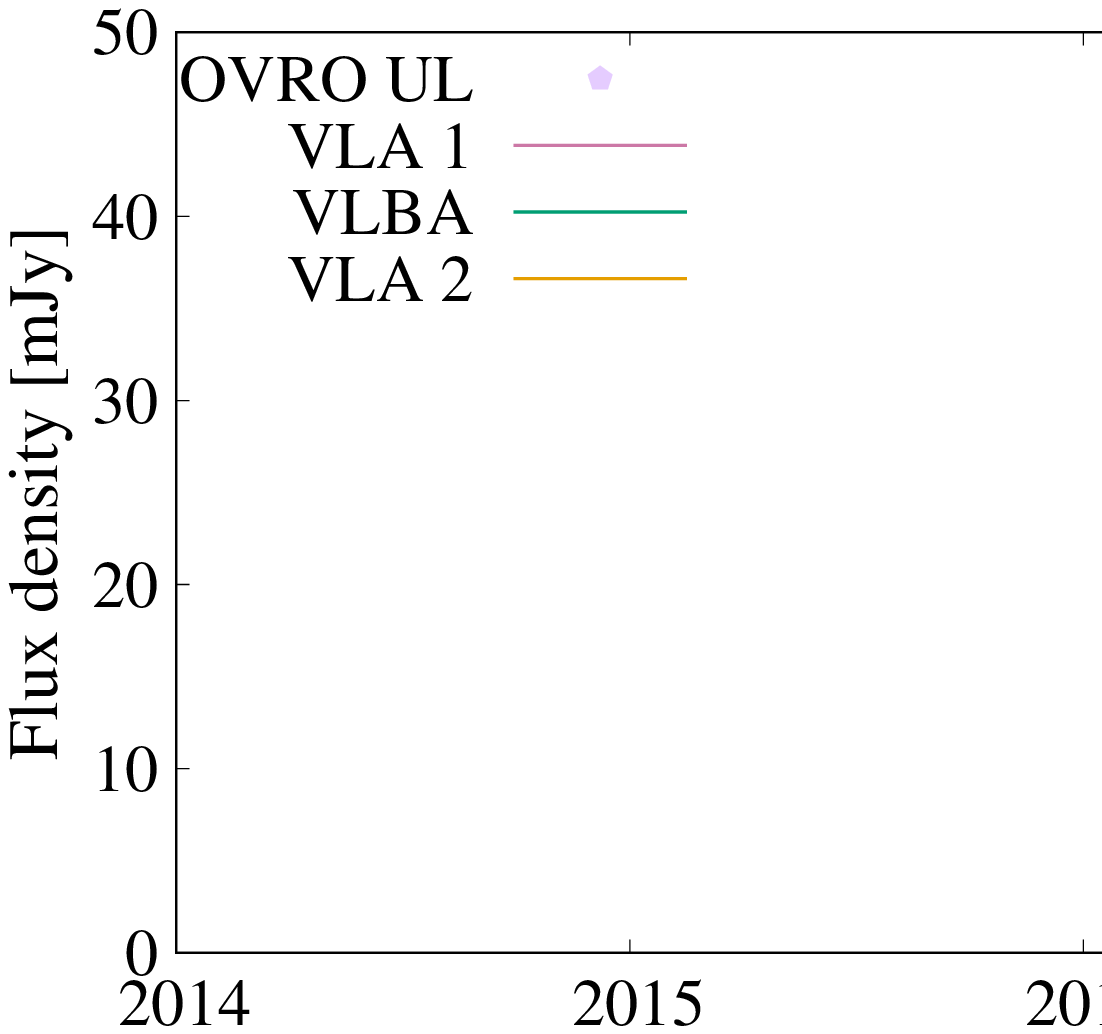}
    \caption{MRO and OVRO light curves of J1509+6137. Symbols and colours explained in the figure. Symbols with downward arrows denote upper limits, for the JVLA and the VLBA only the epochs of the observations are marked.}
    \label{fig:J1509lc-ul}
\end{figure*}


\begin{figure*}
    \centering
    \includegraphics[trim={0 1.6cm 0 0}, clip,width=16cm]{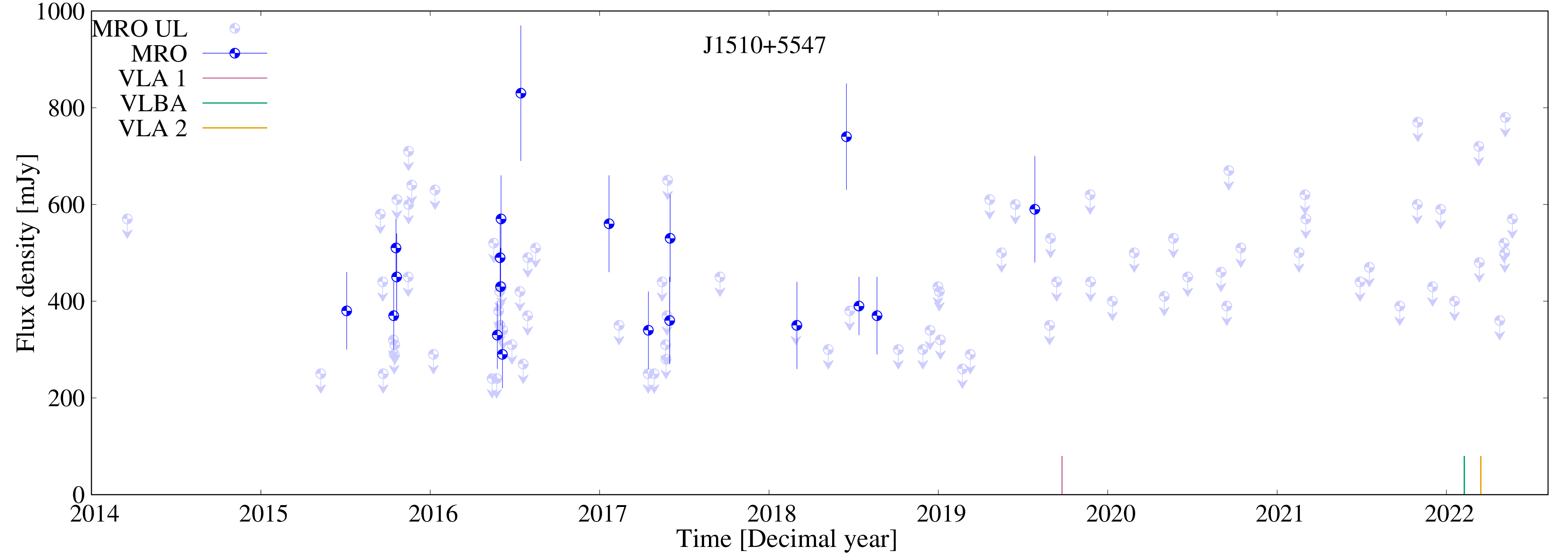}
    \end{figure*}

\begin{figure*}
    \centering
    \includegraphics[width=16cm]{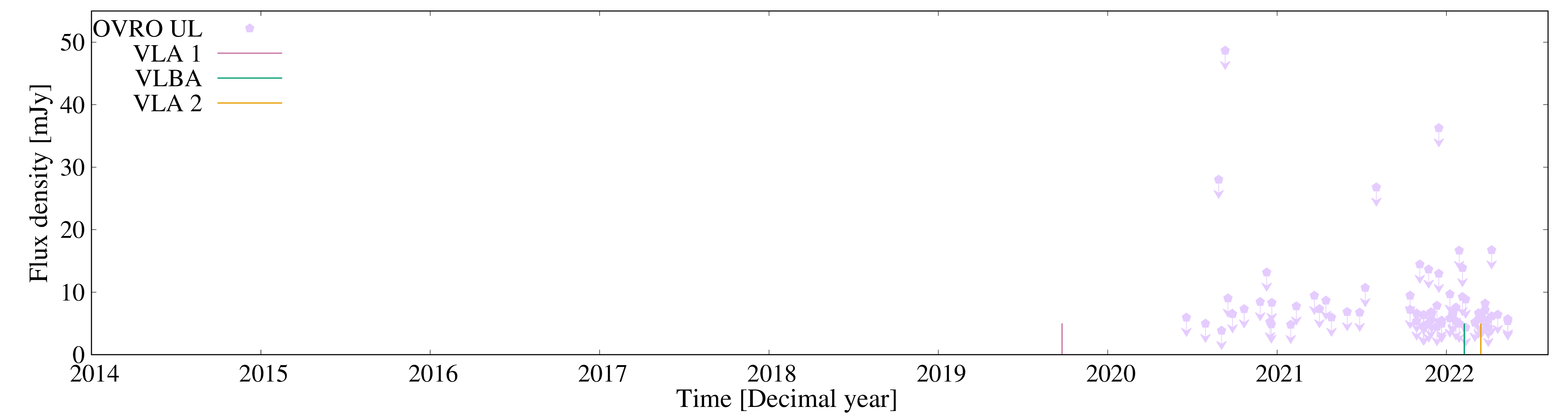}
    \caption{MRO and OVRO light curves of J1510+5547. Symbols and colours explained in the figure. Symbols with downward arrows denote upper limits, for the JVLA and the VLBA only the epochs of the observations are marked.}
    \label{fig:J1510lc-ul}
\end{figure*}


\begin{figure*}
    \centering
    \includegraphics[trim={0 1.6cm 0 0}, clip,width=16cm]{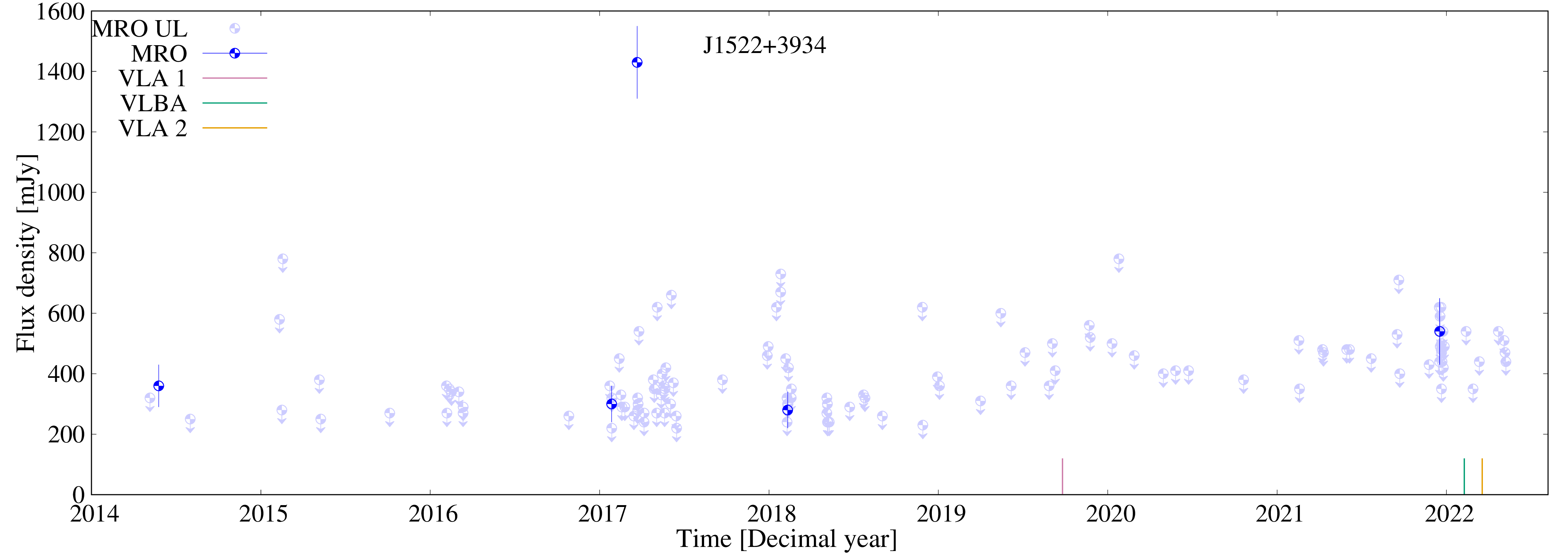}
\end{figure*}

\begin{figure*}
    \centering
    \includegraphics[width=16cm]{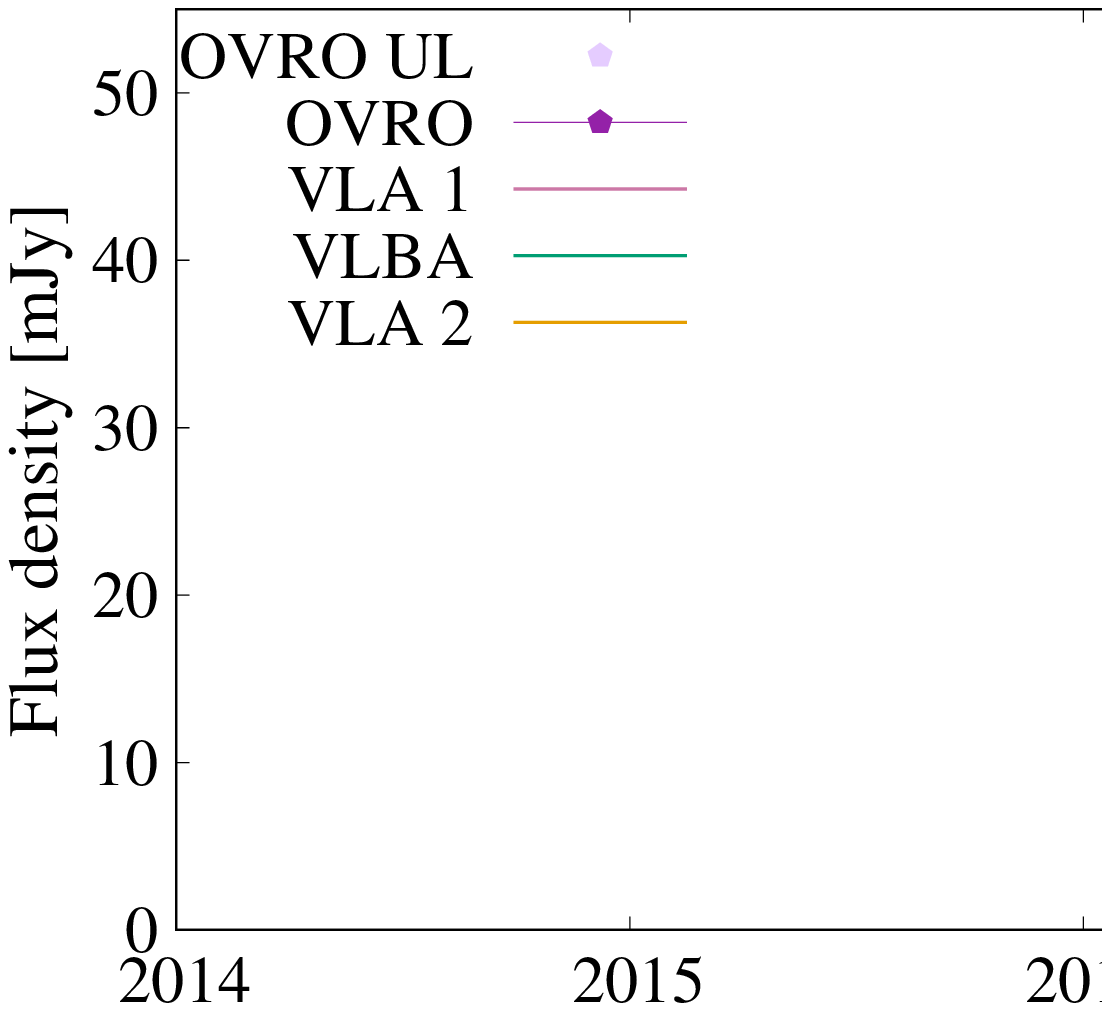}
    \caption{MRO and OVRO light curves of J1522+3934. Symbols and colours explained in the figure. Symbols with downward arrows denote upper limits, for the JVLA and the VLBA only the epochs of the observations are marked.}
    \label{fig:J1522lc-ul}
\end{figure*}


\begin{figure*}
    \centering
    \includegraphics[trim={0 1.6cm 0 0}, clip,width=16cm]{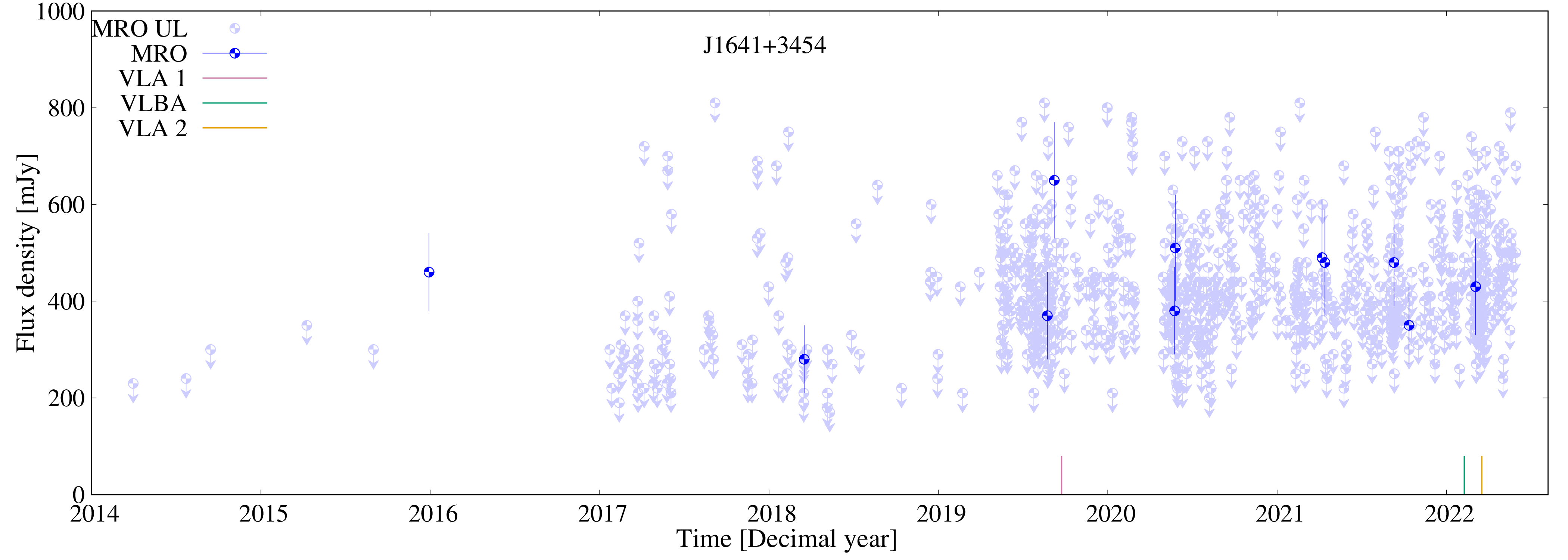}
\end{figure*}

\begin{figure*}
    \centering
    \includegraphics[width=16cm]{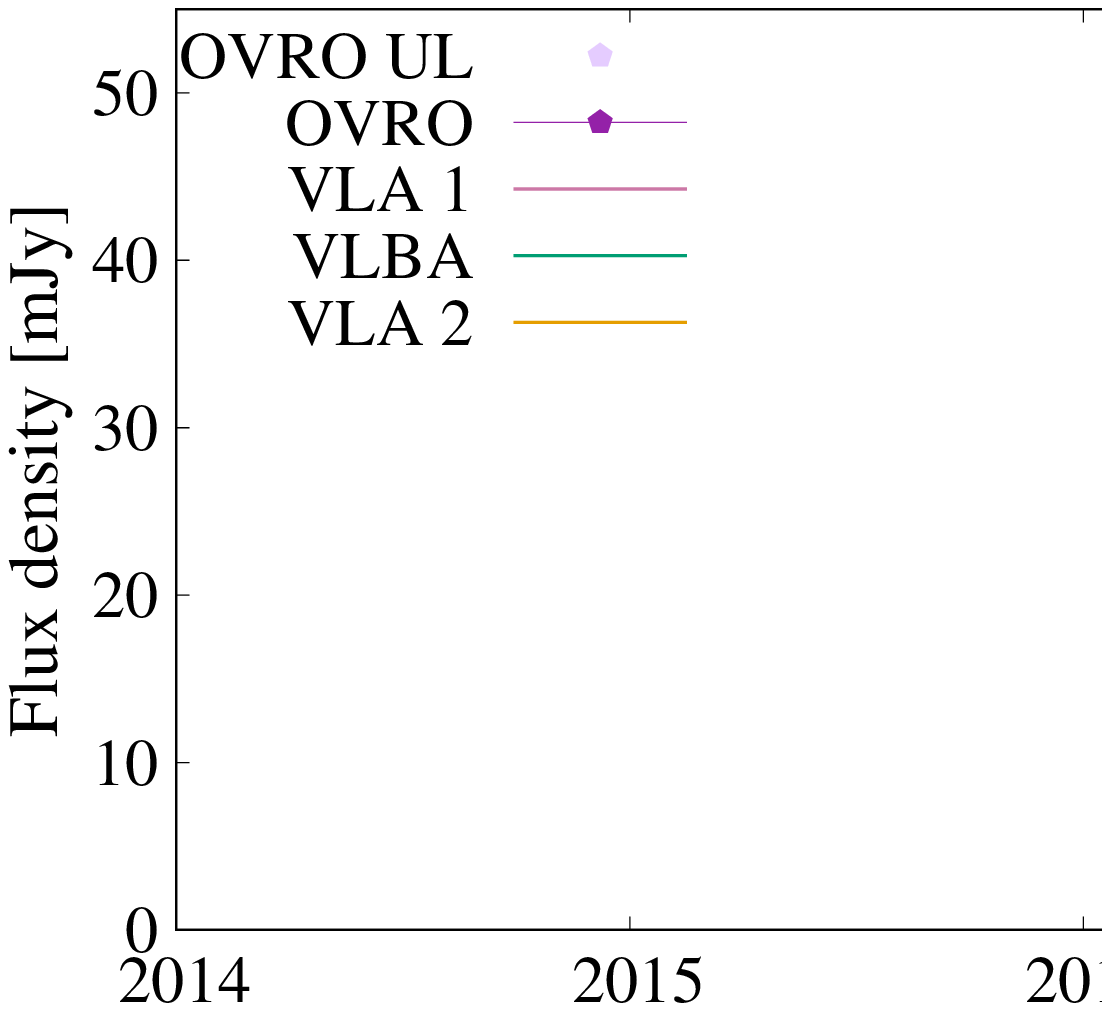}
    \caption{MRO and OVRO light curves of J1641+3454. Symbols and colours explained in the figure. Symbols with downward arrows denote upper limits, for the JVLA and the VLBA only the epochs of the observations are marked.}
    \label{fig:J1641lc-ul}
\end{figure*}

\bsp	
\label{lastpage}
\end{document}